\documentclass[useAMS,usenatbib,times,letter,amssymb]{mnras}
\usepackage{epsfig,times,amssymb,amsmath,verbatim,xspace}
\usepackage{widetext}
\usepackage[usenames,dvipsnames,svgnames,table]{xcolor}

\usepackage{todonotes}
\usepackage{tabularx}

\usepackage{hyperref}
\usepackage{float}

\usepackage{hhline}
\usepackage{lineno}
\usepackage{natbib,hyperref,ifthen,soul} 

\DeclareRobustCommand{\PRF}[3]{#2} 

\usepackage{MnSymbol,wasysym}

\newcommand{\omegam}{\ensuremath{\Omega_\mathrm{m}}}
\newcommand{\omegab}{\ensuremath{\Omega_\mathrm{b}}}

\newcommand{\as}{\ensuremath{A_\mathrm{s}}}
\newcommand{\ns}{\ensuremath{n_\mathrm{s}}}

\newcommand{\rp}{\ensuremath{r_\mathrm{p}}}
 \newcommand{\mpc}{\ensuremath{h^{-1} \mathrm{Mpc}}}

\newcommand{\inprep}{\emph{in preparation}}
\newcommand{\blockfont}[1]{{\textsc{#1}}\xspace}

\newcommand{\mbii}{\blockfont{MassiveBlack-II}}
\newcommand{\illustris}{\blockfont{Illustris-1}}
\newcommand{\tng}{\blockfont{IllustrisTNG}}

\newcommand{\cosmosis}{\blockfont{CosmoSIS}}

\newcommand{\fidsim}{\blockfont{IllustrisTNG}}

\title[IA Constraints from Hydrodynamic Simulations]{Advances in Constraining Intrinsic Alignment Models with Hydrodynamic Simulations}

\author[Samuroff et al]{
S.~Samuroff$^{1}$\thanks{ssamurof@andrew.cmu.edu},
R.~Mandelbaum$^{1}$ and
J.~Blazek$^{2,3}$
\\
$^{1}$McWilliams Center for Cosmology, Department of Physics, Carnegie Mellon University, Pittsburgh, PA 15213, USA\\
$^{2}$Department of Physics, Northeastern University, Boston, MA, 02115, USA\\
$^{3}$Laboratory of Astrophysics, \'{E}cole Polytechnique F\'{e}d\'{e}rale de Lausanne, CH-1290 Versoix, Switzerland
}

\begin{document}

\maketitle

\begin{abstract}
We use galaxies from the \tng, \mbii~and \illustris~hydrodynamic simulations to investigate the behaviour of large scale galaxy intrinsic alignments. Our analysis spans four redshift slices over the approximate range of contemporary lensing surveys $z=0-1$. We construct comparable weighted samples from the three simulations, which we then analyse using an alignment model that includes both linear and quadratic alignment contributions. Our data vector includes galaxy-galaxy, galaxy-shape and shape-shape projected correlations, with the joint covariance matrix estimated analytically. In all of the simulations, we report non-zero IAs at the level of several $\sigma$. For a fixed lower mass threshold, we find a relatively strong redshift dependence in all three simulations, with the linear IA amplitude increasing by a factor of $\sim 2$ between redshifts $z=0$ and $z=1$. We report no significant evidence for non-zero values of the tidal torquing amplitude, $A_2$, in TNG, above statistical uncertainties, although \mbii~favours a moderately negative $A_2\sim-2$. Examining the properties of the TATT model as a function of colour, luminosity and galaxy type (satellite or central), our findings are consistent with the most recent measurements on real data. We also outline a novel method for constraining the TATT model parameters directly from the pixelised tidal field, alongside a proof of concept exercise using TNG. This technique is shown to be promising, although the comparison with previous results obtained via other methods is non-trivial.

\end{abstract}

\begin{keywords}
cosmology: theory — gravitational lensing: weak – large-scale structure of Universe — methods: numerical
\end{keywords}

\section{Introduction}\label{sec:intro}
It is now well established that the weak lensing of distant galaxies 
by foreground mass provides a relatively clear window onto
the large scale structure of the Universe.
This is true whether that foreground mass is in the form of
discrete matter concentrations, as traced by galaxies
(i.e. galaxy-galaxy lensing; 
\citealt{mandelbaum13}; \citealt{leathaud17}; \citealt*{y1ggl}; \citealt{joudaki18}; \citealt{blake20}),
massive dark matter halos (cluster lensing; 
\citealt{melchior17}; \citealt{y1clusters}),
or the continuous large scale matter distribution
(cosmic shear;
\citealt{heymans13}; \citealt{svcosmology}; \citealt{y1cosmicshear}; 
\citealt{hildebrandt18}; \citealt{chang19}; \citealt{hamana19}; \citealt{asgari20}; 
see also the forthcoming DES Y3 analyses \citealt{amon20} and \citealt{secco20}).
Though the measurement method and the exact form of the theory predictions differ
slightly in the three cases, they are all fundamentally probes of the growth of structure at low redshift.   
Similarly cross correlations between galaxy lensing and other observables can be powerful probes in their own right;
recent examples include galaxy lensing $\times$ CMB lensing \citep{schaan17},
voids correlated with CMB lensing \citep{vielzeuf19}
and galaxy weak lensing crossed with gamma ray emission \citep*{ammazzalorso20},
each of which provide probes of dark matter with slightly different sensitivities. 
A measurement of cosmological weak lensing, however, is subject to a range of systematic effects; that is, observational effects that mimic a cosmological lensing signal, and so bias cosmological inference if one neglects them. Depending on the systematic in question, the most effective mitigation strategy may be quite different. In broad terms, however, the standard approach is to either (a) mitigate systematics where possible, either by applying a calibration to the data, or discarding the data points most strongly affected or (b) marginalise over them with a parametric model. Often a combination of the two is appropriate, and the prior used in (b) is informed by additional data or simulations, and detailed testing of the calibration step in (a).

This work focuses on one particular source of systematic bias,
which enters all of the weak lensing measurements described above: galaxy intrinsic alignments (IAs). 
The fact that the projected shapes of galaxies residing in the same local region of the cosmic web are correlated has
been known for many years now \citep{catelan01, heymans03}. For pairs of galaxies at the same redshift, the physically localised intrinsic shape-shape correlations can persist even on relatively large angular scales. Fortunately, in practice this signal, commonly referred to as the II contribution, is typically weak;
it is also absent, by construction, from a measurement of galaxy-galaxy lensing, which reduces the sensitivity to II further in the context of a multiprobe analysis. Often more dangerous are what are known as GI correlations, which arise due to the fact that foreground mass causes both local gravitational interactions in foreground galaxies and lensing in background objects \citep{hirata04}.

Unfortunately, many of the avenues available for understanding other lensing
systematics are not feasible in the case of intrinsic alignments. For example image simulations, which
have become an invaluable tool for quantifying shear calibration errors (\citealt*{y1shearcat}; \citealt{mandelbaum18,kannawadi19,sanchez20}) cannot be used for understanding IAs due to their fundamentally astrophysical nature. For quite different reasons, the various sophisticated methods that the lensing community has developed for
calibrating photometric redshift errors in recent years (e.g. \citealt{choi16,gruen17}; \citealt*{gatti18}; \citealt*{prat19}; \citealt{alarcon19}, \citealt*{myles20}; \citealt*{giannini20})
have limited potential for cross-use as IA mitigation tools.
Although direct mitigation methods have been proposed in the
literature \citep{heymans03,joachimi10}, 
to date these have been limited in their applicability, in
large part because they tend to rely on having good single-galaxy redshift information. They also
often focus on the (typically subdominant) II contribution (although the \citealt{joachimi10} method here can include both). The standard approach in cosmological lensing studies is to model IAs using a (semi-) physically motivated parametric model, and marginalise over its (typically $2-5$) parameters with wide flat priors. 
Given this background, hydrodynamic simulations are one of a small number of possible routes to 
understanding intrinsic alignments in cosmological lensing surveys, either for model building, or
deriving informative priors for the existing models. 
Although analytic models are relatively well motivated on very large physical scales, this is much less true on small to intermediate scales. Extending beyond this regime, then, either requires simulations or the addition of extra terms to the model, controlled by new parameters \citep{schneider10, blazek15, blazek17, fortuna20}. 
Although not the focus of this paper, another route is to use real galaxies to make a direct IA measurement (see e.g. \citealt{hirata07,mandelbaum10,joachimi11,blazek12,singh15,johnston18}). This approach avoids questions about
the realism of simulations. It does, however, have its own challenges, not least the need for
accurate per-galaxy redshift information, and the typically fairly restricted galaxy selections (often bright, red, low redshift samples)

Although a substantial amount of literature exists on the subject of IAs in hydrodynamic simulations, it is fair to say that there is significant variation in focus and methodology. For example, a series of studies by a group working on the \blockfont{Horizon-AGN} simulation have looked in detail at the alignment of two- and three-dimensional subhalo
shapes with their local large scale structure and the cosmic web
(e.g. \citealt{dubois14,codis15a,soussana19}).
Intriguingly, \citet{codis15a} found hints that blue galaxy IAs could survive in projection at a level detectable by future surveys. A number of papers based on \mbii~(e.g. \citealt{tenneti14,tenneti15,bhowmick19}) and 
\illustris~\citep{hilbert17} have explored similar themes. Minor discrepancies in the details of the IA signal, and its dependence on galaxy properties, have been uncovered; thus far, however, the interpretation of these differences has been complicated by both the relatively low signal-to-noise on large scales, and methodological
differences.

This work is intended as a step towards a more complete understanding of intrinsic alignments in hydrodynamic simulations, building on these earlier studies. We present a unified analysis of samples from various recent simulations, with measurement methods and selection functions matched in order to make a meaningful quantitative comparison. Unlike many previous studies, we focus on two-point intrinsic-galaxy and intrinsic-intrinsic alignment statistics $w_{g+}$ and $w_{++}$, which are commonly used in observational studies; this is primarily because one can derive well defined analytic predictions for them, which directly correspond to the IA modelling used in cosmological lensing analyses. This is less true of statistics like the 3D $EE$ and $ED$ correlations (e.g. \citealt{tenneti15,chisari15}), and halo misalignment statistics \citep{tenneti14,codis15b}, all of which have been used in many simulation-based studies. 
In this work we perform a simultaneous analysis of these $w_{g+}$ and $w_{++}$, alongside the equivalent galaxy-galaxy correlations, in order to fully exploit the large scale IA information in these simulated data sets.
 
The paper is structured as follows. In Section \ref{sec:data} we outline the properties of the three simulated datasets used in this work, \tng, \mbii~and \illustris, and describe the selection used to construct comparable object catalogues. Section \ref{sec:measurements} then sets out the pipeline taking us from public (stellar and dark matter) particle data and \blockfont{SubFind} group tables to shape catalogues, and eventually to two-point measurements.
The theory calculations, which we use to connect these measurements to IA models, are described in Section \ref{sec:modelling}. In Section \ref{sec:results:main} we present the results of our baseline likelihood analyses using the two-point alignment data, and then in Section \ref{sec:results:extensions} we discuss a series of extensions. We fit one of the more sophisticated alignment models in the literature, and consider the dependence of its parameters on various galaxy properties.
In addition to the two-point constraints, Section \ref{sec:daff} presents a novel method for extracting alignment information directly from the simulated matter field. We develop the basic principles, and present an example using \tng. Finally, we conclude and briefly discuss our results in the context of the field in Section \ref{sec:conclusion}.

\section{Data}\label{sec:data}
We consider three discrete cosmological simulation volumes in this study. Of these, \tng~is chronologically the most recent, and so benefits from the improvements derived from the analysis of earlier simulation efforts. The simulation runs are evolved according to Newtonian dynamics and assume similar but non-identical cosmologies, which are set out in Table \ref{tab:data:simulations}, with particles evolved from a set of initial conditions at high redshift.  In each redshift snapshot, groups are identified using the \blockfont{SubFind} friends-of-friends (FoF)
group finding algorithm \citep{springel01}.

\begin{table}
\begin{tabular}{cclc}
\hline
Simulation            & Volume              & Cosmology & Mean Gas Particle    \\
                      & / $h^{-3}$ Mpc$^3$  &           & Mass / $10^6 M_\odot$ \\
\hline
\hline
\tng                  &  $205^3$            & $\as = 2.13\times10^{-9}$ & 11.0 \\
                      &                     & $\omegam = 0.31$ &  \\
                      &                     & $\omegab = 0.05$ &  \\
                      &                     & $\ns = 0.97$ &  \\
                      &                     & $h = 0.68 $ &  \\
                      &                     & $(\sigma_8 = 0.816)$ &  \\
MBII                  &  $100^3$            & $\as = 2.43\times10^{-9}$ & 2.2 \\
                      &                     & $\omegam = 0.28$ &  \\
                      &                     & $\omegab = 0.05$ &  \\
                      &                     & $\ns = 0.97$ &  \\
                      &                     & $h = 0.70 $ &  \\
                      &                     & $(\sigma_8 = 0.816)$ &  \\
\illustris            &  $75^3$             & $\as = 2.23\times10^{-9}$ & 1.3 \\
                      &                     & $\omegam = 0.27$ &  \\
                      &                     & $\omegab = 0.05$ &  \\
                      &                     & $\ns = 0.96$ &  \\
                      &                     & $h = 0.70 $ &  \\
                      &                     & $(\sigma_8 = 0.809)$ &  \\

\hline
\end{tabular}
\caption{Properties of the simulation volumes used in this work.
The particle mass quoted in the right-most column is the mean of
gas particles. Note that in \mbii~all particles are equally weighted,
while \tng~they cover a range (see \citealt{nelson19}). $\sigma_8$ is shown in 
parentheses as it is a derived parameter.
}\label{tab:data:simulations}
\end{table}

\subsection{MassiveBlack-II}

\mbii~has been used in various previous studies, and is described in a number of existing publications; details about the approximations and modelling can be found in \citet{khandai15} and \citet{dimatteo12}. The simulation has a comoving volume of $(100 \mpc )^3$, and was generated using \blockfont{P-GADGET}, which is a version of \blockfont{GADGET3} \citep{springel05}. Initial conditions were generated with a transfer function generated by \blockfont{CMBFAST} at $z = 159$. Star formation is modelled as a binary phase process, triggered when a region of gas reaches some threshold density. Stellar particles are generated randomly from gas particles with a probability determined by their star formation rate. Stellar winds are modelled using the parametrisation of \citet{hernquist03}. AGN feedback, which is particularly relevant in high mass galaxy populations, where IAs are also strong, is also included; details of the black hole growth and AGN feedback models see \citet{khandai15}'s Sec 2.3.

\subsection{Illustris}

\illustris~is another hydrodynamic simulation whose data are now public\footnote{\url{http://www.illustris-project.org/data}}. The smallest of the three considered in this work, the box has a total comoving volume of $V=(75 \mpc )^3$, which was evolved using the moving mesh grid code, \blockfont{Arepo} \citep{weigberger19}. The various physical processes approximated in \illustris, in brief, include radiative cooling (both primordial and due to heavy elements) with self-shielding corrections; star-formation in dense regions of gas; stellar evolution with associated metal enrichment; supernova feedback and quasar-mode, radio-mode, and radiative mode AGN feedback. The above prescriptions have $\sim15$ tunable parameters, which were fixed to values obtained using a significantly smaller volume, higher resolution, set of simulations. Details of these models can be found in \citet{vogelsberger14}. 

\subsection{IllustrisTNG}

\tng~is the most recent hydrodynamic simulation included here. The particle and group data are described in the release papers \citep{springel18,nelson19}, and are available for download\footnote{\url{http://www.tng-project.org/data}}.
The \tng~data are generated using \blockfont{Arepo}. A Monte Carlo tracer particle scheme is used to to follow the Lagrangian evolution of baryonic matter. 
The hydrodynamic element comprises prescriptions for a handful of different physical processes, including emission line radiative cooling; stochastic star formation; supernova feedback and AGN feedback. The latter has two
modes (referred to as ``quasar" and ``kinetic wind" modes), depending on the accretion rate.
Details of these prescriptions can be found in \citet{pillepich17}.
It is worth remarking that \tng~is tuned explicitly to match observations at $z=0$ using a number of
statistics;
specifically the galaxy stellar mass function, the total gas mass content within the virial radius of massive groups, 
the stellar mass-stellar size and the black hole - galaxy mass relations, and the overall shape of the cosmic star formation rate density at high redshift.

\subsection{Sample Selection}\label{sec:data:cuts}

\subsubsection{Fiducial Catalogues}

To obtain a galaxy sample from which we can draw useful conclusions for each of the simulated datasets, we impose additional quality cuts. Although our measurements are not subject to the usual observational biases (due, for example, to fitting ellipticities in the presence of pixel noise, or imperfect PSF modelling), they are affected by convergence bias (e.g. \citealt{chisari15}). That is, subhalos with an insufficient number of particles to provide a meaningful shape measurement alter the ensemble ellipticity distribution of the sample. To avoid such effects, we impose a selection based on the number of particles in a galaxy (dark matter and stellar). This translates into a slightly different mass cut for each simulation due to the respective mass resolutions of the three datasets. We thus additionally impose a direct cut on stellar mass, such that the samples all have the same lower bound on $M_*$. The final selection is then:

\begin{align}\label{eq:data:selection}
&n_{\rm DM} > 1000 \notag \\
&n_{*} > 300\\
&M_{*} > 1.6 \times 10^9 h^{-1} M_{\odot}. \notag
\end{align}  

\noindent
This leaves a total of $\sim 15,000$, $35,000$ and $170,000$ usable galaxies in \illustris, \mbii~and \tng~samples respectively. Note that the cut in Eq.~\eqref{eq:data:selection} is imposed on each snapshot independently, resulting in the per-redshift numbers shown in Table~\ref{tab:data:catalogues}.

\begin{figure}
\includegraphics[width=\columnwidth]{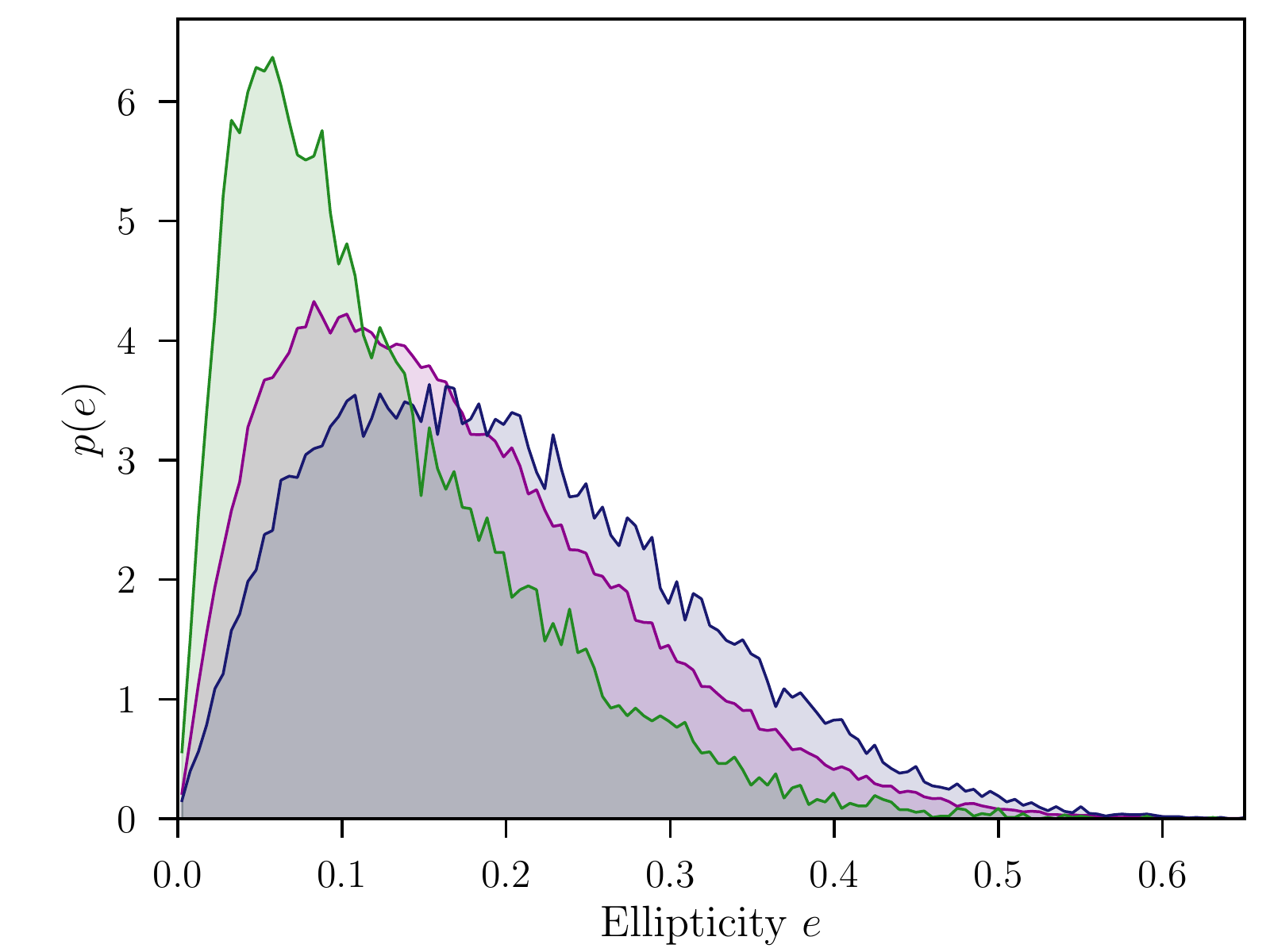}
\includegraphics[width=\columnwidth]{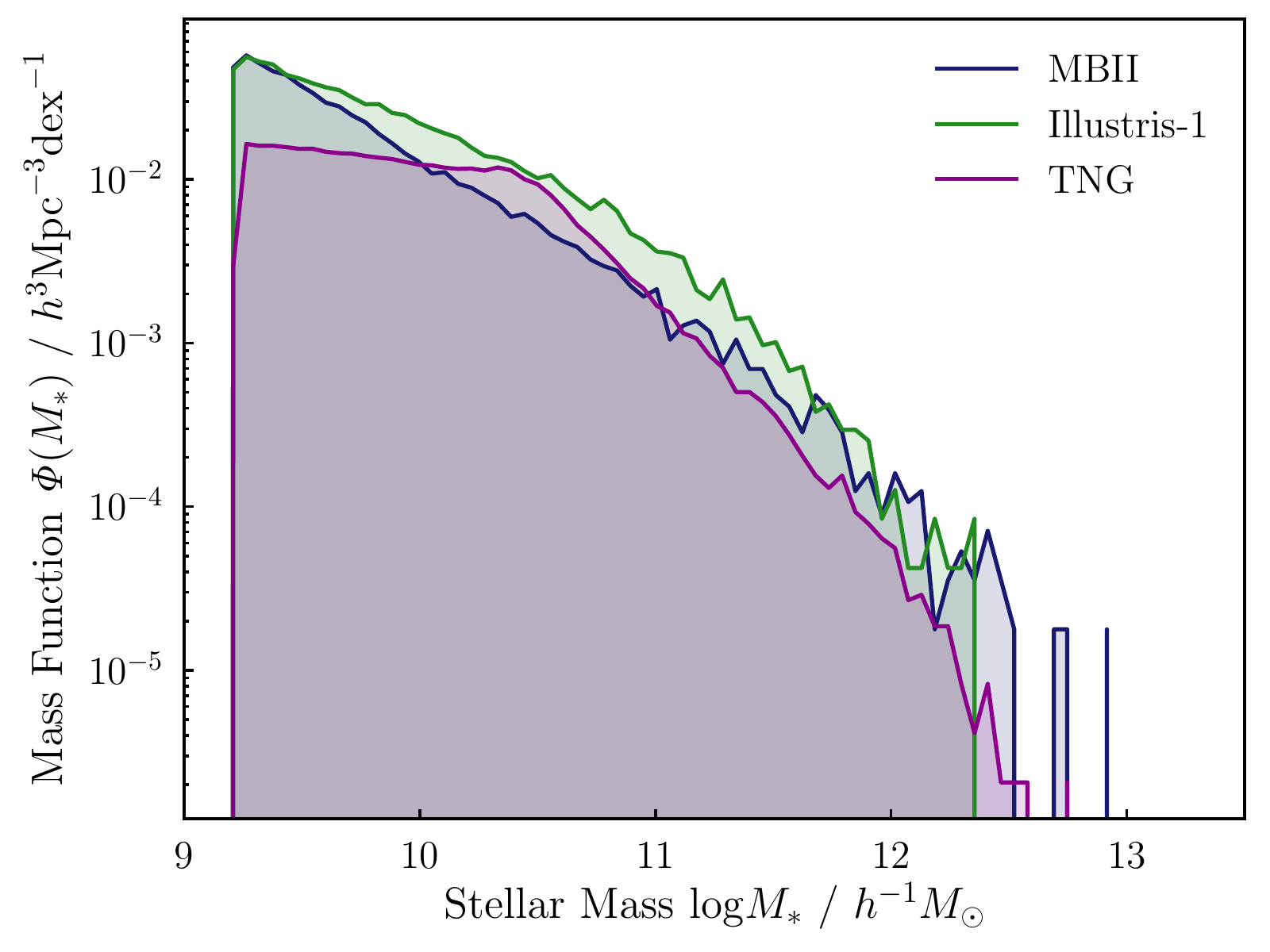}
\caption{\emph{Upper:} Normalised distributions of projected ellipticity for the 
$z=0$ samples used in this work, with the cuts described in Section \ref{sec:data:cuts}. 
Shown are \tng~(purple), \mbii~(dark blue) and \illustris~(green). 
\emph{Lower:} the stellar mass functions for the same samples.
}\label{fig:data:distributions}
\end{figure}

In Figure~\ref{fig:data:distributions} we show the ellipticity distribution and stellar mass function for each of the samples. ``Ellipticity" in this context is defined as the magnitude of the spin-2 complex ellipticity defined in Section \ref{sec:measurements:shapes}. As discussed there, the exact value for a given galaxy is dependent on
the details of the measurement method (i.e. the relative weighting of stellar matter at different radii).
Given that the measurement pipeline is applied consistently to the different simulations, however, Figure~\ref{fig:data:distributions} does allow a meaningful comparison.  
The striking discrepancy in the upper panel has been noted elsewhere (see, for example, \citealt{tenneti16}'s Figure 2); galaxies in \illustris~are significantly rounder than both comparable simulations and real data.
The differences in the mass function (lower panel) mean that, even with a common lower bound, the
mean stellar mass of the samples differs slightly (see the right-hand column in
Table~\ref{tab:data:catalogues}). At given redshift, the mean masses are ordered
(descending) \tng, \illustris, \mbii. 

\begin{figure}
\includegraphics[width=\columnwidth]{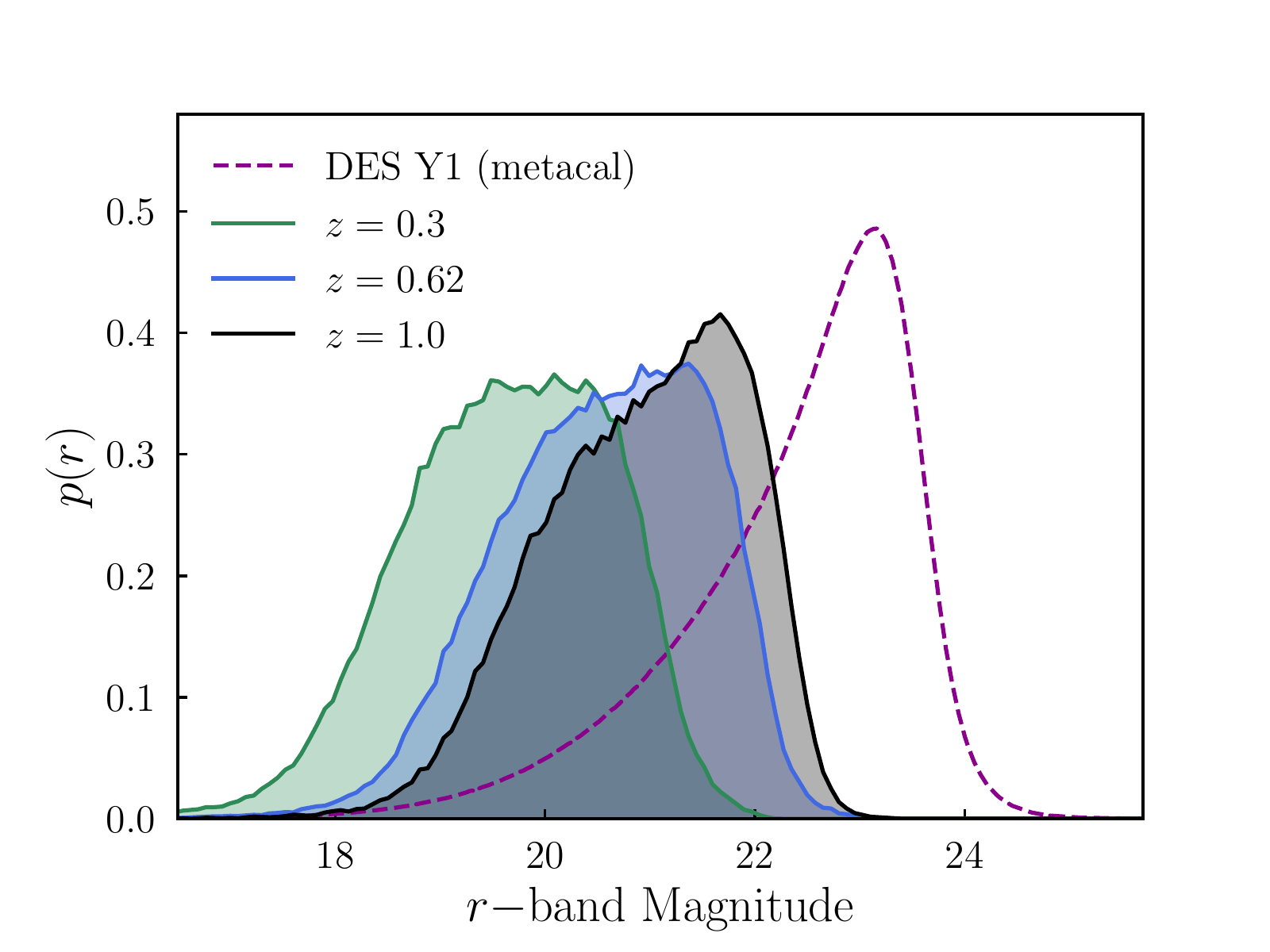}
\caption{Normalised distributions of $r-$band apparent magnitude for our \tng~sample. We convert the absolute magnitude in the catalogues to apparent magnitude at each snapshot, assuming the correct cosmology of the simulation. 
A detailed description of how the simulated absolute magnitudes 
are computed can be found in \citet{nelson18}.
A power-law approximation for the SED is used to compute the $k-$corrections
for the apparent magnitudes; although this is not rigorously correct, it is sufficient
for our purposes, given that the $k-$corrections are comfortably subdominant to the
distance modulus, and that we are only attempting a qualitative comparison here.
For reference, the unfilled curve shows the equivalent distribution for 
the fiducial Dark Energy Survey Year 1 shape catalogue, after
quality cuts (c.f.~\citealt{y1shearcat}'s Fig.~3).
}\label{fig:data:mag_distributions}
\end{figure}

\tng~is unusual amongst hydrodynamic simulations, in the sense that it has realistic galaxy magnitudes, integrated over a number of different pass bands. We include the SDSS $griz$ band magnitudes in our processed catalogues, and will use them in the following sections. Briefly, these are evaluated by summing the luminosity of star particles in a particular subhalo, and the appropriate filter band-pass is applied. More detail on this calculation can be found in \citet{nelson18}'s Sec.~3. The distribution of apparent $r-$band magnitudes in three \tng~snapshots is shown in Figure \ref{fig:data:mag_distributions}. For reference, the observed magnitude distribution from the DES Y1 \blockfont{Metacalibration} catalogue is also included (dashed purple). It is worth remembering here that, unlike the simulated data, DES is a flux-limited imaging survey, with galaxies distributed across a range of redshifts (ensemble median redshift $z\sim0.59$; \citealt{y1shearcat}), and so direct comparison is not useful; they are shown here to illustrate that the simulated galaxy samples here not representative of those in a typical lensing survey, but are a brighter subset.

\begin{table*}
\begin{center}
\begin{tabular}{ccccccc}
\hline
Simulation            & Redshift  &  Number of galaxies  & $n_c$ / $h^3 \mathrm{Mpc}^{-3}$  & Red Fraction & Satellite Fraction & Mean Stellar Mass / $10^9 M_\odot$  \\
\hline
\hline
\tng                  & 0.00      & 171,684     &  0.020      & 0.34            & 0.33 & $20.0$ \\
\tng                  & 0.30      & 168,399     &  0.020      & 0.22            & 0.32 & $18.8$ \\
\tng                  & 0.62      & 159,925     &  0.019      & 0.18            & 0.30 & $17.5$ \\
\tng                  & 1.00      & 145,394     &  0.017      & 0.12            & 0.27 & $16.0$ \\     

\mbii                 & 0.00      & 33,578      & 0.033       &  N/A            & 0.45 & $15.0$ \\
\mbii                 & 0.30      & 34,646      & 0.035       &  N/A            & 0.46 & $13.2$ \\
\mbii                 & 0.62      & 35,523      & 0.036       &  N/A            & 0.48 & $11.6$ \\
\mbii                 & 1.00      & 35,482      & 0.036       &  N/A            & 0.49 & $10.0$ \\   

\illustris            & 0.00      & 18,489      & 0.044       &  N/A            & 0.32 & $17.6$ \\
\illustris            & 0.30      & 17,203      & 0.041       &  N/A            & 0.31 & $16.8$ \\
\illustris            & 0.62      & 15,181      & 0.036       &  N/A            & 0.29 & $16.1$ \\
\illustris            & 1.00      & 12,881      & 0.031       &  N/A            & 0.27 & $15.0$ \\

\hline
\end{tabular}
\caption{Physical properties of the galaxy samples considered in this work.
The object selection is as set out in Section \ref{sec:data:cuts}, 
and is applied independently at each redshift.
Here $n_c$ (fourth column) is the comoving galaxy number density
of the sample.  
The methods used to separate red/blue and satellite/central galaxies 
are described in Sections 
\ref{sec:results:sc_split} and \ref{sec:results:colour_split} respectively. 
}\label{tab:data:catalogues}
\end{center}
\end{table*}

\subsubsection{Central Flagging}\label{sec:data:cs_split}

\begin{figure}
\includegraphics[width=\columnwidth]{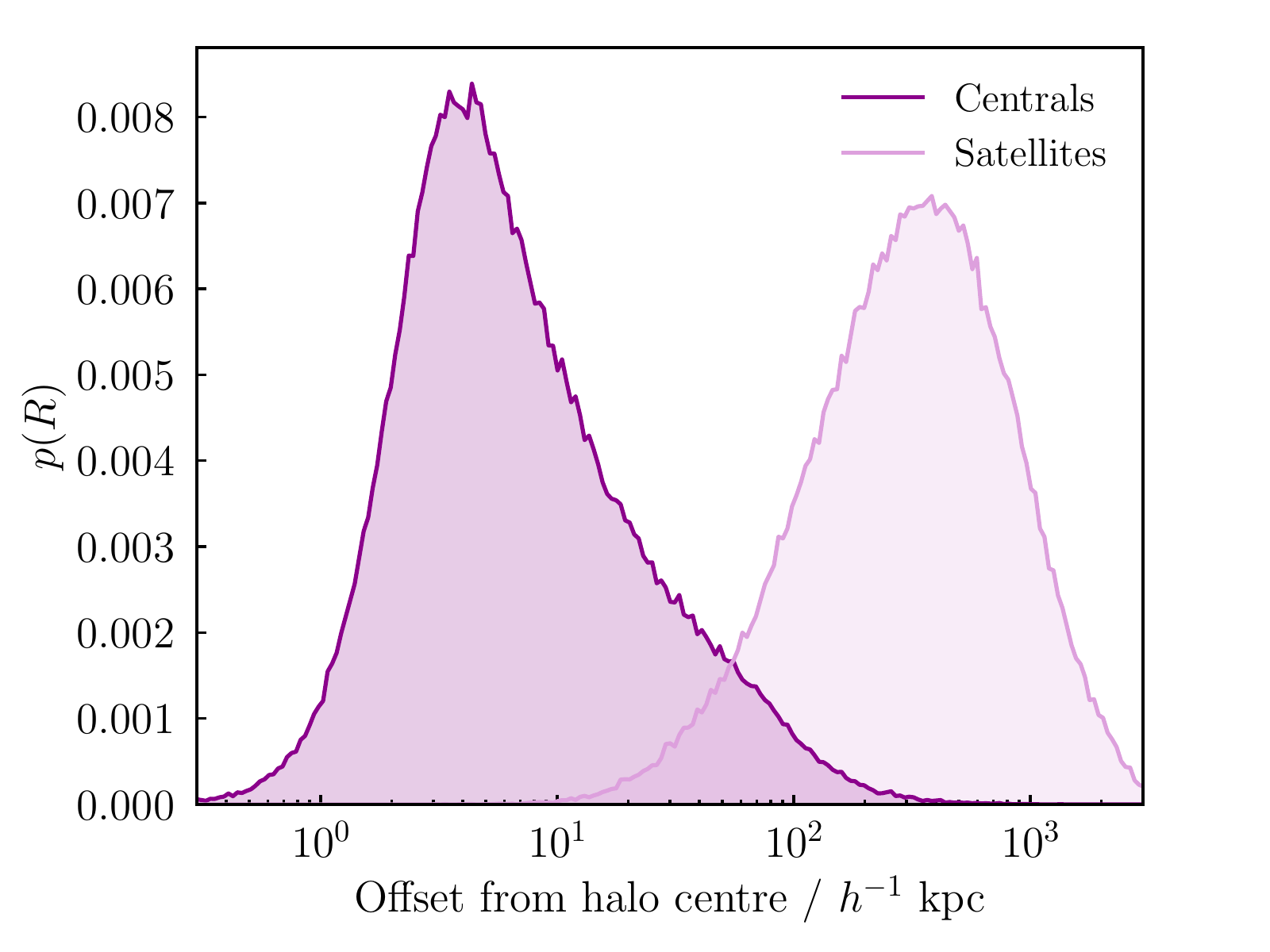}
\caption{Normalised distributions of galaxy offsets from the centre of mass of their host halos, in the \tng~simulation at $z=0$. The central flag used is defined in Section \ref{sec:data:cs_split}. Note that the distribution labelled ``satellites" is boosted by a factor of 15 for visibility.}\label{fig:data:hao_offsets}
\end{figure}

Key to halo model-based descriptions of galaxy alignments is the ability to split galaxies cleanly into satellites and centrals (see \citealt{fortuna20} for a recent example). Galaxies residing at the centres of their halo tend to be older and more massive than the satellites in the same halo; in the halo model picture, the clustering and shape properties of these two sets of galaxies is fundamentally different. For this reason it is, then, interesting to explore the behaviour of satellites and centrals separately.
For this work we simply designate the most massive galaxy in each FoF group as the central\footnote{In the nomenclature of the TNG data release,
\url{http://www.tng-project.org/data/docs/specifications/},
the central in each group is identified using the ``GroupFirstSub" flag.}.
Although noisy, this definition is less prone to misclassification than one based on geometry,
particularly in high mass groups in which the region around the bottom of the potential well is
relatively crowded. We show the distribution of galaxy-halo separations for centrals and satellites
at $z=0$ in Figure \ref{fig:data:hao_offsets}. Although not shown here, a similar pattern is seen in the higher redshift snapshots. 
That the mass based classifier is a strong indicator of galaxy position in the halo offers some reassurance that the central flagging is, in fact, literally selecting central galaxies. This is a relatively old problem, and various previous studies have explored different ways to flag central galaxies (see, for example, \citealt{rykoff16}).

\subsubsection{Galaxy Colours}\label{sec:data:colour_split}

There is much evidence in the literature to indicate that IAs are strongly dependent on galaxy 
colour \citep{joachimi11,heymans13,singh15,y1coloursplit,johnston18}. 
Clearly photometric colour is a proxy for a host of other physical properties, which ultimately determine how
strongly the galaxy sample is aligned, and one could equivalently use other properties such as morphology and bulge/disc ratio.
Although crude, a binary type split is often useful, given that mixed galaxy samples commonly exhibit a
clear bimodality in colour (or colour-magnitude) space (e.g. \citealt{baldry04, valentini18}), and that this maps roughly onto differences in IA properties.
That said, the IA signal in the simulations (or indeed any galaxy sample) is a complex function of many correlated quantities (e.g. colour, morphology, dynamical properties). 
Although it is useful to study IAs in subpopulations defined using proxies, it is worth proceeding with care, and bearing in mind that the full picture is more complicated.

Whereas quantities like stellar mass and subhalo shapes are relatively simple to obtain from hydrodynamic simulations, mapping them onto observable quantities like fluxes and colours is non-trivial. 
This has historically been a challenging problem, and there are documented deficiencies in the galaxy photometry for
\mbii~and \illustris; the equivalent quantities for \tng~are, however, thought to be fairly realistic (see e.g.~\citealt{nelson18}). In brief, in \tng~a stellar synthesis model is used to predict the stellar population of each particle in a subhalo as a function of metallicity and age. This process includes basic models for dust emission and nebular line emission. The predicted stellar spectrum is multiplied by the SDSS optical/near IR $ugriz$ band-passes (airmass 1.3), producing a magnitude in each filter. The per-particle magnitudes are then summed over the ensemble bound to the subhalo. This process is explained in more detail in \citealt{nelson18}'s Section 3 (see their ``Model (A)").

\begin{figure}
\includegraphics[width=\columnwidth]{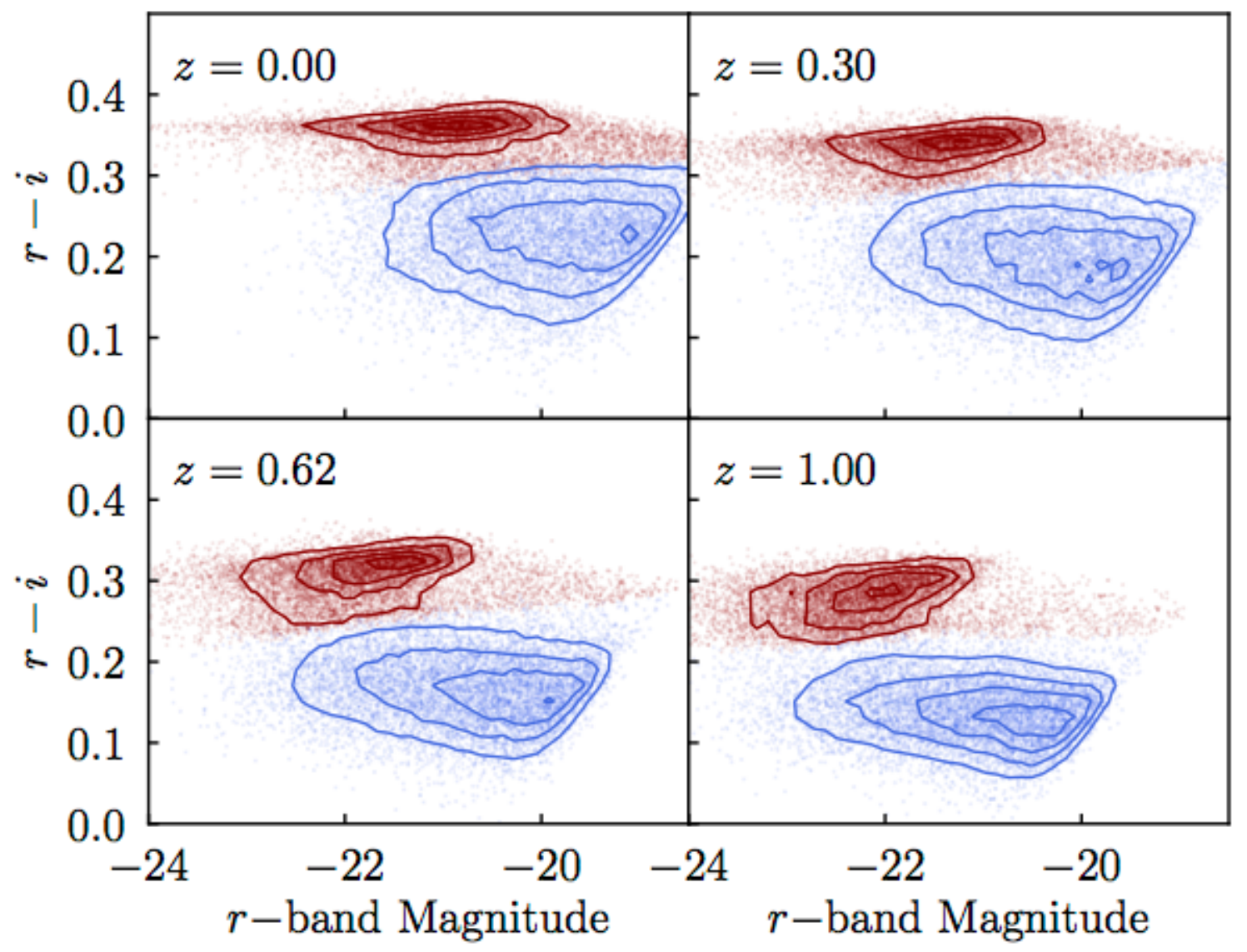}
\caption{Colour magnitude diagrams for our \tng~sample at four redshifts (labelled upper left). The two sets of contours show the distributions of the red and blue samples, as outlined in the text. Note that the split is imposed in $g-i$ versus $r$ space, which is why the division is not sharp. The fine points show a randomly downsampled selection of galaxies from each population. Note that unlike in Figure \ref{fig:data:mag_distributions}, the magnitudes used here (including for estimating colours) are absolute, not apparent, ones.
}\label{fig:results:ri_rmag_rbsplit}
\end{figure}

We inspect the colour-magnitude diagrams and make a linear division in $g-i$ colour space

\begin{equation}
(g-i) = m_{\rm gi} \times r + c_{\rm gi},
\end{equation}

\noindent
to roughly mimic the green valley division. The colour magnitude diagram evolves with redshift, and so
we carry out this process independently in each snapshot, giving $\mathbf{m}_{\rm gi} =
(0.045,0.045,0.055, 0.022)$, $\mathbf{c}_{\rm gi} = (1.84,1.84,1.95,1.19)$.
The red fraction resulting from this split at each redshift is shown in Table \ref{tab:data:catalogues}. The numbers
here are roughly consistent with those seen in real data, and change with redshift in an intuitively correct way 
(i.e., the low redshift Universe has a larger abundence of massive red elliptical galaxies compared with $z=1$).
The $r-i$ colour magnitude diagram for our split sample is shown in Figure \ref{fig:results:ri_rmag_rbsplit}.
Given that the split is imposed in $g-i$, it is somewhat reassuring that we see clearly defined well separated samples in this space.

\section{Measurements}\label{sec:measurements}
\subsection{Galaxy Shapes}\label{sec:measurements:shapes}

In a three dimensional cosmological volume, the most natural way to quantify a galaxy's shape is via its intertia tensor. Analogous to projected ellipticities, which are constructed from the moments of a galaxy light profile, the most general form for the inertia tensor is:

\begin{equation}
I_{ij} = \frac{1}{W} \sum^{N_{\rm p}}_{k=1} w_{k} x_{i,k} x_{j,k},
\end{equation}

\noindent
where the indices $ij$ indicate one of the three spatial coordinate axes $i,j \in (x, y, z)$, and the sum runs over the number of particles within the subhalo. For our purposes, this means star particles, but one could equivalently estimate the shape of the dark matter subhalo using the same equation. The prefactor $w_{k}$ is the weight allocated to particle $k$, and $W$ is the sum of the weights;
in the case of \mbii, all of the star particles have the same mass, and so the weights are flat. In \tng~and\illustris, this is not the case, and each particle is weighted by its mass. 
An alternative, known as the reduced inertia tensor (see \citealt{chisari15},
\citealt{tenneti16}), weights particles by their inverse square distance from the subhalo
centroid. This process is known to bias the measured ellipticities low, necessitating a further
iterative correction procedure. Although we mention this here for context, since it has been used a
handful of times in the literature, it is not used in this work. 
Note also that we have reason to think the IA signal is, in reality, dependent on the radial weighting of the shape measurements, an effect that has been observed in real data \citep{singh16}.

By performing an eigenvalue decomposition on $I$, one can obtain three dimensional axis vectors and lengths,
which in turn can be projected into the 2D second moments $Q_{xx}, Q_{yy}, Q_{xy}$. The recipe is
set out by \citet{piras18} (see their Eq 13-15), and we refer the reader to that paper for the
mathematical detail.
Although for technical reasons our pipeline goes via three dimensional shapes, it is also worth noting that one could also simply measure the projected two dimensional moments of a subhalo directly.  
With the projected moments, one can then construct the spin-2 ellipticity of a galaxy as

\begin{equation}
(e_1,e_2) = \frac{(Q_{xx}-Q_{yy}, 2Q_{xy})}{Q_{11} + Q_{22} + 2 \sqrt{|\mathbf{Q}|}}.
\end{equation}

\noindent
It is worth bearing in mind that there are in fact two common ellipticity definitions used for weak lensing. 
The one defined above is equivalent to an ellipticity magnitude, written in terms of (projected) axis ratios, $e=(a-b)/(a+b)$; for detailed discussion of both this and the alternative ellipticity definition, and their respective advantages, see \citet{melchior12}.
Note that this is a Cartesian projection along one axis of the simulation box, not a lightcone projection with conversion to angular coordinates. The positive and negative $e_1$ direction, then, is defined by the $x,y$
coordinate directions of the square simulation volume. Although this measurement does not correspond directly to what one could do in reality, the difference is not thought to be significant, given the statistical size and other limitations of the samples considered in this
work.

\subsection{Two-Point Correlations}\label{sec:measurements:2pt}

All correlation functions used in this paper are computed using the public \blockfont{halotools} 
package\footnote{\url{https://github.com/duncandc/halotools_ia}}\footnote{\url{https://halotools.readthedocs.io; v0.7}} \citep{hearin17}. 
The most straightforward (and highest signal-to-noise) two-point measurement one could
make is that of galaxy clustering in three dimensions.
We adopt a common estimator of the form \citep{landy93}:

\begin{equation}
\xi^{ij}_{gg}(r_{\rm p}, \Pi) = \frac{D^i D^j - D^iR^j - D^jR^i + R^iR^j}{R^iR^j},
\end{equation}

\noindent
where $DD$, $RR$ and $DR$ are weighted counts of galaxy-galaxy, random-random and galaxy-random pairs, binned in perpendicular and line-of-sight separation, \rp and $\Pi$. The indices $i,j$ denote a pair of catalogues (either galaxy positions, or random points), which are correlated together. In both cases above, $R$ represents the positions of a set of random points drawn from a flat distribution within the simulation volume. 

The cross correlation of galaxy positions and intrinsic ellipticities, $\xi_{g+}(r_\mathrm{p}, \Pi)$, can similarly be estimated, as a function of $r_\mathrm{p}$ and $\Pi$. We use a modified Landy-Szalay estimator of the form:

\begin{equation}
\xi^{ij}_{g+} (r_\mathrm{p}, \Pi) = \frac{S^i_+D^j - S^i_+R^j}{R^iR^j}
\end{equation}

\noindent
(see \citealt{mandelbaum10}). One can similarly measure the shape-shape correlation:

\begin{equation}
\xi^{ij}_{++} (r_\mathrm{p}, \Pi) = \frac{S^i_+S^j_+}{R^iR^j}.
\end{equation}

\noindent
The terms in the numerator represent shape correlations and are defined as

\begin{equation}
S^i_+D^j \equiv \frac{1}{2} \sum_{\alpha\neq \beta} w_{\alpha} w_{\beta} e_{+}(\beta|\alpha),
\end{equation}
\begin{equation}
S^i_+S^j_+ \equiv \frac{1}{4} \sum_{\alpha\neq \beta} w_{\alpha} w_{\beta} e_{+}(\alpha|\beta) e_{+}(\beta|\alpha),
\end{equation}

\noindent
where the indices $\alpha,\beta$ run over galaxies and $e_+(\beta|\alpha)$ is the tangential ellipticity
of galaxy $\beta$,
rotated into the coordinate system defined by the separation vector with galaxy $\alpha$.
For the fiducial catalogue, \tng, the weights are equal and normalised to the number
of galaxies. 
In order to make a direct comparison of the different samples, galaxies in the \mbii~and \illustris~catalogues are assigned weights, such that the host halo mass distributions of the three match. For detail about the weighting scheme, which we refer to as halo-mass reweighting, and discussion about the impact on our results, we refer the reader to Appendix~\ref{app:weights}. IAs are known to be dependent on cosmology and the host halo mass distribution, and this process should remove differences due to discrepancies in these factors. It is also true, however, that other properties such as the details of the galaxy-halo connection and the properties of the galaxies themselves also potentially have an impact. Such differences represent a systematic uncertainty (since we cannot say with certainty which of the simulations, if any, represents reality, nor straightforwardly homogenise them), and so any resulting differences should be treated as such.

In lensing studies it is also common to assign galaxies something approximating inverse variance (shape noise + measurement uncertainty) weights (see, for example, \citealt{y1shearcat}). Since this weighting tends to upweight bright, high $S/N$ galaxies, it seems likely it would also boost the IA signal. That said, in practice lensing weights tend to be shape noise dominated, and so relatively uniform across the sample, meaning the magnitude of this effect is expected to be small.

From these three dimensional measurements, obtaining the two dimensional projected correlations is a case of integrating along the line of sight. One has,

\begin{equation}\label{eq:xi_to_w}
w_{ab} (r_\mathrm{p}) = \int_{-\Pi_\mathrm{max}}^{\Pi_\mathrm{max}} \mathrm{d}\Pi \xi_{ab} (r_\mathrm{p}, \Pi),
\end{equation}

\noindent
Here the lower indices $ab$ denote a type of two-point correlation, $a,b\in(g,+)$. 
$\Pi_\mathrm{max}$ is an integration limit, which is set by the simulation volume.
For this study we adopt a value equal to a third of the box size, For this study we adopt a value equal to a third of the box size, 
or $\Pi_\mathrm{max}=68 h^{-1}$ Mpc for \tng,
$\Pi_\mathrm{max}=33 h^{-1}$ Mpc for \mbii,
and $\Pi_\mathrm{max}=25 h^{-1}$ Mpc for \illustris. 
In practice, for our purposes we wish to maximise $\Pi_\mathrm{max}$; although it is true that very long baselines will eventually harm the signal-to-noise by including uncorrelated pairs, on scales of a few tens of Mpc we are well within the regime where extending $\Pi_{\rm max}$ helps to access additional large scale signal modes (see \citealt{joachimi11}'s App. A2 for further discussion).

 \begin{figure*}
\includegraphics[width=2\columnwidth]{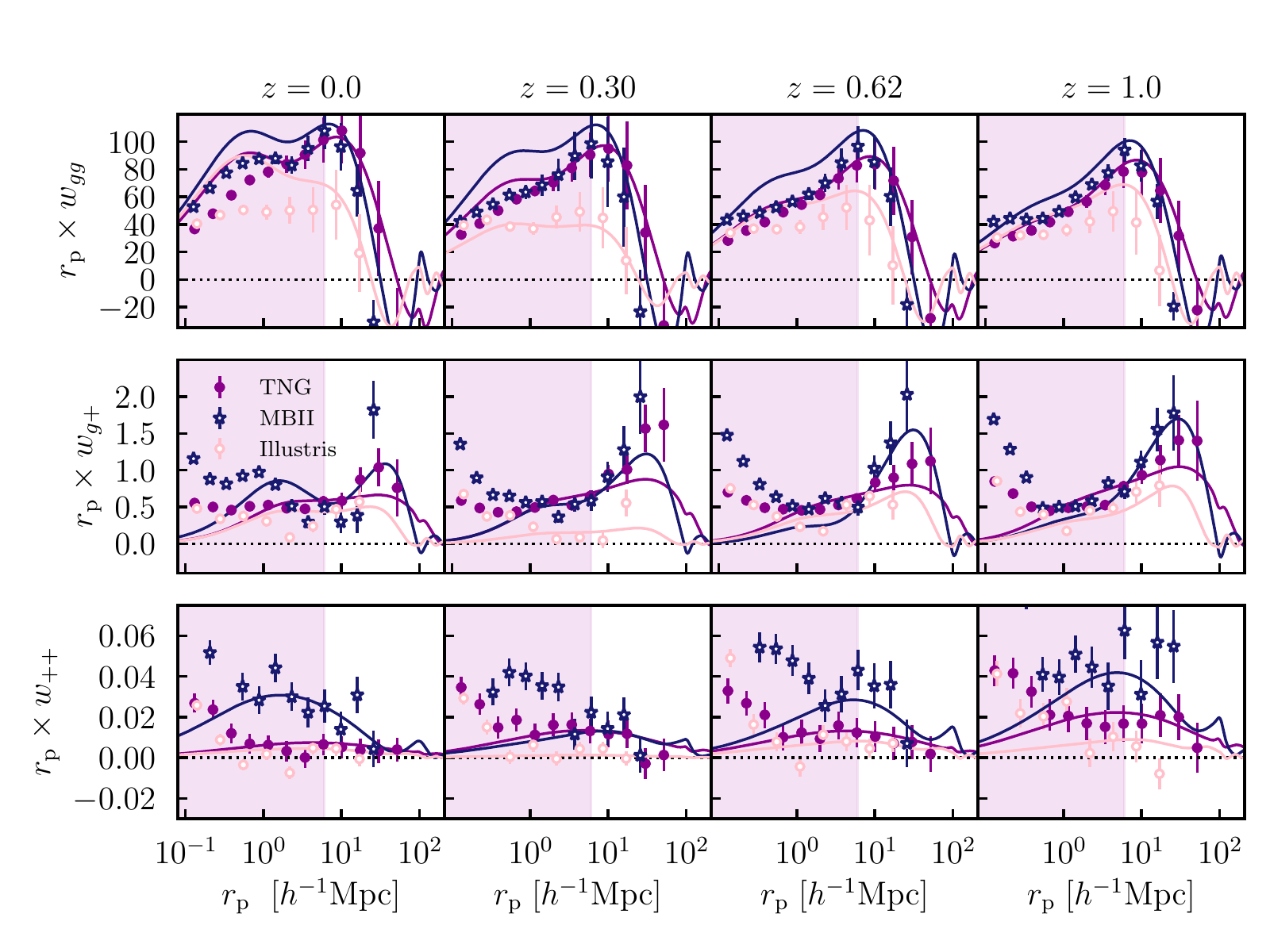}
 \caption{The fiducial data vectors used in this work. Shown from top are
 galaxy-galaxy, galaxy-shape and shape-shape two-point correlations, at
 four discrete redshifts (left to right, as indicated).
The point styles indicate measurements made on \mbii~(dark blue stars), \illustris~(pink open circles) and \tng~(purple filled circles). The solid lines are the theory predictions, evaluated at the best fitting point in the TATT parameter space for each data set. Scales within the shaded regions ($\rp<6\mpc$) are excluded from the fits,
 using both the TATT and NLA models.
 }\label{fig:data:fiducial_datavector}
 \end{figure*}
\subsection{Tidal \& Shape Fields}\label{sec:measurements:tidal}

\begin{figure*}
\includegraphics[width=2\columnwidth]{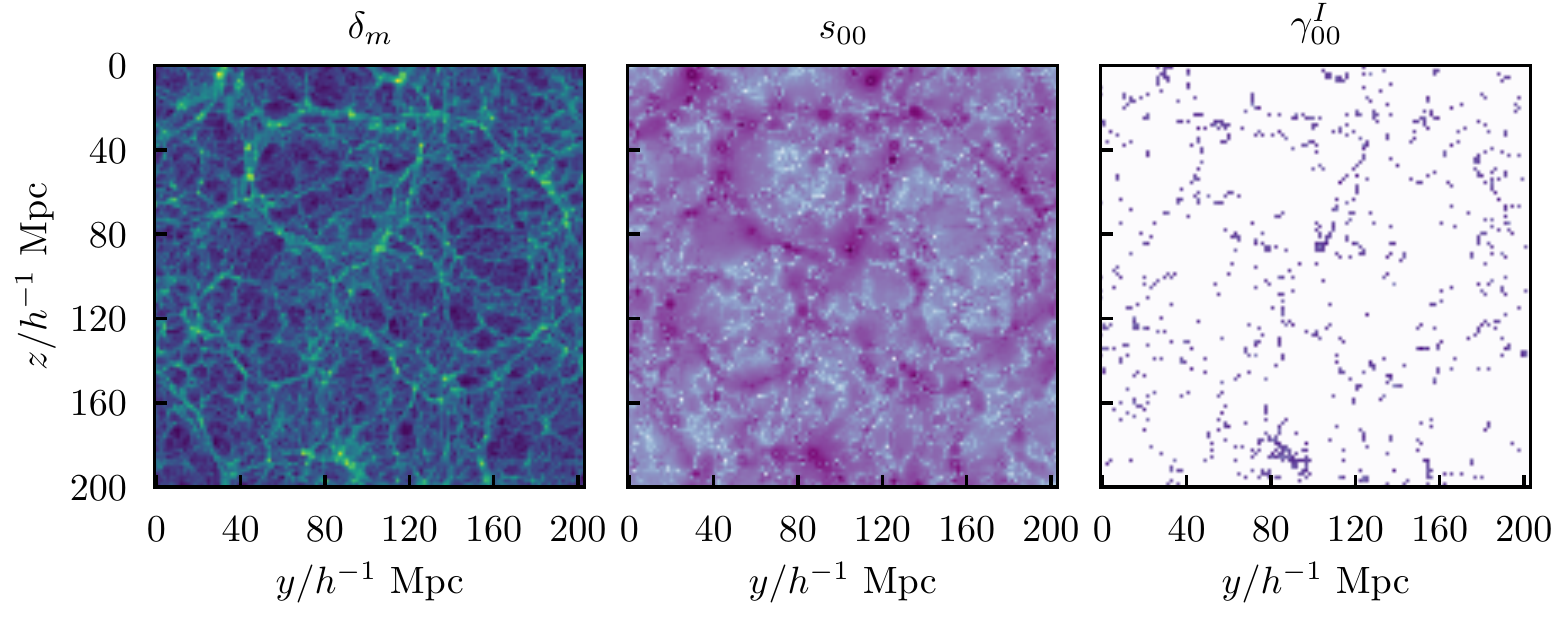}
\caption{The $z=0$ dark matter overdensity field and associated quantities 
from \tng.
Here we show (left to right)
the matter overdensity $\delta_m$, 
the $0,0$ component of the $3\times3$ dark
matter tidal tensor 
and the same component of the $3\times3$ galaxy
shape tensor $\gamma_I$. 
The pixel resolution is 128/side, resulting
in a physical pixel scale of $1.59 h^{-1}$ Mpc,
which is at the finer end of the range of pixel
scales presented in this work.   
While the full simulation box is clearly three dimensional,
for illustrative purposes
we choose here to show a 2D slice through the centre. 
In the right-hand panel, white pixels indicate those containing
no galaxies that pass cuts into our final \tng~shape catalogue. 
   }\label{fig:results:1pt:density_field}
\end{figure*}

In addition to the two-point measurements described above, we
also implement a new method to derive IA constraints at the field
level. We refer the reader to Section \ref{sec:daff} for details, but
the method involves deriving constraints on IA parameters 
via a comparison of the (pixelized) three dimensional tidal field 
and the intrinsic galaxy shape field (see also \citealt{hilbert17}, who also use the tidal field directly to measure
IAs, albeit via two-point functions).
To this end, we need an estimate of that tidal tensor as a function of position; we obtain this from the gridded particle data as follows.  

Starting with the table of particle positions at fixed redshift, we divide the 
simulation box into 3D cubic pixels. 
The pixel size $L$ is an unconstrained analysis variable, and affects the
physical interpretation of the eventual results. 
We choose to perform our measurements using three different scales, 
$16$ pixels across ($L\sim12.8 \mpc$), 
$32$ pixels across ($L\sim6.4 \mpc$) and 
$64$ pixels ($L\sim3.2 \mpc$).
Within each pixel $p$ in the grid, centered at position $\mathbf{x}_p$, we measure the overdensity of matter and stars,

\begin{equation}
\delta (\mathbf{x}_p) = \frac{N_{p}} {\langle N_p \rangle_p} - 1.
\end{equation}
\noindent
That is, the total number of dark matter particles in pixel $p$, divided by the mean occupation across all pixels. 
In the case of dark matter, all particles in \tng~are weighted equally, and the values in the equation above are raw number counts, rather than sums of masses.
Using the Fourier space version of the Poisson equation, one can show that 
the traceless tidal tensor can be obtained from the overdensity field 
as:

\begin{equation}\label{eq:fourier_tidal}
s_{ij}(\mathbf{k}) = 
\left ( 
\frac{k_i k_j}{k^2} - \frac{1}{3} \delta_{ij}
\right )
\delta(\mathbf{k}),
\end{equation}

\noindent
where $k^2 = k_1^2 + k_2^2 + k_3^2$.
For more details about the mathematics see \citet{catelan01a,alonso16}.
The two indices here $ij$ denote a single element of the $3 \times 3$
tensor matrix within pixel $p$.

We also obtain a noisy estimate for the intrinsic shear in pixel $p$,
$\gamma^I_{ij}(\mathbf{x}_p)$, by averaging the trace-free inertia tensors of galaxies 
within it.
That is,

\begin{equation}
\gamma^I_{ij} (\mathbf{x}_p) = \left \langle 
I_{ij,k} - \frac{1}{3} \delta_{ij} 
\mathrm{Tr} \left [ \mathbf{I}_{k} \right ] 
\right \rangle _k,
\end{equation}

\noindent
where the subscript $k$ denotes a particular galaxy
from pixel $p$, and the angle brackets $\langle \rangle_k$ indicate averaging
over those galaxies.
We estimate the per-element variance of 
the $3\times3$ matrix $\gamma^I$
directly by computing the RMS over all galaxies;
that is, we assume that shape noise dominates, such 
that the covariance matrix is diagonal,
and can be written as
$C^{-1}_{ij,p} = \delta_{ij} \sigma^{-2}_{\mathrm{SN} \mu}$, 
or the inverse square shape variance for component $\mu\in(1,2)$. Note that this is a global quantity, computed across pixels and applied to each of them. We confirm that the covariance scales with pixel size roughly as one might expect from geometric arguments as
$\sigma_{\mathrm{SN} \mu}\propto L^{-3/2}$.
A 1D slice of the three fields described here, 
as measured in the $z=0$ \tng~snapshot, can be seen in Figure~\ref{fig:results:1pt:density_field}.
Shown are (left to right): dark matter overdensity, the upper diagonal element of the dark matter tidal
tensor and the smoothed galaxy shape field.  
It is apparent from Figure~\ref{fig:results:1pt:density_field} that there is an obvious qualitative 
correspondance between the raw matter field and the tidal tensor (compare the left-most and middle panels).
The sampling of galaxies is much sparser, which is evidenced by the amount of white space in the
right-most panel.
Depending on the pixel scale, the fraction of unoccupied pixels is between
$20\%$ and $80\%$.
Although striking in this Figure, and worth noting, the impact of this sampling is explicitly incorporated into our IA modelling, as described in Section \ref{sec:modelling}.

\subsection{Covariance Matrix of Two-Point Functions}\label{sec:measurements:cov}

\subsubsection{Analytic Covariance Matrix}

\begin{figure}
\includegraphics[width=1.1\columnwidth]{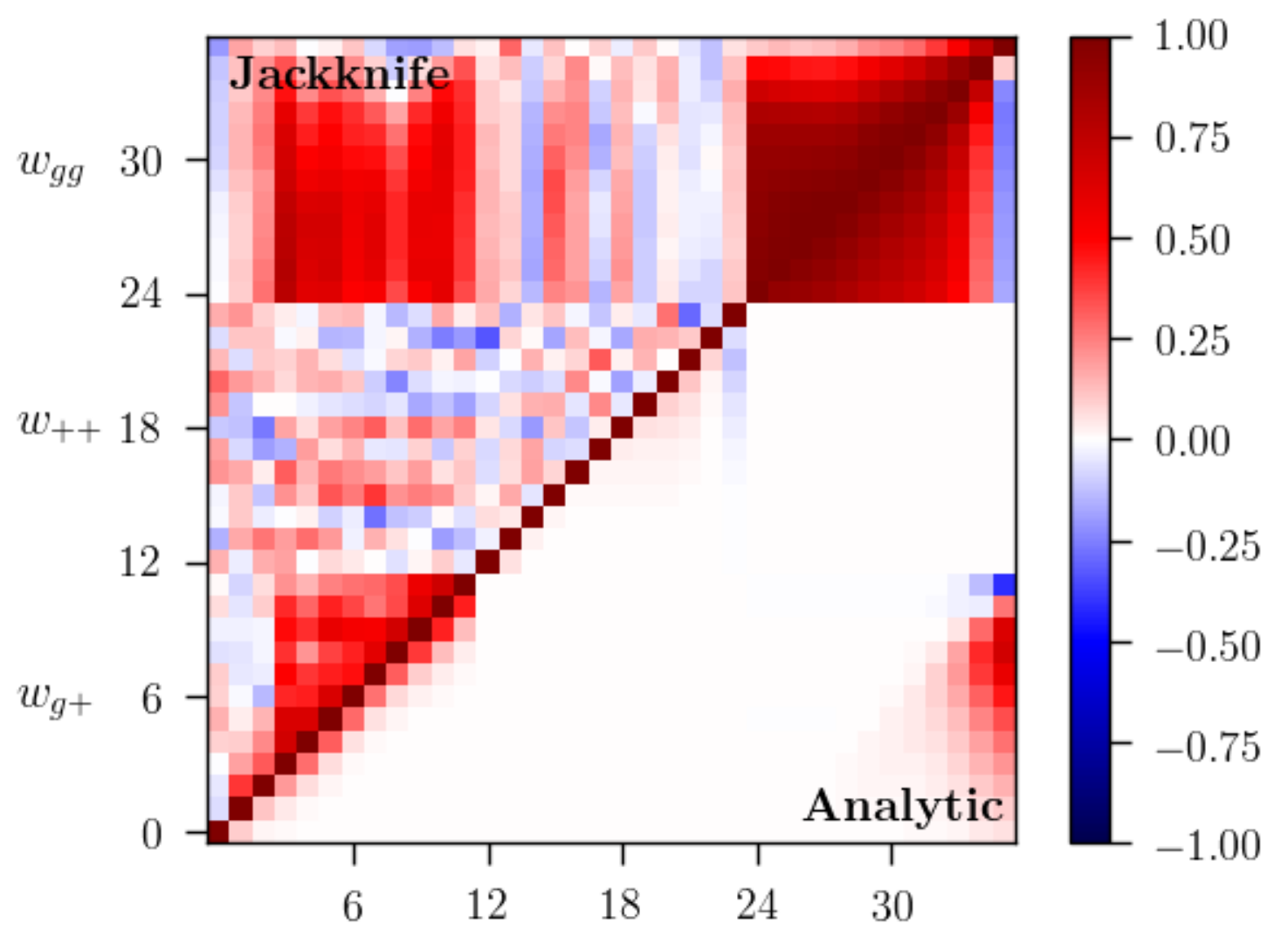}
\caption{The correlation matrix for our fiducial \fidsim~two-point
measurements, as estimated using jackknife resampling 
(upper left triangle), and an analytic Gaussian approximation
(lower right).
Note that the two covariance matrices are symmetric about the
diagonal; the triangle configuration is shown here for illustrative
purposes only.
}\label{fig:data:correlation_matrix}
\end{figure}

In order to derive robust parameter constraints from our measurements we need a representative, numerically stable, estimate for the covariance matrix of those measurements. The full data vector consists of three two-point measurements for each of four snapshots; this gives us $N = 4 \times 3 \times N_{rp}$ data points for each simulated
galaxy sample (96, 144 and 192 in the case of \illustris, \mbii, and \tng~respectively). Our fiducial covariance estimate is calculated analytically, a detail of this analysis that differs from many previous studies, most of which have opted for an internal covariance estimator such as jackknife resampling. The analytic approach has a number of advantages, not least the ability to extend to large scales where jackknife estimates break down.
We show a comparison of our fiducial correlation matrix, calculated using the method described below, and a jackknife estimate in Figure~\ref{fig:data:correlation_matrix}.

Although in principle the covariance has higher order contributions resulting from mode mixing (e.g. \citealt{krause15}), given the limited statistical power of the simulations, and the fact that shot and shape noise tend to dominate on the scales we fit, the dominant Gaussian contribution is considered sufficient for our purposes. 
In the Gaussian approximation, a given element is the sum of a noise term and a cosmic variance contribution:

\begin{equation}
\mathrm{Cov}\left[ w^{z_{\rm s}}_{\alpha \beta}(r_{\mathrm{p},j}), w^{z_{\rm s}}_{\delta \gamma}(r_{\mathrm{p},k}) \right ] = 
\mathrm{C}^{\mathrm{SN},z_{\rm s}z_{\rm s},kj}_{\alpha\beta\delta\gamma} +
\mathrm{C}^{\mathrm{CV},z_{\rm s}z_{\rm s},kj}_{\alpha\beta\delta\gamma},
\end{equation}

\noindent
where the Greek indices denote correlation types 
i.e. $\alpha,\beta,\delta,\gamma \in (g,+)$;
$z_{\rm s}$ identifies a particular redshift slice and $j,k$ are comoving scale bins.
The cross correlations between snapshots is potentially complicated, given that the
galaxy properties are strongly (but not fully) correlated.
However, since we will not attempt a fully simultaneous analysis across redshifts, but rather
restrict our inference to one snapshot at a time, we will neglect these additional covariance terms.   
One can write each element as:

\begin{multline}
\mathrm{Cov}\left[ w^{z_{\rm s}}_{\alpha \beta}(r_{\mathrm{p},i}), w^{z_{\rm s}}_{\delta \gamma}(r_{\mathrm{p},j}) \right ] =\\
\delta_{ij} \frac{2\pi}{\mathcal{A}_p r_{\mathrm{p},i} \Delta r_\mathrm{p} } 
\int k \,\mathrm{d}k\, 
\Theta_{\alpha \beta}(kr_{\mathrm{p},i}) \Theta_{\delta \gamma}(kr_{\mathrm{p},i}) \\
\left[ \tilde{P}_{\alpha \delta}(k, z_{\rm s})\tilde{P}_{\beta \gamma}(k, z_{\rm s})
+ \tilde{P}_{\alpha \gamma}(k, z_{\rm s})\tilde{P}_{\beta \delta}(k, z_{\rm s})
\right ],
\end{multline} 

\noindent
with the kernels

\begin{equation}
\Theta_{\mu\nu}(x) = 
\begin{cases}
J_2(x)          \;\;\;\;\;\;\;\;\;\;\;\;\;\;\;\;\;\; \mu\nu=g+ \\
J_0(x)          \;\;\;\;\;\;\;\;\;\;\;\;\;\;\;\;\;\; \mu\nu=gg \\
J_0(x) + J_4(x) \;\;\;\;\;                           \mu\nu=++ \\
\end{cases},
\end{equation}

\noindent
where $J_\nu$ is a Bessel function of the first kind of order $\nu$.
In IA measurements on real data $\mathcal{A}_p$ is a function of
redshift, and accounts for the survey mask;
in our case it is simply the cross sectional area of the simulation
box in $h^{-2}$ Mpc$^{-2}$.
One should also note that the power spectra here are subject to a 
noise contribution, 

\begin{equation}
\tilde{P}_{\alpha \beta}(k, z_{\rm s}) = P_{\alpha \beta}(k, z_{\rm s}) + N^{z_{\rm s}}_{\alpha\beta},
\end{equation}

\noindent
where $N^{z_{\rm s}}_{\alpha \beta}=1/n^{z_{\rm s}}_{\rm c}$ for $\alpha \beta=gg$, 
$N_{\alpha \beta} = (\sigma^{z_{\rm s}}_e)^2/n^{z_{\rm s}}_{\rm c}$ for $\alpha \beta=++$,
and $N_{\alpha\beta}=0$ for $\alpha \beta=g+$.
The denominator $n_{\rm c}$ is the comoving volume number density of the
sample at $z=z_{\rm s}$ in $h^{3}$ Mpc$^{-3}$, and $\sigma^{z_{\rm s}}_e$ is the projected ellipticity dispersion.

As is apparent from the above,
the analytic covariance matrix is sensitive to some extent on the input parameter values (cosmology, galaxy bias, and IAs). As stated before, cosmological parameters are fixed to those 
appropriate for the simulation in question, as per Table \ref{tab:data:simulations}.
For the other (IA and bias) parameters, we generate the fiducial matrix for each sample using an iterative procedure similar to that of \citet*{y1methodology}. That is, we repeatedly fit the data to obtain IA and galaxy bias parameter constrains, update the covariance matrix and fit again.
Our convergence criteria are that (a) the marginalised 1D parameter 
posteriors are not systematically different between iterations, and 
(b) the $\chi^2$ and evidence values are stable to within a few
percent. In all samples, the covariance matrix converges within $2-4$ iterations.

We also test our fiducial analytic covariance matrix against a version computed using jackknife resampling. In brief, the jackknife method involves dividing the data into $N$ spatial subregions, and repeating the measurement $N$ times, each time removing one of them. The validity of this approach relies on various (potentially strong) assumptions; not least it assumes the subregions are statistically independent (see \citealt{anderson03,hartlap07} for discussion), and that the scales of interest are much smaller than scale of the subregions.
These factors, combined with the relatively small number of subregions
allowed by even \tng~(the largest of the simulations considered here), are the primary reason we consider jackknife as an approximate test of, and not a viable alternative to, our analytic predictions. 
In the fiducial case (\tng), we divide the three dimensional box into $N_{\rm jk} = 4^3 = 64$ cubic subvolumes.
A visual comparison of the correlation matrices can be found in Figure \ref{fig:data:correlation_matrix}.
We also compare the root diagonals of the two covariance matrices (see Figure~\ref{fig:app:covariance_comaprison}).
Although there is approximate agreement between the two, the jackknife method
tends to underestimate the variance on virtually all scales in the three correlations.
On the relevant scales for our fits ($>6\mpc$), the differences are at the level of up to $\sim 25 - 50\%$ in $w_{g+}$.  

\section{Theory}\label{sec:modelling}

Our analysis pipeline is built within \cosmosis\footnote{\url{https://bitbucket.org/joezuntz/cosmosis}, v1.6; master branch}~\citep{zuntz14}.
The new modules introduced in this paper has been validated against older free-standing code. Although we will not discuss this process in detail here, a longer discussion can be found in Appendix \ref{app:validation}. 
Sampling is performed using \blockfont{MultiNest} \citep{feroz19}, and in the subset of chains where the Bayesian evidence is needed,
we also run using \blockfont{polychord} \citep{handley15}, with more stringent accuracy 
settings\footnote{$\mathrm{live\_points}=250, \mathrm{tolerance}=0.1, \mathrm{num\_repeats}=60$}.
In all cases, we fix the cosmology to the input for the relevant simulation, with the parameters given in Table \ref{tab:data:simulations} and zero neutrino mass. The matter power spectrum is generated using \blockfont{CAMB}
with nonlinear modifications from \blockfont{halofit} \citep{takahashi12}. A simulation of finite box size (i.e. any simulation) has an effective $k$ limit, at which the power spectra are truncated (see \citealt{power06} and \citealt{bagla09} for discussion and quantification), an effect that primarily impacts large physical scales, but potentially has ramifications at smaller separations too. In order to avoid biasing our results, we explicitly include this
truncation in our modelling. Given the box sizes, the actual effective small$-k$ cutoff is at $k_{\rm min}=2\pi/L$, or $\sim0.03\mpc$, $\sim0.06\mpc$ and $\sim0.08\mpc$ in the cases of \tng, \mbii~and \illustris~respectively.
We assess the impact of this detail by repeating our fits with fixed $k_{\rm min}=0.06 h^{-1}$ Mpc
for the three simulations. The resulting biases, arising from ignoring the small-$k$ cut off, is potentially quite significant ($\sim 20 \%$) in both the galaxy bias and IA parameters.

Our fiducial analysis includes physical scales in the range $6 < \rp < L/3$ \mpc, where $L$ is the length of the simulation box. Unlike in real survey data, an upper cut is necessary to avoid edge effects due to the finite simulation size. The lower cut follows several other studies \citep{joachimi11, singh15, johnston18}, and is intended to be conservative in removing data affected by nonlinear bias. We explicitly test this choice in Section \ref{sec:results:scales}.

\subsection{Modelling Intrinsic Alignments}

We consider two different IA scenarios in our fits, discussed in more detail below.
While it is useful to think of these as entirely separate models, and indeed we will refer to them as such, it is worth bearing in mind that they are nested. That is, the more complex model reverts to the simpler one when a subset of its parameters are zero. For reference, the free parameters in each of these models and the associated
priors in each case are shown in Table~\ref{tab:theory:models}. 

\begin{table}
\begin{center}
\begin{tabular}{c|cc}
\hline
Model                 & Parameter                     & Prior   \\
\hhline{=|==}
NLA                   & $A_1$                          & $\mathrm{U}[-6, 6]$ \\
                      & $b_g$                          & $\mathrm{U}[0.05, 8]$ \\
\hline
TATT                  & $A_1$                          & $\mathrm{U}[-6, 6]$ \\
                      & $A_2$                          & $\mathrm{U}[-6, 6]$ \\
                      & $b_{\rm TA}$                   & $\mathrm{U}[-6, 6]$ \\
                      & $b_g$                          & $\mathrm{U}[0.05, 8]$ \\
\hline
\end{tabular}
\caption{Free parameters for the IA model fits implemented in this work.
All fits are performed on a single snapshot, with the same priors 
applied irrespective of redshift.
Note that the linear galaxy bias $b_g$ is not an IA parameter 
(i.e. it does not enter either the GI or II power spectra),
but it is included in the modelling and so is shown here.
The choice of priors here is designed to be conservative, and well 
clear of the posterior edges. 
We discuss the possible impact of this choice, and demonstrate robustness
to projection effects in Section \ref{sec:results:main}.
}\label{tab:theory:models}
\end{center}
\end{table}

\subsubsection{Nonlinear Alignment Model}\label{sec:theory:nla}

One common predictive IA model is the Nonlinear Alignment (NLA) model; in essence, it is an empirically motivated modification  (see \citealt{bridle07}, \citealt{hirata07}) 
to a physically motivated (at least partially, in certain regimes) prescription known as the 
Linear Alignment (LA) model \citep{catelan01, hirata04, hirata10, blazek11}.
Under the assumption of linear alignments, one can write the intrinsic shape of a galaxy in terms of the background gravitational potential at the time of galaxy formation as:

\begin{equation}
(e^I_{+}, e^I_{\times}) = 
- \frac{\bar{C}_1 }{4 \pi G} 
\left ( \frac{\partial^2 }{\partial x^2}-\frac{\partial^2 }{\partial y^2} ,
2 \frac{\partial^2}{\partial x \partial y} \right ) 
\phi(\chi_{*}),
\end{equation}

\noindent
where $\bar{C}_1$ is a normalisation constant, typically fixed at a value of $5 \times 10^{-14} M_{\odot}^{-1} h^{-2}$ Mpc$^3$ \citep{brown02}. Following \citet{hirata04}, the GI and II power spectra have the form:

\begin{equation}\label{eq:theory:nla_gi}
P_{\rm GI} (k) = - \frac{\bar{C}_1 \bar{\rho}(z)}{D(z)} a^2(z) P^{\rm lin}_\delta (k)
\end{equation}

\noindent
and 

\begin{equation}\label{eq:theory:nla_ii}
P_{\rm II} (k) = \left( \frac{\bar{C}_1 \bar{\rho}(z)}{D(z)} \right )^2 a^4(z) P^{\rm lin}_\delta (k).
\end{equation}

\noindent
Here $\bar{\rho}$ is the (spatially averaged) mean matter density of the Universe and $D$ is the linear growth function. The model also predicts higher order contributions, as well as non-zero B modes arising from galaxy clustering, though these are typically neglected in implementations of the NLA model (\citealt{hirata04}, \citealt{blazek15}; see the next section for further discussion).
We follow many previous analyses in fixing $\bar{C}_1$ to \citet{brown02}'s value, and parameterising deviations in strength of alignment from this baseline with a free amplitude, such that $P_{\rm GI} \rightarrow A_1 P_{\rm GI}$ and $P_{\rm II} \rightarrow A^2_1 P_{\rm II}$.

The feature that defines the NLA is the substitution of the linear power spectrum
in Eq. \eqref{eq:theory:nla_gi} and \eqref{eq:theory:nla_ii} for the nonlinear version. The rationale for this change is as an attempt to capture the nonlinear tidal field, and indeed it does appears to improve the performance on small to intermediate scales (see, for example \citealt{bridle07, blazek15, singh15}), even if it is not necessarily internally consistent.

\subsubsection{Tidal Alignment \& Tidal Torque Model}\label{sec:theory:tatt}

Our second IA model, referred to as the Tidal Alignment + Tidal Torque (TATT) model, was first proposed by \citet{blazek17}
and has been employed a number of times in the context of cosmic shear analyses in the recent past (see \citealt{y1cosmicshear,y1coloursplit}). We will provide a brief overview of the theory, and refer the reader to those papers a more detailed description.  

In this framework, a galaxy's intrinsic shape\footnote{The intrinsic shape here is defined
in an analogous way to the projected
ellipticity; it is the trace-free component of the
moment matrix in three dimensions, 
or equivalently, the eigenvector matrix of the 
3D inertia tensor. As noted in \citet{blazek17},
it is not a uniquely defined quantity, and depends 
on the radial weighting of the measurement algorithm.}
is written as an expansion in the trace-free tidal field tensor $s_{ij}$ :

\begin{equation}\label{eq:theory:expansion}
\gamma^I_{ij} = 
\underbrace{C_1 s_{ij} } _{\text{ Tidal Alignment }}
+ 
\underbrace{C_{1\delta} (\delta \times s_{ij} ) }_{\text{Density Weighting}}
+ 
\underbrace{C_2 \left [ \sum^2_{k=0} s_{ik} s_{kj} - \frac{1}{3} \delta_{ij} s^2 \right ]}_{\text{Tidal Torquing}}
 + \dots,
\end{equation}

\noindent
with both sides of the equation evaluated at a position $\mathbf{x}$, which may be either a Lagrangian or an Eulerian position. The two amplitudes $C_1$ and $C_2$ describe the magnitude of alignment due to tidal alignment and tidal torquing respectively. It is worth bearing in mind, however, that these terms can absorb IAs
due to other mechanisms when fit to real data; 
for example, an effective non-zero $C_1$ can in principle arise in a pure tidal torquing IA scenario, when combined with nonlinear growth \citep{hui02}.
The term, with the coefficient $C_{1\delta}$, is a so-called
density weighting contribution, and arises from the fact that one can
only measure galaxy shapes in a position where there is actually
a galaxy (see e.g. \citealt{hirata04} and \citealt{blazek15} for further discussion). 
Also note that the product of the matter overdensity and tidal
fields $\delta s_{ij}$ implicitly assumes a smoothing scale,
a detail we will return to in Section \ref{sec:daff}.
The real-space dark matter tidal tensor is a $3\times3$ matrix, defined in relation to the overdensity field in Equation~\eqref{eq:fourier_tidal}. 
If the tidal tensor is computed using the nonlinear matter field, then the leading term in Eq. \eqref{eq:theory:expansion} is equivalent to the NLA prediction. If the TATT model parameters are varied together, however, they can enter the data in potentially degenerate ways, meaning that the $A_1$ part of the full TATT space will not \emph{necessarily} match the NLA fit to the same data, if $A_2\neq0$ is preferred.
One then has:

\begin{equation}
C_1 = - A_1 \bar{C}_1 \frac{\omegam \rho_{\rm crit}}{D(z)},
\end{equation}

\begin{equation}
C_2 = 5 A_2 \bar{C}_1 \frac{\omegam \rho_{\rm crit}}{D^2(z)}.
\end{equation}

\noindent
and 

\begin{equation}
C_{1 \delta} = - A_{1\delta} \bar{C}_1 \frac{\omegam \rho_{\rm crit}}{D(z)},
\end{equation}

\noindent
The constant $\bar{C}_1$ is the same as the one discussed in the previous section. 
The IA power spectra (GI and II) are derived from perturbation theory and are given by integrals over the matter power spectrum;  for details see Sections A-C of \citet{blazek17}. Our version of the TATT model is identical to that of \citet{y1cosmicshear}, \citet{blazek17} and \citet{y1coloursplit}. It makes use of the \blockfont{FAST-PT} code \citep{mcewen16,fang17}, and is implemented within \blockfont{CosmoSIS}.

Following \citet{blazek15}, we do not vary $A_{1\delta}$ directly, but rather assume the density weighting term is related to the tidal alignment amplitude via a coefficient (i.e. $C_{1\delta} = b_{\rm TA} C_1$).
The original motivation for this parameterization was that IA correlations scaling with $\delta s_{ij}$ were generated by the density weighting of the IA field, which can only be observed where galaxies are located (see \citealt{blazek15} for a more detailed discussion). As with the other terms, $C_{1\delta}$  can be thought of more generally as describing any alignment physics with large-scale correlations that depend on $ \delta s_{ij} $, and so does not necessarily correspond directly to the galaxy bias constrained by $w_{gg}$, as per the simple density weighting picture. Indeed, in a linear and ``local Lagrangian'' picture of IA formation, in which intrinsic galaxy shapes are a linear function of the local tidal field initially present where the halo (and galaxy) form, a $C_{1\delta}\sim C_1$ term will be generated by the advection of galaxies between the Lagrangian and Eulerian frames \cite{schmitz18}. Given the potential for other physical effects to be captured by the same term, it is safest to allow it to vary as a free parameter over a similar range to the other IA parameters (see Table \ref{tab:theory:models}). Previous studies have chosen to fix it to unity (\citealt{y1cosmicshear}, \citealt{y1coloursplit}, \citealt{blazek17}), based on physical arguments. In these cases, however, the density weighting term has been subdominant, allowing only very broad constraints on $b_{\rm TA}$; \citet{y1coloursplit}, show that the decision to fix it was not a significant source of uncertainty in the context of DES Y1 $3\times2$pt cosmology. This is likely to be less true for our direct IA measurements.

Finally, we note that the TATT model predicts a non-zero IA-induced B-mode term, which enters the II power spectrum, and is sensitive to $C_{1 \delta}$ and $C_{2}$ (see \citealt{blazek17}, eq. 37-39). These contributions are included in our modelling of $w_{++}$. Again, we demonstrated in \citet{y1coloursplit} (Appendix C) that this choice has negligible impact on parameter constraints in the context of a DES Y1 $3\times2$pt analysis. This is not trivially true for the type of measurement considered in this work, and so we include the extra B-mode terms when fitting the TATT model here.

\subsubsection{Modelling Two-Point Correlations}

Given an IA power spectrum from either of the models described, one can predict the projected correlation functions at fixed redshift via Hankel transforms. Under the Limber approximation one has:

\begin{equation}\label{eq:wgp_basic}
w^{z_{\rm s}}_{g+} (r_{\rm p}) = -b^{z_{\rm s}}_g 
\int \frac{\mathrm{d} k k}{2 \pi}  
J_2(k r_{\rm p}) P_{\rm GI}(k,z=z_{\rm s}),
\end{equation}

\noindent
with the $z_{\rm s}$ indicating a particular redshift (snapshot), and $J_2$ being a second order Bessel function of the first kind. We assume linear galaxy bias, $b_g \equiv \delta_g / \delta$, which is marginalised with a wide prior (Table \ref{tab:theory:models}). The range $b_g=[0.05,8]$ is intended to be conservative, and the bias is always well constrained within these bounds. An important thing to note here, however, is that in a high dimensional parameter space typical of cosmological analyses such wide priors can cause shifts in the 2D constraints via projection effects (see e.g. \citealt{joachimi20}, \citealt{secco20} for discussion); in our relatively simple setup we do not expect this to be an issue. We verify this in our fiducial \tng~TATT analysis by reducing the $b_{g}$ prior width to $[0.05,4]$, and confirm it does not alter our results. A similar exercise, halving the volume of the prior on the less well constrained $b_{\rm TA}$ again has no significant impact. 

In real data one would also need to evaluate an integral over a redshift kernel, defined by the sample's redshift distribution (\citealt{mandelbaum10}'s Appendix A); in our case this reduces to evaluating $P_{\rm GI}(k)$ at a particular redshift $z_{\rm s}$.
The other two-point correlations follow by analogy as:

\begin{equation}\label{eq:wgg_basic}
w^{z_{\rm s}}_{gg} (\rp) = b^{z_{\rm s}}_g b^{z_{\rm s}}_g 
\int \frac{\mathrm{d} k k}{2 \pi}  
J_0(k \rp) P_\delta(k,z=z_{\rm s}),
\end{equation}

\noindent
and

\begin{multline}\label{eq:wpp_basic}
w^{z_{\rm s}}_{++} (\rp) =
\int \frac{\mathrm{d} k k}{2 \pi}  
\left [ J_0(k \rp) + J_4(k \rp) \right ] P_{\rm II}(k,z=z_{\rm s}).
\end{multline}

\noindent
In the case where we are including an IA induced B-mode contribution, the above becomes a sum of two integrals (see e.g. \citealt{blazek15}, equation 2.8):

\begin{multline}\label{eq:wpp_bmodes}
w^{z_{\rm s}}_{++} (\rp) = w^{z_{\rm s}, \mathrm{EE}}_{++} (\rp) + w^{z_{\rm s}, \mathrm{BB}}_{++} (\rp) \\
=
\int \frac{\mathrm{d} k k}{2 \pi}  
\left [ J_0(k \rp) + J_4(k \rp) \right ] P^{\rm EE}_{\rm II}(k,z=z_{\rm s})
\; + \\
\;\;
\int \frac{\mathrm{d} k k}{2 \pi}  
\left [ J_0(k \rp) - J_4(k \rp) \right ] P^{\rm BB}_{\rm II}(k,z=z_{\rm s}).
\end{multline}

\noindent
The TATT E and B mode power spectra here are given by \citet{blazek17}'s
equations (38) and (39). In the NLA case $P^{\rm BB}_{\rm II}=0$, and Eq. \eqref{eq:wpp_bmodes}
reduces to Eq. \eqref{eq:wpp_basic}.

Although we do not compute the 3D correlations $\xi_{ab}(\rp,\Pi)$, we do factor in the fact that the line of sight integral in the measurement has a finite limit $\Pi_{\rm max}$ (e.g. Eq.~\eqref{eq:xi_to_w}). The effect of this is to suppress the signal slightly, as correlated pairs are cut off. We can test the magnitude of this by comparing our observables at a fiducial point in parameter space with an external modelling code, which explicitly includes $\Pi_{\rm max}$. Since the impact is found to be independent of \rp~on large scales, to the level of $\sim0.5\%$, we incorporate it into our modelling as a single multiplicative factor $\mu$, which we compute for each correlation function, at each redshift (i.e. 12  numbers per simulation). In the case of \tng, $\Pi_{\rm max}$ is relatively large ($68\mpc$), and so the signal damping is only $1-2\%$ (which is comfortably subdominant to uncertainty). For \mbii~and \illustris~($\Pi_{\rm max} = 33$ and $\Pi_{\rm max} = 25\mpc$ respectively), however, $\mu$ is somewhat larger, which shifts the IA parameters upwards slightly. Although our qualitative conclusions are robust even without this correction, omitting it is seen to bias the $A_1$ and $A_2$ towards low values by $\sim10-50\%$.

\section{IA Constraints From Two-Point Measurements}\label{sec:results:main}

As discussed, our baseline methodology is to fit the joint data vector of $w_{gg}$, $w_{g+}$ and $w_{++}$ simultaneously for a given simulation and at a given redshift.
In this section we present the results of these likelihood analyses. This approach is analogous to cosmological inference using $3\times2\mathrm{pt}$ data, with the significant difference that our parameter space is several times smaller (and does not include cosmological parameters). It carries a number of advantages, not least benefitting from some level of complementarity in the degeneracies of the different data vector elements. 

We perform our IA model fits to each of the four redshift snapshots independently, a choice primarily driven by the covariance matrix; unlike in real data, where each galaxy can be assigned (albeit not necessarily correctly) to a single tomographic bin, here we effectively have one realisation of the galaxy field, which is evolved with redshift. The galaxy population, the shape noise and the cosmic variance are, then, potentially heavily correlated between redshifts, which makes a fully simultaneous analysis complicated. Modelling such correlations is non trivial, and not considered a valuable exercise within the scope of this paper.


Despite their potential, concerns persist around the accuracy of hydrodynamic simulations as an effective model for intrinsic alignments; systematic uncertainties arise largely from the underlying physics models, and are evidenced by longstanding disagreements between different simulations.
Discussion of such differences in the literature have focused on the impact of baryons on the matter power spectrum (see \citealt{vanDaalen11}, \citealt{chisari18}, \citealt{huang19});
discrepancies in the magnitude (and sign) of alignments have been noted \citep{chisari15,codis15b,tenneti16,chisari16}, but these have perhaps received less attention due to the fact that, unlike the baryonic effects, IA measurements in these simulations do not feed directly into cosmological analyses (although they could do, potentially, in future). 
To properly diagnose this systematic uncertainty it is useful to compare the results from multiple simulations using a unified analysis framework, and appropriately weighted samples, as we seek to do in this section. As discussed above, the reweighting is designed only to match the halo mass distributions, and not to fix other differences in, for example, the galaxy formation properties; we consider these more complex differences as sources of systematic uncertainty. Indeed, it is interesting to try to disentangle them from discrepancies due to differences in the analysis details (e.g. the galaxy selection method) of previous studies.


\subsection{NLA \& TATT}

\begin{figure}
\includegraphics[width=\columnwidth]{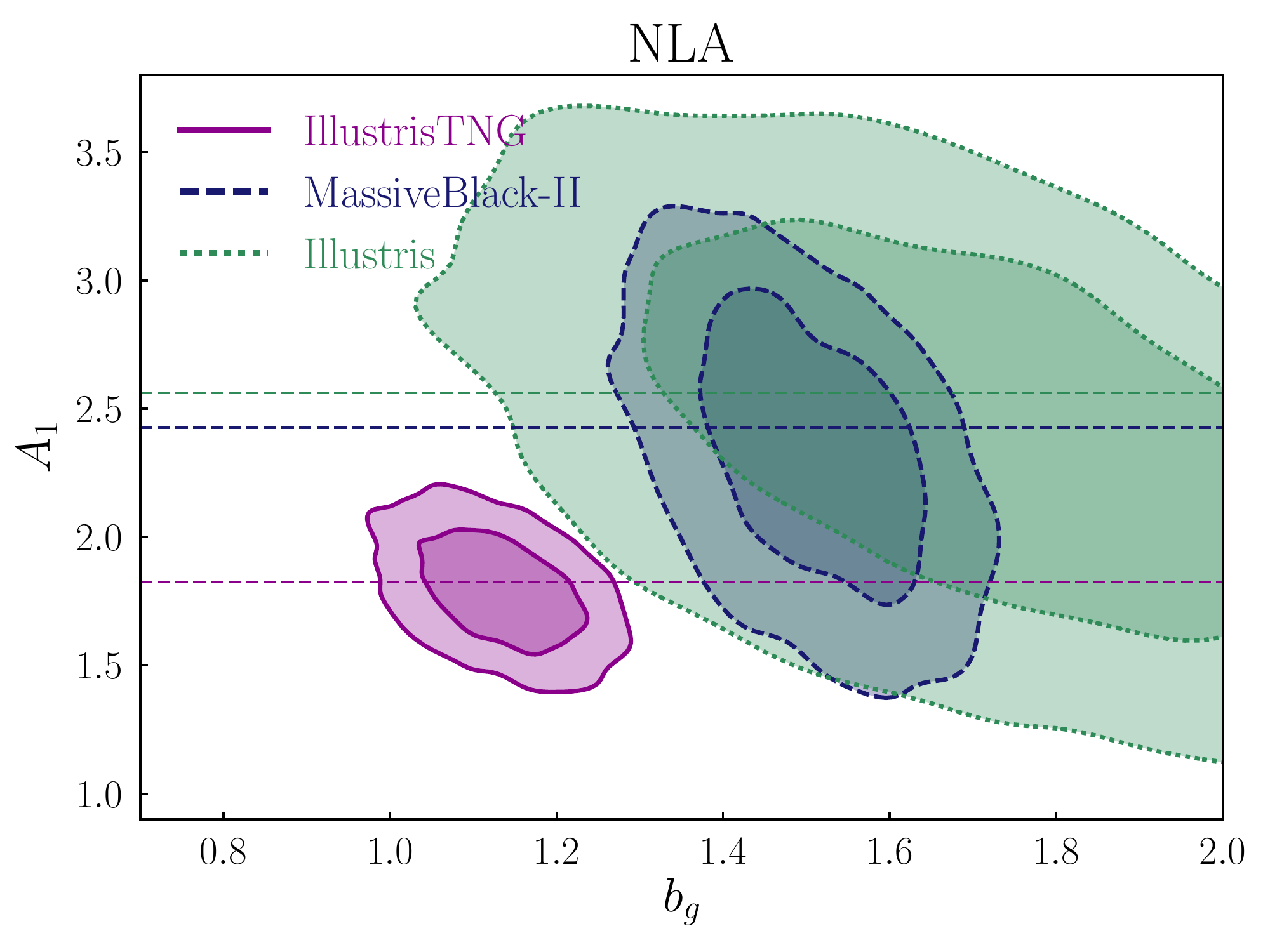}
\includegraphics[width=\columnwidth]{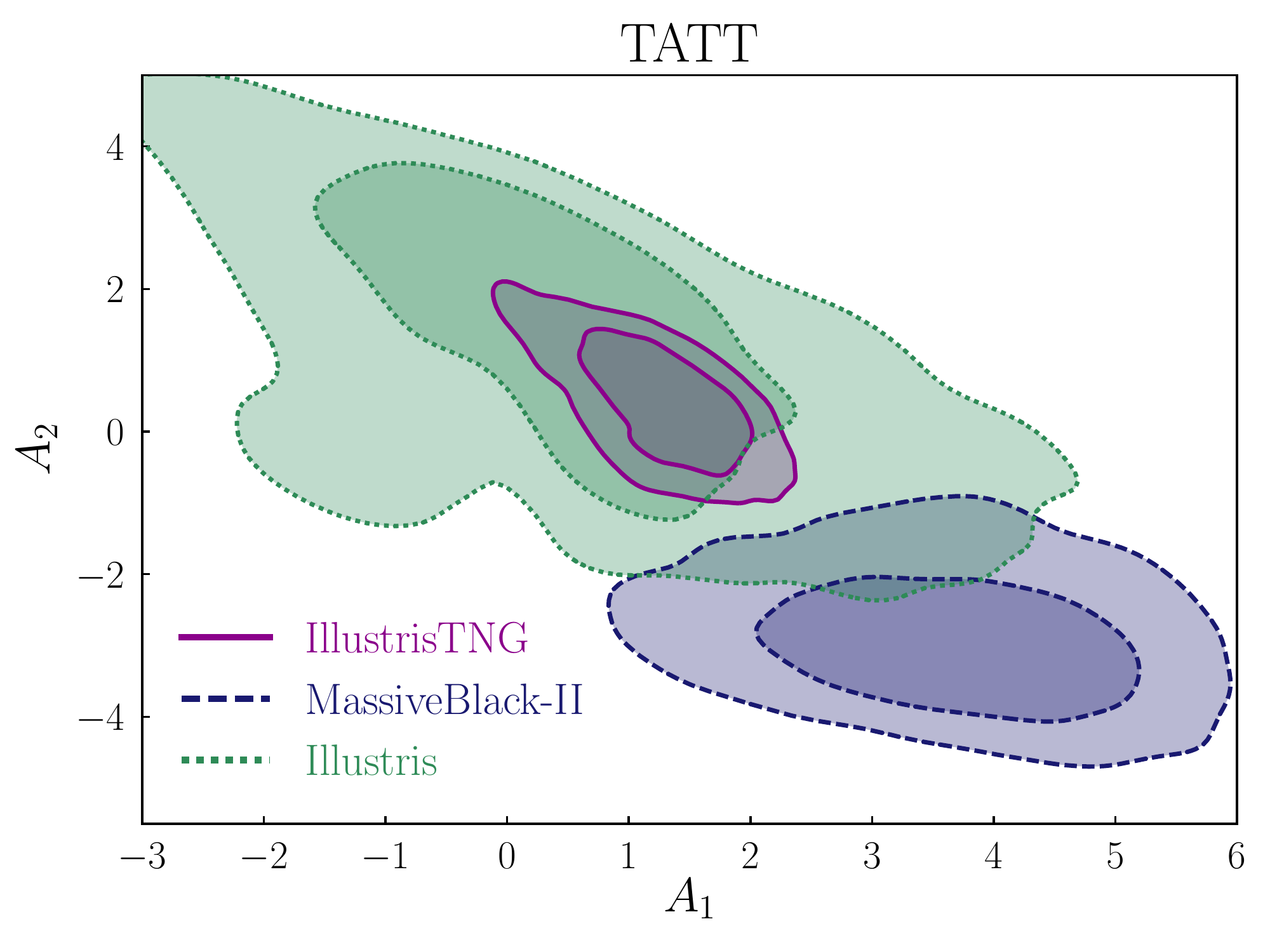}
\caption{$1\sigma$ and $2\sigma$ confidence contours from our
NLA (top) and TATT (bottom) model fits to various hydrodynamic simulations at $z=0$. 
Shown are \tng~(purple, solid), \illustris~(green, dotted) and \mbii~(blue, dashed). 
Note that the \mbii~and \illustris~samples here are weighted, such that the distributions
of host halo masses match between the simulations, in order to allow meaningful comparison with 
\tng~(see Section~\ref{sec:measurements:2pt}).
The three horizontal lines in the NLA panel show the $A_1$ values of the respective posterior peaks; these best-fitting values are $A^{\rm TNG}_1=1.63$, $A^{\rm MBII}_1=2.11$ and $A^{\rm Ill}_1=1.52$. The three simulations are consistent in the NLA space to $\sim1\sigma$, although some mild disagreement is seen in the case of the more complex model.
}\label{fig:results:contours}
\end{figure}

The posteriors from NLA model fits to the various simulations at $z=0$ are shown in the upper panel of Figure~\ref{fig:results:contours}. As described in Section \ref{sec:measurements:2pt}, the \mbii~and \illustris~samples are reweighted, such that the halo mass distributions match (see also the discussion in Appendix~\ref{app:weights}, where we demonstrate the importance of this reweighting). 
This process is designed to allow meaningful comparison between simulations by ensuring that differences in halo mass distribution are not driving the offset in the IA-bias parameter space. As noted above, the halo mass weighting is not guaranteed to eliminate all differences due to sample composition arising from how the galaxy halo connection is implemented in the simulations 
For clarity, we do not show the three other snapshots at $z>0$, but note that very similar qualitative trends are seen out to $z=1$.

Noticeably, the galaxy bias (horizontal axis, upper panel) agrees well between the different simulations;
given the relatively tight relation between halo mass and large scale bias (modulo cosmological parameter-dependence), this is perhaps unsurprising.
Although not shown for TATT, the marginalised posterior on bias is close to independent of the choice IA parameterisation, primarily because $w_{gg}$ dominates the constraint. 
The relative agreement between the detected NLA signal in the different simulations here is interesting, in the context of existing literature. 
It has been observed anecdotally (\citealt{tenneti16}, \citealt{chisari16})
that \mbii~tends to prefer a slightly stronger IA amplitude than \illustris. 
This conclusion is supported at some level here; in the NLA case \mbii~favours slightly larger $A_1$ values than either \tng, although the difference is less than $1\sigma$ at any given redshift. 
The difference is more pronounced in the TATT scenario (lower panel Figure~\ref{fig:results:contours} and also Figure~\ref{fig:results:redshift_dependence} below), although still only at the level of $1-2\sigma$. It is also worth remarking that this is the first time a robust comparison has been attempted with a homogenised sample, using $w_{g+}$ and $w_{++}$ simultaneously, and with an analytic covariance matrix that is numerically stable on large scales.  

The joint posteriors on the TATT model amplitudes $A_1,A_2$ are shown in the lower panel of Figure~\ref{fig:results:contours} (see also Appendix~\ref{app:posteriors} for the full TATT posteriors from the three simulations). These amplitudes can be thought of as controlling the strength of different IA contributions, which are linear and quadratic in the tidal field respectively. Note that the TATT fits also include additional parameters ($b_{\rm TA}$ and linear galaxy bias $b_{g}$), which are marginalised in this 2D representation (see Sec~\ref{sec:theory:tatt} and Table~\ref{tab:theory:models}). In this limited parameter space we do not believe our marginalised results to be significantly affected by prior volume effects (e.g. the discussion in \citealt{joachimi20}). We confirm that rerunning the TATT chains with a reduced $b_{\rm TA}$ prior $\mathrm{U}[0,6]$ does not qualitatively change the TATT posteriors.
In the case of \mbii~and \illustris, the constraint is degraded relative to \tng, to the extent that quite different TATT IA scenarios are allowed within $1\sigma$. That \illustris~offers little-to-no constraint on the extended model is unsurprising; indeed we are fitting a small handful of relatively noisy points in the $>6\mpc$ range, which provide no real information on the shape of the correlation function. Unlike in the NLA case, we now see some level of disagreement between the different simulations; that is, whereas \tng~favours a region of parameter space that resembles NLA (i.e. $A_2\sim0$), \mbii~prefers $A_2<0$ at $3\sigma$.
While this \emph{could} be a sign of a real alignment signal, generated by the physics models of \mbii, it is worth being cautious here; the TATT model will respond to any structure in the data, regardless of physical origin, and \mbii~has known limitations\footnote{In particular, there is a lack of realistic spiral type galaxies, and a relative over-abundance of diffuse elliptical objects compared with data. Due to relatively weak AGN feedback, \mbii~produces an over-predicts the number of massive galaxies at low redshift \citep{khandai15}. 
Another manifestation of this is seen in the ipact of baryons on the nonlinear matter power spectrum, which is significantly different from that in any other hydrodynamic simulation (\citealt{huang19}'s Figure 1).}. 
Inspecting the data vector (Figure \ref{fig:data:fiducial_datavector}) more closely, it seems that the $A_2<0$ is driven by the gradual rise in power between $10-1\mpc$. This feature is seen in both $w_{g+}$ and $w_{++}$, and it does indeed seem to be relatively well fit by the quadratic alignment contribution. It is also notable that there is no corresponding feature at around the same scale in $w_{gg}$, which is somewhat reassuring that this is a real signal, and not an artifact of the simulations. 
%

In all cases we note that the data favour low values of $b_{\rm TA}$, albeit with relatively large uncertainties. 
The region of parameter space where one could reasonably interpret the TATT tidal alignment bias as a pure physical galaxy bias are disfavoured at $\sim1\sigma$, with $b^{\rm TNG}_{\rm TA} = 0.26\pm0.82$, $b^{\rm MBII}_{\rm TA} = 0.25\pm0.73$. Interestingly, in the upper redshift bins \mbii~prefers a weakly negative $b_{\rm TA}$ (Table \ref{tab:results:fullpost}), the physical interpretation of which is not immediately clear. 
Given the sample selection, and the limitations of the simulations, it is not obvious that the low $b_{\rm TA}$ values transfer to real lensing data, but it is interesting, in the sense that the data are (mostly) showing a preference for the simpler IA scenario.

From the \tng~fits, the final posterior mean TATT parameter values at $z=0$ are: 
\begin{multline}
A^{\rm TNG}_1=1.27\pm0.48, \;\;\;\; A^{\rm TNG}_2=0.43\pm0.63, \;\;\;\; b^{\rm TNG}_{\rm TA}=0.26\pm0.74. 
\end{multline} 

\noindent
The $A_1$ constraint here is consistent with the equivalent NLA amplitude from the two-parameter fits ($A_1=1.71\pm0.17$), a conclusion that largely holds across the three simulations. That is, switching to TATT 
leads to a degradation in the uncertainty on $A_1$ (by roughly $50\%$ for \tng~at $z=0$), but no significant shift in the favoured value. Remarkably, although \mbii~favours negative $A_2=-2.3\pm1.0$ at the level of $\sim2-3\sigma$, \illustris~and \tng~are consistent with zero across the redshift range.
The small $A_2$ values differ slightly from recent studies on DES data (\citealt{y1cosmicshear}, \citealt{y1coloursplit}), which report a preference for $A_2<0$ (although our \tng~constraint is still at most $\sim2\sigma$ from the DES Y1 mixed sample; \citealt{y1coloursplit} Figure 12). 
Note however that in such analyses on photometric data like the studies cited above, where the two-point functions are measured in broad redshift bins as a function of angular scale, a significant amount of mode-mixing can occur. That is, one cannot cleanly separate physical scales. In addition to this, it is worth bearing in mind that no analysis on real data can ever be perfect; despite various robustness tests and validation carried out for DES Y1, we cannot altogether rule out the leakage of other modelling errors (e.g. in the photometric redshift distributions) into the IA constraints. For these, amongst other, reasons that it is not trivial to extrapolate from our results to comment on the detectability of higher order IA contributions in real data. 


The lack of a clear detection of higher order alignment terms 
is not altogether surprising, given the relatively conservative scale cuts implemented here ($r_{\rm p} > 6\mpc$; see also Section \ref{sec:results:scales}). Given the difference in physical scaling, naturally the alignment of galaxies on very large scales should resemble the tidal alignment scenario ($A_1>0,A_2=0$). 
Although we do not have a strong first-principles prediction of the scales on which the quadratic terms should become significant, we can make a rough estimate. Based on theory predictions, in scenarios that are consistent with previous observations (\citealt{y1coloursplit}), the regime where the tidal torquing terms are not totally subdominant to tidal alignment is somewhere on the scale of a few \mpc~(see \citealt{blazek15}, \citealt{blazek17}). This places our fits in the marginal regime, where it is possible, but not certain that we might detect a non-NLA-like alignment signal.

\begin{figure}
\includegraphics[width=\columnwidth]{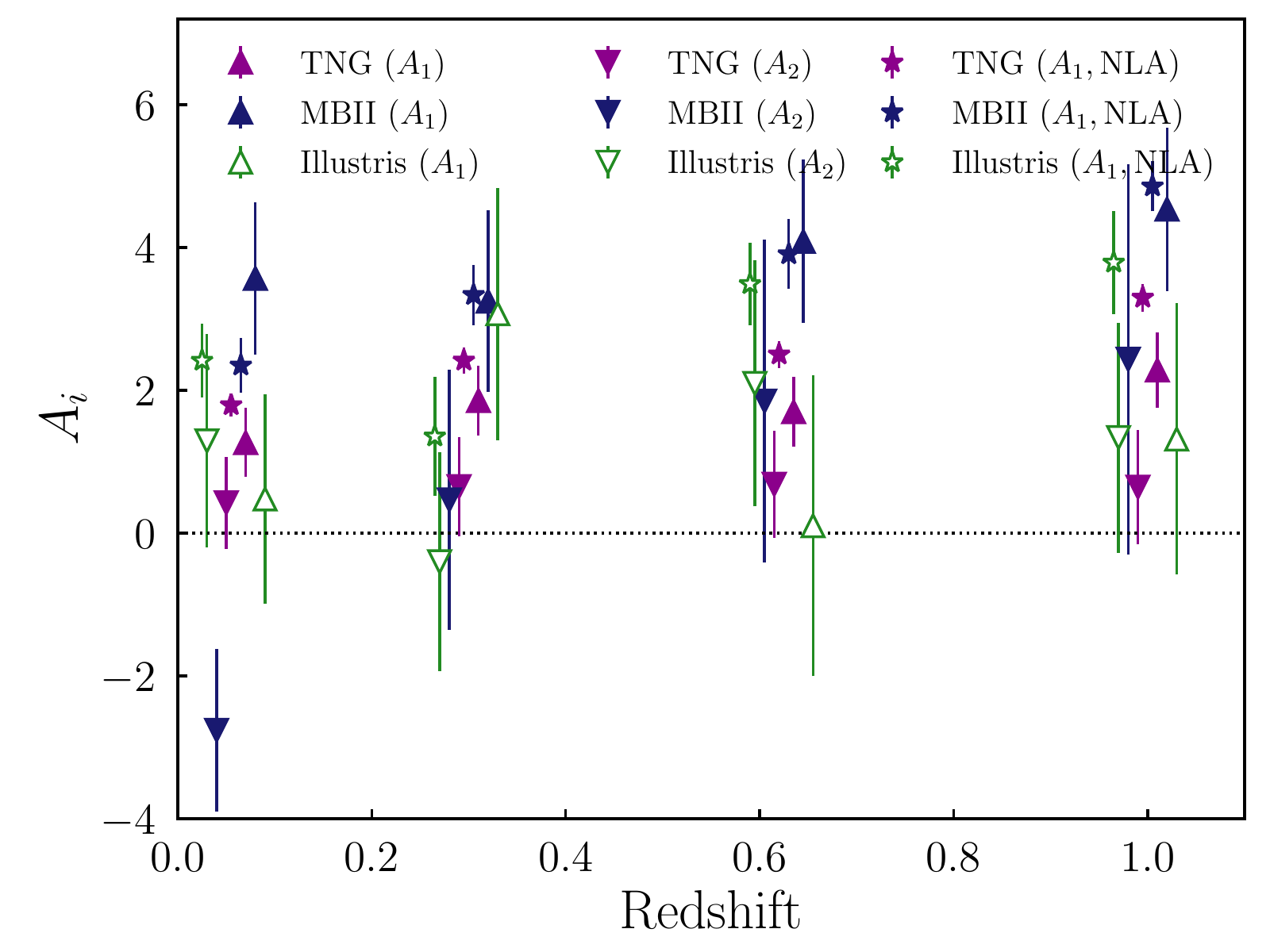}
\caption{
The redshift evolution of the two TATT model amplitudes in our three simulated datasets.
The fits were performed on each redshift slice independently.
The simpler NLA fits are also shown for reference (stars).
The results from \mbii~(dark blue), \illustris~(green) and \tng~(purple) are shown. 
Note that the galaxy samples for a given simulation at different redshifts strongly
overlap, and so the errors are potentially highly correlated, to an extent not
reflected in the $1\sigma$ error bars in this figure.
}\label{fig:results:redshift_dependence}
\end{figure}

\subsection{Evolution with redshift}

To illustrate the redshift evolution of the various IA parameters, we show
the marginalised best-fits and $1\sigma$ uncertainties in Figure~\ref{fig:results:redshift_dependence}. As before, we show all three simulations in purple/blue/green. It is worth keeping in mind here that there is significant overlap between samples at different redshifts, meaning the shape noise is potentially quite strongly correlated. The interpretation of the various trends shown in this figure, then, are not trivial.
That said, the basic patterns noted above are seen to hold across the redshift range. That is, with the partial, weak, exception of \mbii, the $A_1$ values obtained in the NLA and the TATT analyses are consistent with each other for a given simulation (compare the stars with the triangles in Figure~\ref{fig:results:redshift_dependence}).
The TA alignment amplitude rises more or less monotonically in \tng~and \mbii, and the two simulations agree well in the NLA case at all redshifts.
With the extra freedom of the TATT model, however, we see some level of divergence, with \mbii~favouring a higher $A_1$ by a factor of $\sim1.5-2.5$ (although the upwards trend with redshift persists).
This seems to fit with an underlying assumption of the linear alignment model: that IAs are frozen into a population of galaxies at early times (see \citealt{kiessling15}, \citealt{schmitz18} and \citealt{kirk12}, particularly their App. A and references therein). 
As the underlying large scale structure evolves and halos grow, the subhalo mass distribution
shifts upwards. In our case, then, the fixed stellar mass cut is more stringent, and removes a larger fraction of weakly aligned objects at high redshift than at low redshift. 
The net effect of this is an increase in the measured IA signal with increasing $z$. Though physically interesting, we reiterate that the
flat lower mass cut at each redshift is not representative of the selection function in a 
real lensing catalogue. In real data with realistic flux- and shape-based quality cuts, the changes in composition with redshift will have a significant bearing on how the effective IA amplitude evolves.  
A step in this direction (albeit still not capturing the full complexity of a redshift dependent selection in real data) would be to use the simulation merger tree to propagate through a mass cut at given redshift. \citet{bhowmick19} attempt such an exercise for \mbii, with results that are qualitatively consistent with the present study. 
Although in the consistently traced sample (\blockfont{SAMPLE-TREE} in their terminology) increasing halo-satellite misalignment tends to wash out alignments at high $z$, the impact of the changing population opposes, and largely outweighs this trend.

In contrast, $A_2$ is more or less constant with redshift in all simulations (the downward triangles in Figure~\ref{fig:results:redshift_dependence}). Notably \mbii's preference for $A_2<0$ is not seen to persist across snapshots, although the interpretation of this is non-trivial. 
Particularly in the higher redshift slices, the \mbii~posteriors exhibit significant bimodality, which appears to arise from a degeneracy between $A_2$ and $b_{\rm TA}$. Although positive and negative $A_2$ result in quite different $w_{g+}$ predictions, all other parameters held fixed, the combination $b_{\rm TA}\sim0, A_2\sim-3.5$ and $b_{\rm TA}\sim-1, A_2\sim2.5$ both produce theory curves that fit the $z=1$ data adequately on scales $\rp>6$ (see Figure~\ref{fig:app:post2}). The theory predictions differ somewhat on smaller scales, suggesting that pushing below our fiducial scale cut could potentially help to break this degeneracy. 
This distorts the 1D point representation in Figure~\ref{fig:results:redshift_dependence}, shifting the mean towards zero, and also broadening the $1\sigma$ standard deviation significantly.

\subsection{Tensions \& Model Comparison}

Beyond simple posterior constraints, one can also gauge the ability of the data to support
the extended modelling in a quantitative way. A number of goodness of fit metrics exist in the literature,
and we consider a subset of those here. Since \illustris~is relatively unconstraining, and is known to have flaws (in the sense that it over-predicts the strength of baryonic feedback, which is known to interact with IAs; \citealt{soussana19}),  
we compare the results using \tng~and \mbii~only. The simplest metric is the raw shift in $\chi^2$ when switching between models (see, for example, \citealt{krause15}); in the \tng~case, that is 
$\Delta\chi^2=-0.33$, 
marginally favouring the extended model, with similar values obtained at higher redshift. A somewhat stronger preference is seen in \mbii, which gives 
$\Delta\chi^2=-12.92$.
One slightly more sophisticated indicator of model fit is the Bayesian Information Criterion (BIC; \citealt{arevalo17}),
which effectively balances reducing the theory-data residuals against the extra complexity of the model.
For \tng, $\Delta \mathrm{BIC}=4.6$, which translates into a ``positive" preference for NLA. That is, by this indicator, the data do not warrent the additional parameters. In contrast, the \mbii~data, which we recall showed a preference for non-zero $A_2$, 
gives $\Delta \mathrm{BIC}=-3.8$, this time in favour of the TATT model.
Considering finally the Bayes factor \citep{marshall06}, we see a similar picture: $B=Z_{\rm TATT}/Z_{\rm NLA}=0.02$ for \tng, which indicates that the data favour the simpler model 
(or rather, the extra TATT parameters do not provide a sufficiently better fit to outweigh the added model complexity). Again, in the case of \mbii, the results are slightly clearer, with $B=5.65$, which (just) falls into the category of ``substantial" evidence on the Jeffreys Scale.
In summary, these numerical exercises bear out the qualitative picture we saw earlier; while \tng, on the relatively large scales considered, shows no evidence that the NLA model is insufficient, \mbii~does show hints.  

A different, but related, question one could ask is: given our results, and assuming a particular underlying model, to what extent can we say that there is disagreement between the simulations? Do the hints at non-zero TATT parameters in \mbii~point to systematic tension between the underlying physical alignment models, or are they in fact consistent with realisations of the same model? 
We reiterate here that the samples are weighted, such that differences in the underlying halo mass distribution should not be responsible for any differences between the simulations. 
Again, there are a number of metrics available, suited to different scenarios with different caveats (see \citealt*{campos20} for discussion), and we will not attempt a comprehensive comparison. For our purposes, we adopt a slightly different form of the Bayes ratio (see Eq. V.3, \citealt{y1keypaper}),
\begin{equation}
R = 
\frac{p\left(\mathbf{D}_{\rm TNG}, \mathbf{D}_{\rm MBII}| M_{\rm IA}\right )}
{p\left(\mathbf{D}_{\rm TNG}| M_{\rm IA}\right )p\left(\mathbf{D}_{\rm MBII}| M_{\rm IA}\right )}.
\end{equation}

\noindent
The numerator here is the Bayesian evidence obtained from jointly analysing the two-point data from 
the two simulations. The lower terms are those from the separate analyses of \tng~and \mbii~in isolation.
Note that in the joint analysis, we assume the two data sets are independent, with no cross covariance.
In the TATT case, we find $R<0.1$, which constitutes strong evidence for tension on the Jeffreys Scale. Again, this is implied by the differences we saw in the marginalised credibility contours, but it is interesting that it is borne out by the numerical metric. 

\section{Extensions Beyond the Fiducial Two-Point Analysis}\label{sec:results:extensions}

In this section we discuss a series of modifications to our baseline analysis, with the aim of exploring the basic results above in more depth. This includes a series of analyses with less stringent cuts, probing scales down to $1\mpc$. We also examine the dependence of the signal on various physical properties, including colour, type (central or satellite) and luminosity.  

\subsection{Exploring Smaller Physical Scales}\label{sec:results:scales}
As we have seen in Section \ref{sec:results:main}, our fits to the large scale \tng~correlation functions are consistent with the NLA scenario (i.e. pure tidal alignment). While there is a detectable IA signal, the parameters controlling deviations from NLA are consistent with zero. At least in principle, however, there exists a regime where the higher-order corrections are significant (and thus necessary to model the data adequately), but one halo contributions are still subdominant (see \citealt{blazek17}'s Fig.~1). It is this that motivates us to extend our fits below the fiducial cut off at $6\mpc$.

\begin{figure}
\includegraphics[width=\columnwidth]{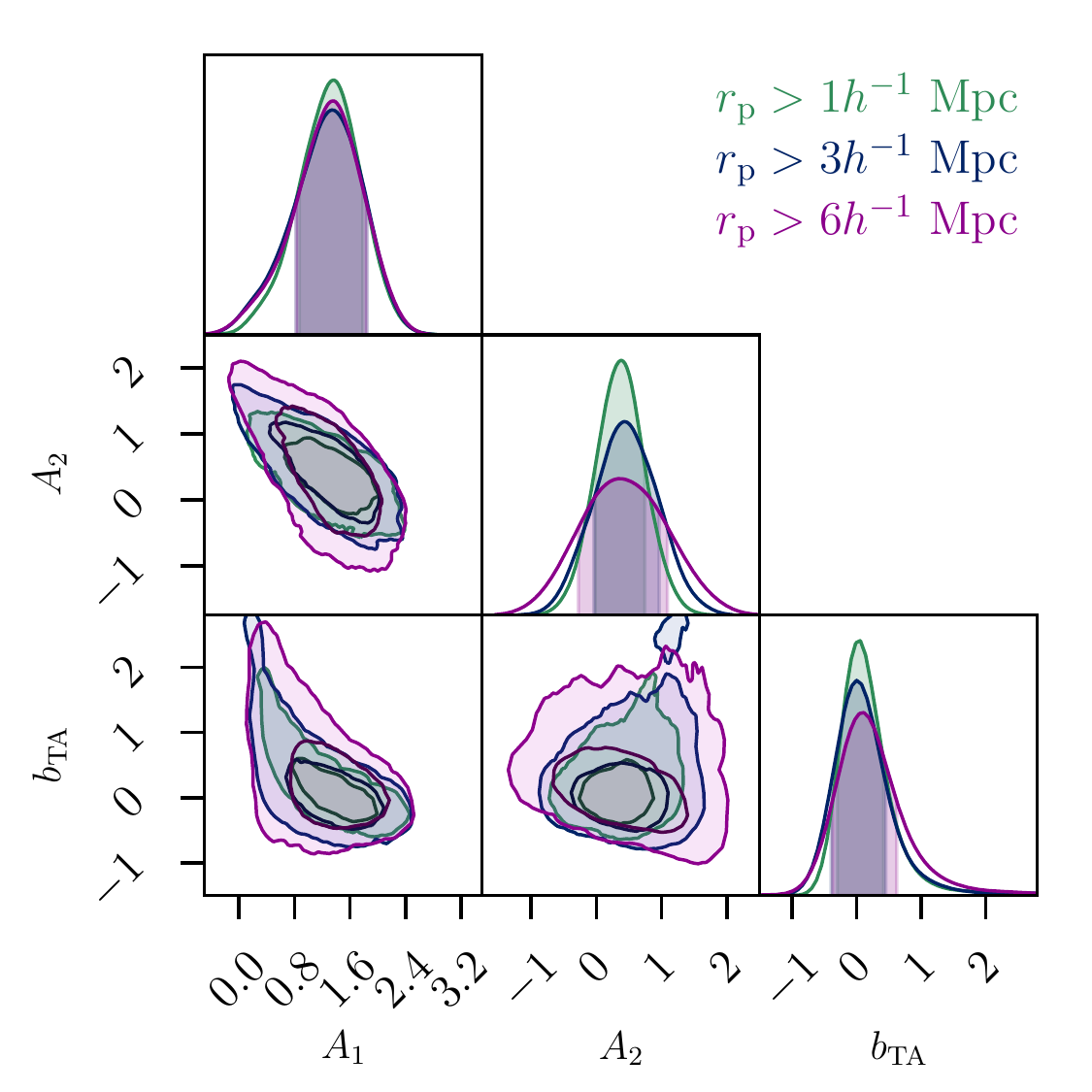}
\caption{TATT parameter constraints from our $z=0$ 
\tng~sample with a selection of lower scale cuts (as labelled). The three analyses favour approximately the same $A_1$, with slightly varying precision. Even in the case of the least stringent cuts, the results are consistent with $A_2=0$.
}\label{fig:results:tatt:scales}
\end{figure}

The fiducial cut follows \citet{joachimi11} and, as discussed there, is conservative by design, 
intended to be well clear of the scales on which nonlinear bias enters the data. The precise scales on which the linear approximation breaks down is, however, somewhat dependent on the galaxy selection, as well as the statistical precision of the measurement. One benefit of using simulated data, however, is that we have access to the dark matter field directly; it is, then, possible to check where exactly nonlinear galaxy bias begins to manifest in our particular measurements. A longer discussion can be found in Appendix \ref{app:bias}, but in brief we estimate the effective scale-dependent bias as a function of \rp~as the ratio 
$b_{g}=(w_{gg}/w_{\delta \delta})^{\frac{1}{2}}$. Based on this exercise, within \tng's statistical uncertainties, we see that the linear bias assumption holds well down to $\sim 1\mpc$. Motivated by this finding, we repeat our fiducial analysis, sequentially relaxing the lower scale cut down to $\rp>1\mpc$. The results can be found in Figure \ref{fig:results:tatt:scales} (see also Table \ref{tab:results:tatt:scales}).

As we can see in Figure~\ref{fig:results:tatt:scales}, all the way down to $1\mpc$, the higher-order TATT model parameters favoured by the \tng~data are consistent with zero. This includes the density weighting term $b_{\rm TA}$ (not shown), as well as the quadratic amplitude $A_2$. Although there appears to be information on the smaller scales, evidenced by the reduction in the size of the posteriors and the slight change in the degeneracy direction, there is no clear sign of deviations from NLA. The added constraining power is particularly clear in the case of the $A_2$ amplitude, although we also see a modest tightening of the uncertainties on $A_1$ and $b_{\rm TA}$ about their central values. It seems reasonable to draw from this that although we have physical reason to think that the additional TATT contributions exist in the Universe, they are small enough on the scales we use to be undetectable, given the statistical precision of \tng. The higher order terms scale rapidly with \rp, and so it is quite possible that they dominate in a similar regime to nonlinear galaxy bias.
This is also consistent with the conclusions one might draw from naively looking at the data vectors in Figure \ref{fig:data:fiducial_datavector}; the purple points are reasonably fitted by the purple lines (the best fitting NLA model), even down to scales $\sim 1\mpc$. This is true of both $w_{g+}$ and $w_{++}$ and, while deviations do exist, they are at slightly smaller scales. Fitting IAs on even smaller scales, where nonlinear bias becomes non-negligible, is possible, given that perturbation theory predicts higher order bias contributions in much the same way as the higher order IA contributions in TATT. It is, however, complicated by the presence of nonlinear bias - nonlinear IA cross terms, which we cannot safely assume are negligible. Although we do not attempt such an analysis here, implementing a consistent perturbative model, including the cross terms, is the focus of ongoing work.

We perform a similar exercise with \mbii, fitting the $z=0$ correlation functions down to $1\mpc$. Again, the constraints tighten significantly; now, however, the contours shift in the negative $A_2$, positive $A_1$ direction ($A_1=5.1\pm0.6$, $A_2=-3.9\pm0.3$, $b_{\rm TA}=-0.1\pm0.1$). 

\begin{table*}
\begin{tabular}{ccccccccc}
\hline
Cut / \mpc & Model                & $N_{\rm pts}$ & $A_1$ & $A_2$  &  $b_{\rm TA}$   &  $b_g$    \\
\hline
\hline

$\rp > 6$  & NLA        & 14      & $1.71\pm0.17$ & $0.0$ & $0.0$ & $1.11\pm0.07$ \\ 
$\rp > 6$  & TATT       & 14      & $1.29\pm0.49$ & $0.32\pm0.65$ & $0.21\pm0.86$ & $ 1.10\pm0.07$ \\ 
$\rp > 3$  & TATT       & 17      & $1.26\pm0.49$ & $0.45\pm0.47$ & $0.18\pm0.86$ & $1.11\pm0.05$ \\ 
$\rp > 2$  & TATT       & 20      & $1.30\pm0.44$ & $0.37\pm0.36$ & $0.21\pm0.68$ & $1.09\pm0.03$ \\ 
$\rp > 1$  & TATT       & 23      & $1.58\pm0.39$ & $-0.01\pm0.28$ & $0.28\pm0.38$ & $0.98\pm0.02$ \\ 

\hline
\end{tabular}
\caption{Quality metrics for TATT model fits to \tng~at $z=0$.
The second column, labelled $N_{\rm pts}$ indicates the total 
number of points included in the joint fit to $w_{gg}$, $w_{g+}$ and $w_{++}$,
after scale cuts.
}\label{tab:results:tatt:scales}
\end{table*}

\subsection{Dependence on Galaxy Properties}
In this section we impose a series of catalogue level splits, with the aim of understanding how our results depend on galaxy properties. For two main reasons, we only consider the fiducial \tng~catalogues in this section. First, the larger volume allows some leeway, such that sub-divisions can be made without degrading the constraining power beyond the point of usefulness. Second, and more importantly, only in \tng~do we have sufficiently realistic galaxy photometry (see Section \ref{sec:data:colour_split}). Although some of the properties considered here are correlated, we seek to disentangle the impact of each insofar as we can. For each of the cases discussed below, the new data vectors are recomputed using the same pipeline as before. For each subsample, we also repeat the iterative covariance matrix calculation discussed in Section \ref{sec:measurements:cov} with the appropriate galaxy densities and ellipticity dispersions.

\subsubsection{Galaxy Colour}\label{sec:results:colour_split}

\begin{figure}
\includegraphics[width=\columnwidth]{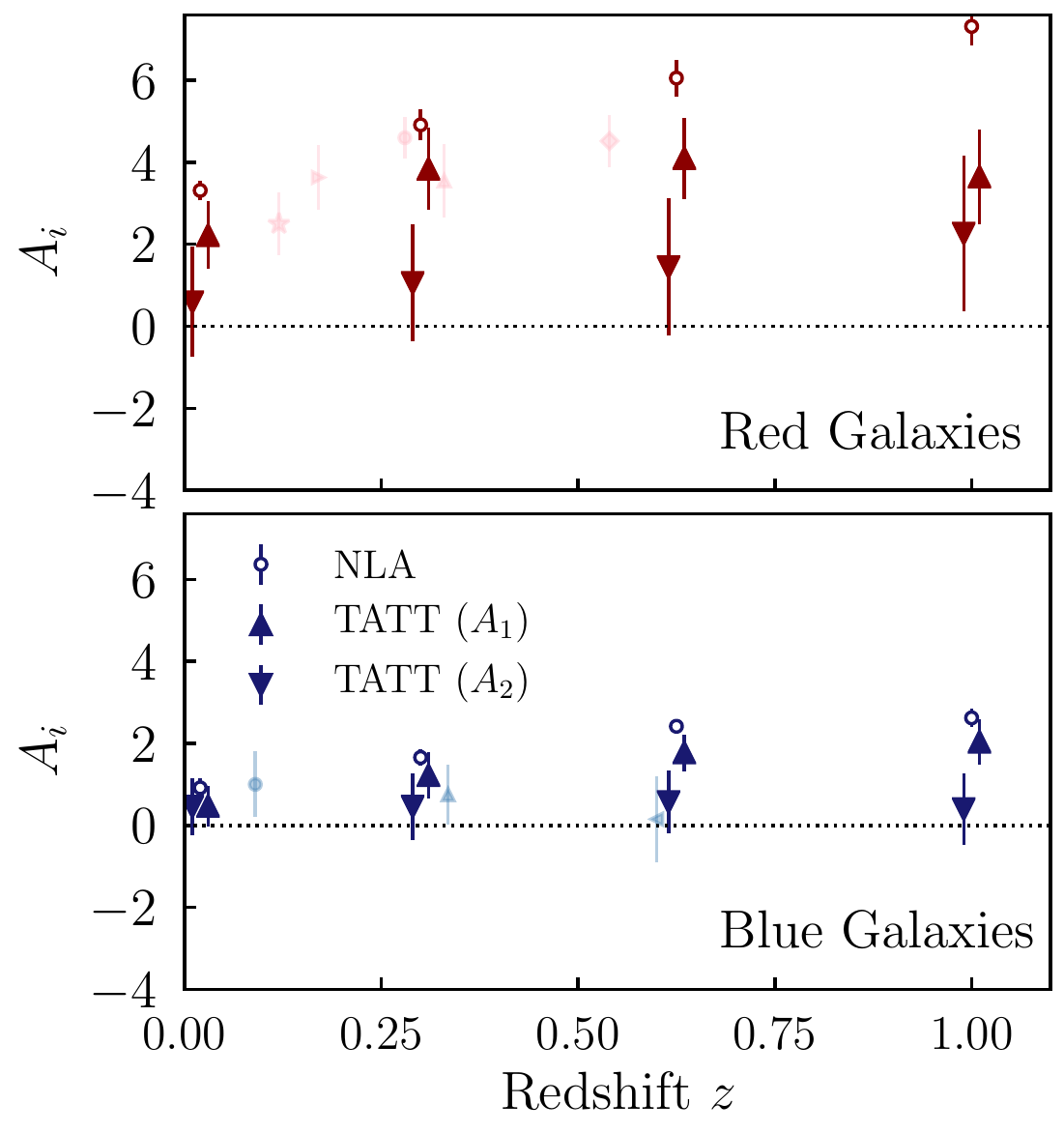}
\caption{Best fitting IA model parameters as a function of redshift for our 
colour split \tng~samples. Note that two TATT model amplitudes are fit simultaneously
for each of the two samples.
The pink points in the upper panel are measurements of the NLA model amplitude 
in red galaxies from the literature. Specifically, we show 
SDSS Main Sample ($z=0.12$; \citealt{johnston18}),
BOSS LOWZ ($z=0.28$; \citealt{singh15}),
GAMA red sample ($z=0.17$ and $z=0.33$; \citealt{johnston18})
and
MegaZ ($z=0.54$; \citealt{joachimi11}).
Similarly, the light blue points in the lower panel represent published blue galaxy
constraints: 
SDSS Main Sample (at $z=0.09$; \citealt{johnston18}),
GAMA Z2B ($z=0.34$; \citealt{johnston18}), 
and WiggleZ ($z=0.51$; \citealt{mandelbaum10}).  
}\label{fig:results:colour_split}
\end{figure}

The first split we examine is in colour-magnitude space. The ability to perform a colour cut, and retain a significant number of red and blue objects, is a marked difference between this work and previous direct IA measurements on real data, which have focused on bright red samples at low redshift. We recompute the correlation functions and covariance matrices for the red and blue subsamples described in Section~\ref{sec:data:colour_split}. As in all of our large scale fits, the full unsplit catalogue is used for the density part of the correlations. This gives us an analogous two new data vectors, $\mathbf{D}^{\rm red} = (\mathbf{w}^{RR}_{++}, \mathbf{w}^{R}_{g+}, \mathbf{w}_{gg})$ and $\mathbf{D}^{\rm blue} = (\mathbf{w}^{BB}_{++}, \mathbf{w}^{B}_{g+}, \mathbf{w}_{gg})$, with the superscripts $R$ and $B$ denoting the red and the blue samples. Note that the density tracer sample is not split, and so $w_{gg}$ here is the same in the two data vectors (and the same as that analysed in Section \ref{sec:results:main}). We fit both IA models using each data vector, with the results shown in Figure~\ref{fig:results:colour_split}. For the sake of clarity and to aid comparison, rather show the full parameter contours, we have condensed the IA amplitude parameters into 1D posterior means and 68\% error bounds. While this is useful for illustrative purposes, it can be reductive in cases where the posterior is non Gaussian, as we will discuss below. 

As before, the single NLA amplitude approximately agrees with the $A_1$ amplitude from the TATT fits in almost all cases; the exception to this is the high $z$ red sample, which favours a combination with nonzero TT contribution and a correspondingly lower TA amplitude, although the significance of $A_2\neq0$ is still only $\sim1-2 \sigma$. Although the details of the redshift distribution and the sample selection make direct comparison non-trivial, it is interesting to note that this disagrees mildly with the findings of \citet{y1coloursplit},  which are based on fits to real cosmological lensing measurements from DES Y1, where positive values of $A_2$ in a red source sample were disfavoured at the level of $\sim 2 \sigma$ (see their Figure~16). We also plot a number of previous direct IA measurements in Figure~\ref{fig:results:colour_split} (the pastel coloured points in both panels), from BOSS LOWZ \citep{singh15}, KiDS, GAMA and SDSS \citep{johnston18}, MegaZ \citep{joachimi11} and WiggleZ \citep{mandelbaum10}. Although red galaxy measurements are more numerous, there are a handful of comparable studies on blue galaxies. As one can see from Figure~\ref{fig:results:colour_split}, our fits on \tng~are largely consistent with the measurements on data. The only slight deviation from this is WiggleZ, which is lower than our results at equivalent redshift (albeit only by $\sim 2\sigma$). It is, however, worth bearing in mind that WiggleZ is atypical in terms of sample, comprising a bright starburst population, rather than a simple colour-selected blue sample. 

A notable, and perhaps worrying, feature of Figure \ref{fig:results:colour_split}
is the relatively strong IA signature in blue galaxies. The amplitude of $w^{B}_{g+}$, while significantly lower $w^{R}_{g+}$, is persistently non-zero at $z>0.5$.
To aid in understanding this observation, we repeat the two-point measurements
and NLA fits on the upper redshift snapshot, with an additional mass cut,
considering only galaxies in the lower $25\%$, $M_* < 2.1\times10^{10} h^{-1} M_\odot$
(mean stellar mass $M_*=1.3\times 10^{10} h^{-1} M_\odot$).
Even here, we see non-zero alignments at several $\sigma$, $A_1=2.1\pm0.6$.
Although lower than both the blue and unsplit samples at $z=1$, it is still a relatively
strong signal.  
Remarkably, we find that the high redshift blue IA feature persists under further mass
splitting, down to $M_* < 1.9\times10^9h^{-1} M_\odot$; at this point, there are only
$\sim 1000$ blue galaxies in the shape sample, such that although the measurement
is consistent with null, the errorbars still encompass significant non-zero values.
Although interesting, it is not clear whether this is a function of the relatively 
stringent convergence cut $(M_*>1.6\times10^9 h^{-1} M_\odot)$, and if so how far down
in mass the alignment signal continues. 
It is also not obvious whether this transfers to a significant IA lensing contaminant
+in a more realistic setup; implementing a redshift-dependent selection function, typical of real cosmic shear is a topic we will explore in future work.

Given that the role of a galaxy within its halo is a significant factor in determining its alignment, we also repeat the high $z$ blue measurements with an additional satellite/central split. The results here are less ambiguous: the residual blue galaxy signal is generated almost entirely by central galaxies. 
That is $w_{g+}$, as measured using satellite galaxy shapes is consistent with zero on scales $\rp>6\mpc$.
This result seems to support, at least in our case (which is simplified relative to real data in a number of ways),
the findings of \citet{johnston18}, which suggest colour alone is an imperfect determinant of IA properties.
\citet{singh15} also note similar, although consider only LRGs (that is, their results were a statement on the relative homogeneity of IAs in red sequence galaxies of given luminosity, rather than on the efficacy of colour based splits). In the absence of blue high $z$ alignment measurements in real data, it is difficult to say whether this is a fault in the simulations, generating an artificially strong IA signal in blue centrals, or a real feature of the Universe.

\subsubsection{Stellar Mass}

Although stellar mass is not, in general, an observable quantity it is an important one; this is true 
both in that IAs (and other galaxy properties) significantly depend on it, 
and that it is a proxy for actual observables. Indeed, this link is key to approaches such as Halo Occupation Distribution (HOD) modelling. 
We compute each galaxy's stellar mass as an unweighted sum over the stellar particles assigned to its subhalo. Unlike with colour and centrals/satellites there is no natural dividing line for this split, and so we choose to divide galaxies into equal number mass bins. 
For the moment we will consider a simple upper/lower mass division, but will consider a more complex binning in what follows. Again, the split is applied to the shape sample only, leaving the density tracer intact (and so $w_{gg}$ unchanged).

\begin{figure}
\includegraphics[width=\columnwidth]{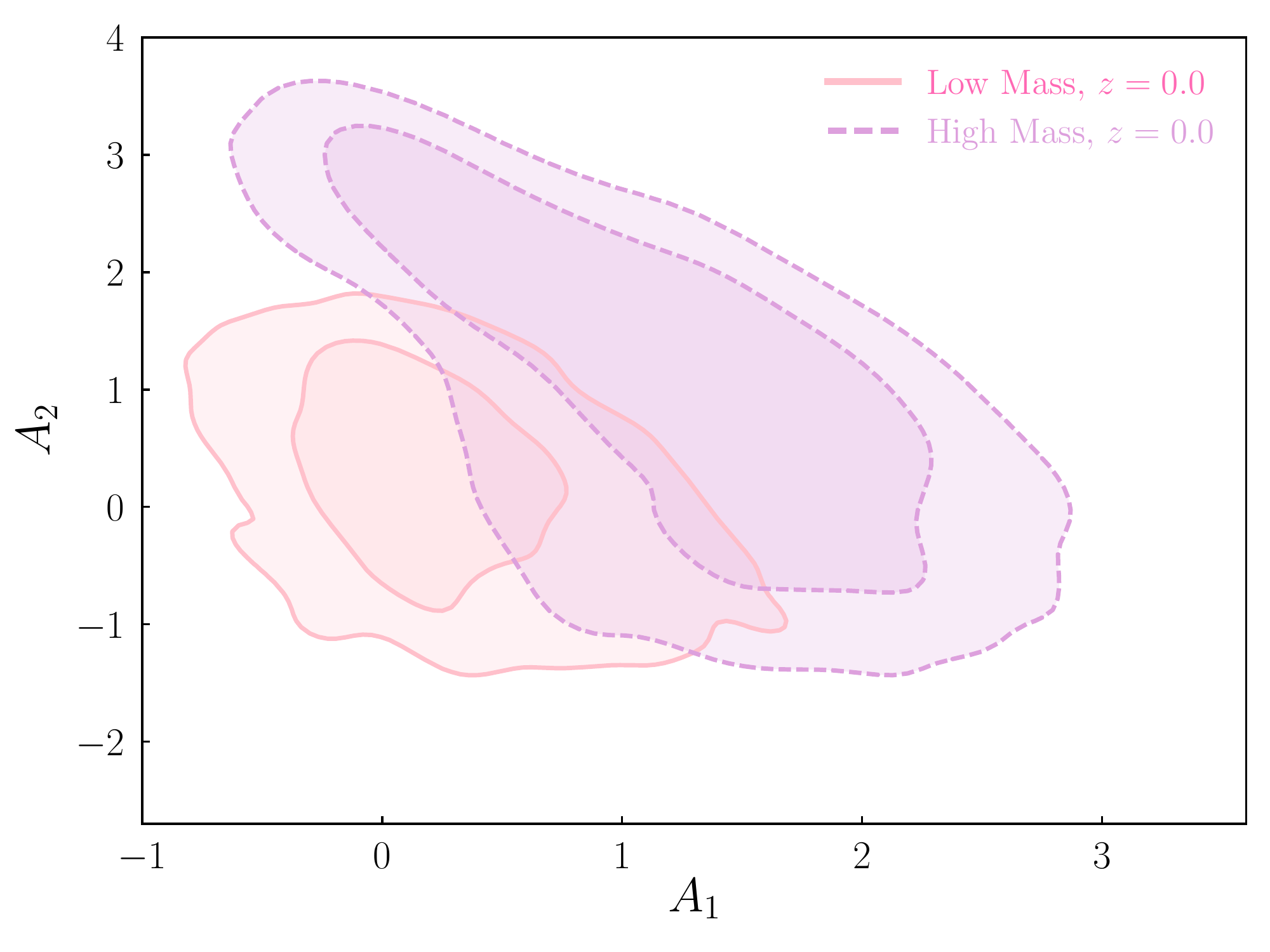}
\caption{TATT model posterior constraints from \tng, under a binary
high/low mass split. The sample is divided about the median stellar mass,
$M_{*} = 7.8\times 10^9 h^{-1} \mathrm{M}_{\odot}$, and the two subsamples
are fit independently.
}\label{fig:results:tatt:mass_split}
\end{figure}

Though the satellite fraction is not systematically changed by the division in any of the snapshots, we do see a shift in the abundence of red galaxies. That is, the red fraction of the high mass sample is boosted relative to the full sample, from $\sim35\%$ to $\sim 60\%$ at $z=0$ and from $\sim12\%$ to $24\%$ at $z=1$. 
This qualitative trend, that the red fraction increases with mass, and declines with redshift,
is consistent with the patterns seen in real data (see e.g. \citealt{prescott11}). In the case of the NLA constraints we have a relatively simple picture from the mass-split reanalysis;
at a given redshift, high mass galaxies are both more biased, and more strongly 
aligned, as illustrated in Figure \ref{fig:results:tatt:mass_split}; 
the direction of the shift in bias-IA amplitude parameter space when going from the high to
low mass is roughly the same, irrespective of redshift. 
The redshift trend can perhaps be understood as follows:
if we are to believe the basic LA model premise, then 
intrinsic alignments are imprinted at early times, and persist
into the low redshift universe. In this picture, at least, high mass red galaxies at
$z=0$ are strongly aligned, and so we can extrapolate from this that the objects that
\emph{become} bright red high mass galaxies are also strongly aligned.
In other words, as redshift increases, even if the mean subhalo mass declines
(which it does, in Table \ref{tab:data:catalogues}) the more massive, redder section of 
the galaxy population will be strongly aligned.

What does stay fixed, however, is the lower mass threshold we impose on our
catalogues. As the whole mass distribution shifts downwards, then, we are
preferentially cutting more of the lower part of the mass distribution, and so
discarding a larger fraction of weakly aligned objects.
The gradual evolution in IA and bias parameters covers the range from
virtually unaligned low mass galaxies at $z=0$ to $A_1\sim3.5$ in the high
mass high redshift bin.
It is worth bearing in mind that, although physically interesting, 
this pattern does not trivially carry over into real data, because in such cases
other observational effects become relevant.
The fact that we typically use flux-limited galaxy samples for lensing measurements, 
for example, means that the mean stellar mass tends to \emph{increase} with 
redshift, not decline as in our case.
Fully separating out these effects, of sample composition and evolution of intrinsic alignments, would require a more careful exploration using merger trees of the sort presented by \citet{bhowmick19}.

\begin{figure}
\includegraphics[width=\columnwidth]{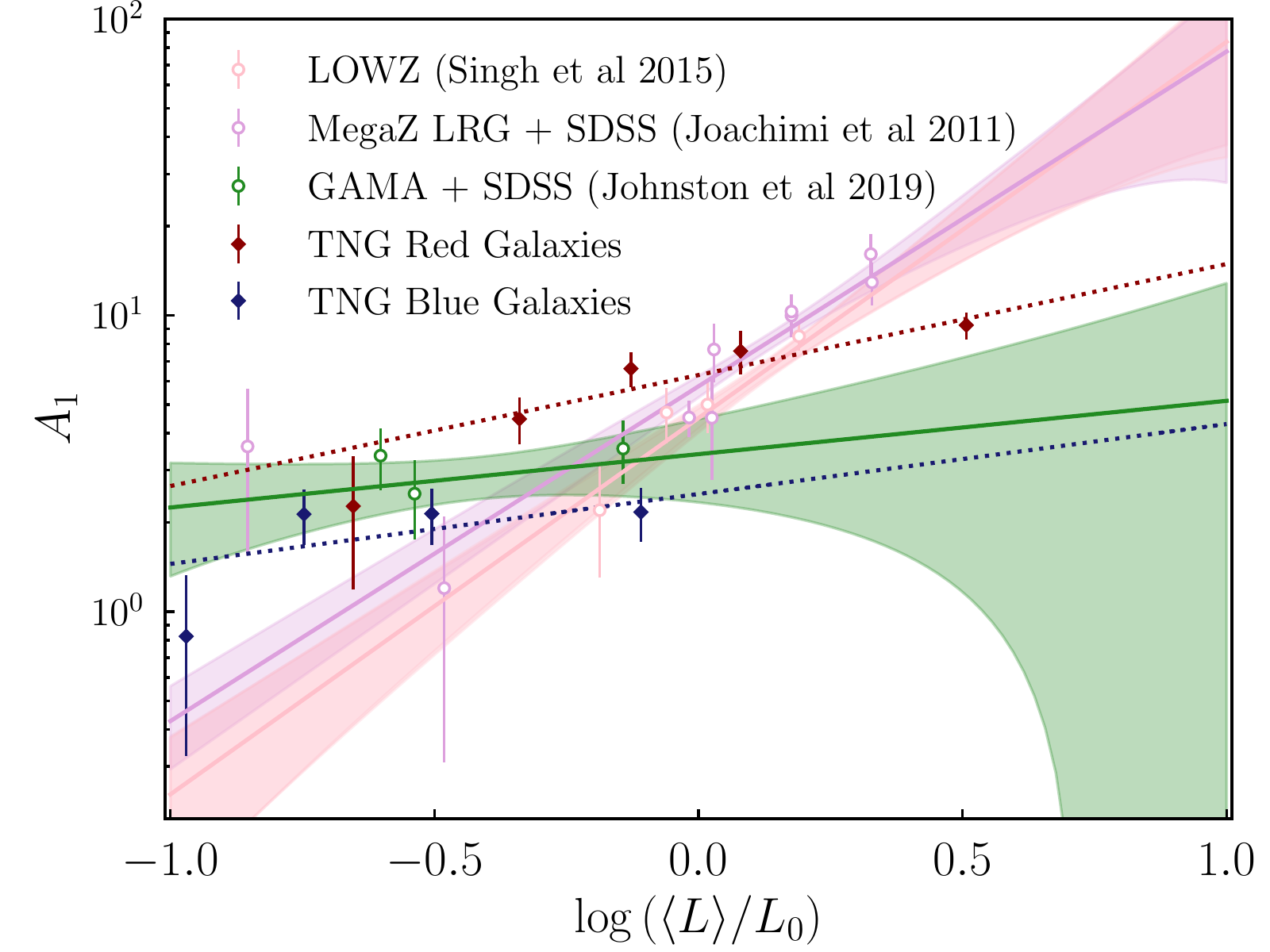}
\caption{Luminosity dependence of the measured NLA intrinsic alignment amplitude.
The red/blue diamonds show the two colour subsamples of \tng~at $z=0.3$, and the
dotted lines of the same colour show power law fits to these data.
For reference, we also show comparable measurements from MegaZ + SDSS LRG + L4 + L3 \citep{joachimi11},
BOSS LOWZ \citep{singh15}
and KiDS$\times$GAMA+SDSS \citep{johnston18} in purple, pink and green. The shaded bands represent their
fits and the corresponding uncertainties.
}\label{fig:results:nla:luminosity_dependence}
\end{figure}

\begin{figure}
\includegraphics[width=\columnwidth]{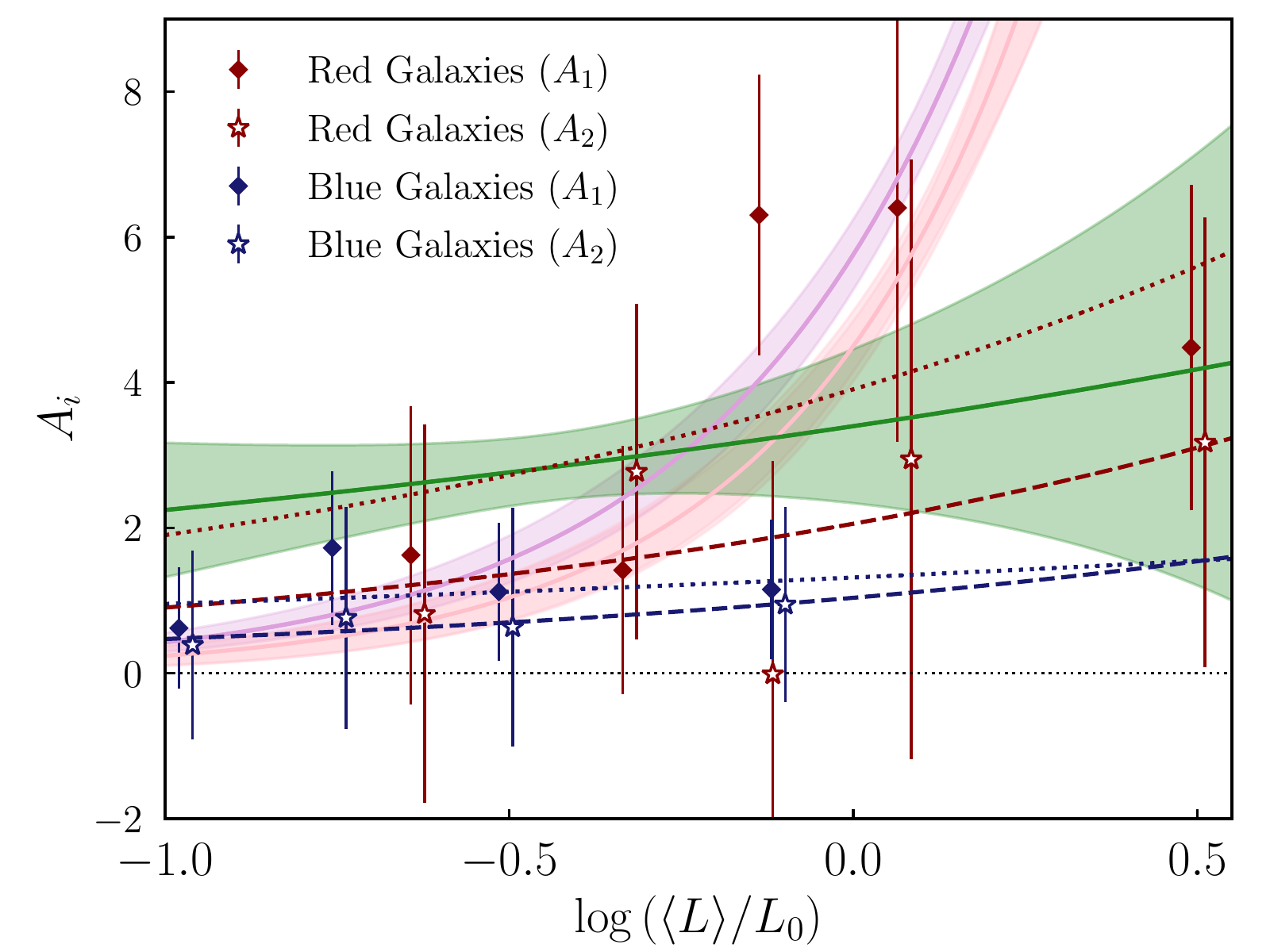}
 \caption{Luminosity dependence of the measured TATT model intrinsic alignment parameters. 
 As in Figure \ref{fig:results:nla:luminosity_dependence} the shaded bands show fits to
 MegaZ, LOWZ and SDSS+GAMA (points now omitted). The diamonds show the tidal alignment amplitude
 $A_1$, while the stars show the tidal torquing contribution $A_2$. We show the best fitting power laws, parameterised $A_i(L)=A_{i,z} (L/L_0)^{\beta_i}$, for $A_1$ (dotted) and $A_2$ (dashed). The numerical values of the power law slopes are quoted in Section \ref{sec:results:luminosity_dependence}.
}\label{fig:results:tatt:luminosity_dependence}
\end{figure}

We also rerun our TATT analysis on the mass-split data vectors, giving the 
marginalised parameter constraints shown in Figure \ref{fig:results:tatt:mass_split}.
For clarity, we show only $z=0$ here, but find similar patterns in all snapshots.
As before, $A_1$ gradually increases over the range $z=0-1$, and the $1\sigma$ contour 
encompasses $A_2=0$ in all cases. That said, there is a relatively strong anti-correlation between the two IA amplitudes, such that a range of scenarios with $A_2>0$, combined with slightly reduced $A_1$, are also equally favoured.  

\subsubsection{Luminosity}\label{sec:results:luminosity_dependence}

In addition to the binary mass cut above, we also consider directly the luminosity
dependence of the measured IA signal, defining four equal-number bins in $r-$band luminosity.
To separate actual luminosity dependence and changes in the red fraction between bins, we impose the red/blue colour split described in Section \ref{sec:results:colour_split}. The signal-to-noise in the upper red bin is particularly high, which motivates a further equal number subdivision, slightly extending our luminosity coverage. 
We then have the luminosity bins
$L_{r,\mathrm{red}}/L_0 = [(0.030-0.351), (0.351-0.575), (0.575-0.977), (0.977-48.091)]$
and
$L_{r,\mathrm{blue}}/L_0 = [(0.028-0.131),(0.131-0.224),(0.224-0.424),(0.424-16.341)]$,
which roughly, but not exactly, correspond to mass bins.
In each one we recompute the correlation functions and the covariance matrix, then fit using both IA models. As before we impose the split only on the shape sample, which is correlated with the full density sample. The results, as a function of $r-$band luminosity, are shown for NLA and TATT respectively in Figures \ref{fig:results:nla:luminosity_dependence} and \ref{fig:results:tatt:luminosity_dependence}. For the purposes of comparison with the literature, we consider the second snapshot, $z=0.3$ only here. This choice does not significantly change the conclusions of this section. 

Figure~\ref{fig:results:nla:luminosity_dependence} shows the the NLA case, with open points indicating previous constraints on the IA luminosity relation using data (\citealt{joachimi11}, \citealt{johnston18}; see also \citealt{fortuna20}'s Figure 5 and \citealt{singh15}'s Figure 10). All of these represent direct IA measurements at low redshift, using relatively bright red samples.
The MegaZ + SDSS LRG + L4 + L3 fit \citep{joachimi11} in particular has been widely used in the literature to extrapolate the luminosity dependence to fainter samples (see e.g. \citealt{krause15}).
In addition, we show our new results from \tng, both red and blue samples. The blue simulated subsample lies towards the fainter end of this plot, going fainter than any of the data measurements. The red, on the other hand, covers a wider luminosity range, spanning both KiDS$\times$GAMA+SDSS \citep{johnston18} and the MegaZ + SDSS LRG + L4 + L3 fit \citep{joachimi11} points.
It is worth remarking here that unlike Figure~\ref{fig:results:colour_split}, the points each represent a different set of galaxies. Whereas there the different snapshots strongly overlap, and so are subject to highly correlated errors, the noise realisations should now be independent, making fitting a trend relatively simple.
The luminosity dependence is parameterised as

\begin{equation}\label{eq:luminosity_power_law}
A_1(L,z) = A_z \left ( \frac{L}{L_0} \right )^{\beta_1},
\end{equation}

\noindent
where $L_0$ is a pivot luminosity, corresponding to an absolute magnitude $M_r = -22$. The amplitude $A_z$ and power law index $\beta_1$ are left as free parameters in our fits. Doing a simple least-squares fit to Equation~\eqref{eq:luminosity_power_law}, we obtain $\beta_{1,\mathrm{red}} = 0.38 \pm 0.08, A_{z, \mathrm{red}} = 6.3\pm0.5$ for the \tng~red sample. Notably, this is somewhat shallower than both MegaZ + SDSS LRG + L4 + L3 ($\beta_{1, {\rm MegaZ}} = 1.13^{+0.25}_{-0.27}$; shaded purple in Figure~\ref{fig:results:nla:luminosity_dependence}) and LOWZ ($\beta_{1, {\rm LOWZ}}=1.27\pm0.27$; pink shaded); it is slightly steeper than, but consistent to $\sim1\sigma$ with, the KiDS$\times$GAMA+SDSS red sample ($\beta_{1, {\rm GAMA}} = 0.18^{+0.20}_{-0.22}$; green shaded). For the most part, this fits with the broken power law picture painted by the existing datasets (i.e. a relatively steep slope at high $L$, turning into a much flatter function below $L/L_0\sim0.8$). Our uppermost $L$ bin, however, indicates something slightly different; the measured IA amplitude in this bin is both relatively well constrained, and below the extrapolated MegaZ power law prediction by several $\sigma$. Taken together with the third point $A_1(L/L_0=0.74) = 6.6\pm0.9$, which is slightly above the data, this seems to hint at some level of disagreement between simulations and data. 

One caveat here is that the $x$ axis positions are point estimates from luminosity distributions, which have finite width. In the case of the blue sample, the distributions are relatively compact and Gaussian; in the case of the higher $L$ red sample luminosity bin, this is not the case, and the $p(L)$ distribution is broad, with a trailing upper tail,
reaching a maximum luminosity of $\mathrm{log}(L/L_0)=1.6$.
Using the modal luminosity as our point estimate, the rightmost point in Figure~\ref{fig:results:nla:luminosity_dependence} shifts slightly to the left, thereby reducing the apparent tension with the earlier results. 

Another complicating factor here is the evolving satellite fraction in both our, and the published, samples. In our case, the \tng~red sample satellite fraction changes significantly from $f_s = 0.7$ in the lowest luminosity bin ($\mathrm{log}\langle L \rangle /L_0 = -0.53$) to $f_s = 0.18$ in the brightest bin ($\mathrm{log} \langle L \rangle /L_0 = 0.5$). In contrast, the satellite fraction of the \tng~blue sample is quite stable at $f_s \sim 0.3$ across the $L$ range. We explore the impact of this directly by repeating the measurements using red centrals only; as one might expect from the numbers above, the amplitude in the lower bins shifts upwards slightly (to $A_1=3.9\pm1.0$), to a value which is consistent with the central only GAMA measurement (the yellow point in \citealt{fortuna20}'s Fig. 14). The upper luminosity bins are almost completely unchanged (since the IA signal in those bins is heavily dominated by centrals anyway).
That is, the central/satellite trend does not seem to be sufficient on its own to explain the discrepancy between our results on \tng~and the steeper slope seen in LOWZ and MegaZ.

The trend in blue galaxies, $\beta_{1,\mathrm{blue}} = 0.24 \pm 0.21, A_{z, \mathrm{blue}} = 2.5 \pm 0.7$. is also interesting, particularly given the lack of existing blue sample measurements. Our results are consistent with no luminosity evolution in blue galaxies, at least at the faint end of the luminosity function. 
More troubling, perhaps, is the persistence of a relatively strong blue IA signal. This is in accordance with our previous findings, but it is particularly striking here that even in the faintest blue galaxies at $z=0.3$, there is a non-negligible IA signal, $A_1=1-2$. Although the topic clearly warrants caution, and further investigation in real data, if it bears out this could have significant consequences for future cosmic shear anlayses. 

An important point to bear in mind here is the choice of pivot luminosity. By convention, and to facilitate comparison with previous results on real data, we choose a pivot $L_0$ corresponding to an $r-$band absolute magnitude of $M_r = -22$ (see e.g. \citealt{joachimi11,singh15,johnston18}). This is appropriate for those studies, and for our red galaxy sample, in that $L_0$ is more or less in the centre of the luminosity range. In the case of the blue \tng~sample, however, the bulk of the sample is below $L/L_0=1$. Although this is a valid analysis choice, and indeed useful for comparison with the literature, it does mean that $A_z$ and $\beta$ are likely non-trivially correlated in this case. 

Finally, we repeat this exercise using the TATT model, again at $z=0.3$, with the results shown in Figure \ref{fig:results:tatt:luminosity_dependence}. Again, we show the best fits to MegaZ, LOWZ and KiDS$\times$GAMA+SDSS, but now for clarity we omit the corresponding data points. Although we fit power law slopes as before, the constraints are degraded relative to the NLA case. Fitting to the red sample, we find $\beta_{1,{\rm red}} = 0.31 \pm 0.31$, $\beta_{2,{\rm red}} = 0.36 \pm 0.35$.
As before, the red galaxy TA amplitude $A_1$ increases from faint to bright galaxies ($\beta_{1,\mathrm{red}}>0$), as one would naively expect. In all but the brightest two bins in the red galaxy sample, the TT amplitude $A_2$ remains consistent with zero to $\sim 1 \sigma$. Although we report weak positive $\beta_2$ here, the fits are extremely noisly, such that very different scenarios are allowed within the uncertainties. At the current precision, then, there is little hope of distinguishing between power law and non power law forms of luminosity evolution, at least for $A_2$.
In the blue sample, we find $\beta_{1,{\rm blue}} = 0.14 \pm 0.31$, $\beta_{2,{\rm blue}} = 0.34 \pm 0.14$,
consistent with no coherent variation with $L$ across the range.

\subsubsection{Centrals \& Satellites}\label{sec:results:sc_split}

As well as luminosity and colour, the IA signal is known to depend on galaxy type, and so we next consider a satellite/central spit. We divide the \tng~shape catalogues into centrals and satellites
using the method described in Section~\ref{sec:data:cs_split}, 
before remeasuring the two-point functions and repeat the covariance matrix calculation.
This gives us two new data vectors 
$\mathbf{D}^{c} = (\mathbf{w}_{++}^{cc}, \mathbf{w}_{g+}^{c}, \mathbf{w}_{gg})$,
$\mathbf{D}^{s} = (\mathbf{w}_{++}^{ss}, \mathbf{w}_{g+}^{s}, \mathbf{w}_{gg})$.
In all cases the shape part of the correlation is either central or satellite, and the density part uses the unsplit catalogue. Note that we repeat this exercise with  split density samples, and confirm that we return consistent (albeit slightly degraded) IA constraints when fitting on large scales.  

\begin{figure}
\includegraphics[width=0.96\columnwidth]{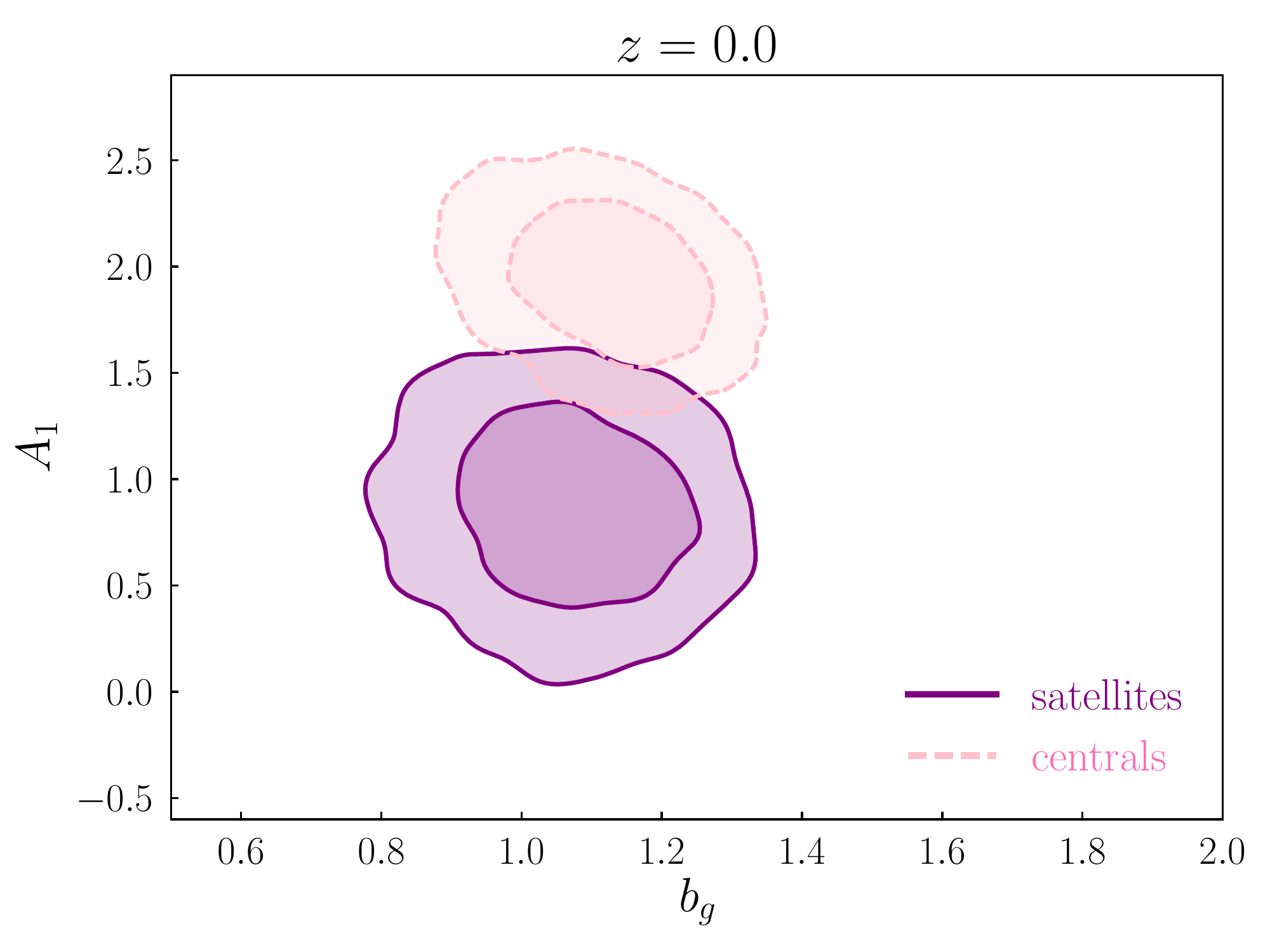}
\caption{NLA model constraints on our \tng~sample, split into
satellites and centrals. 
In both cases the split is imposed on the shape sample;
the density tracer sample used in $w_{g+}$ and $w_{gg}$ is the
full $z=0$ \tng~catalogue.
As shown, the large intrinsic alignment of satellites is 
weaker than that of centrals by a factor of $\sim 2$.
}\label{fig:results:nla:cs_split}
\end{figure}

The NLA analysis on these new satellite/central disaggregated data vectors are shown in Figure \ref{fig:results:nla:cs_split}. As expected, the galaxy bias is consistent between the two, and constrained primarily by $w_{gg}$, which is the same in the two data vectors. By construction the large scale fits to these data are each sensitive to a particular combination of two halo IA power spectra. Specifically 
$\mathbf{D}^{c}$ is sensitive to
($P^{2h, s}_{\rm GI}$, $P^{2h, ss}_{\rm II}$),
and 
$\mathbf{D}^{s}$
probes 
($P^{2h, c}_{\rm GI}$, $P^{2h, cc}_{\rm II}$).
Again, we assume that on two halo scales, the satellite/central composition of the density tracer is not relevant. 
Notably, the amplitude of large scale central alignments in Figure \ref{fig:results:nla:cs_split} is stronger than that of satellites by a factor of $\sim2$, at the level of a few $\sigma$. The subject of satellite alignments has been discussed quite extensively in the literature, and the overall picture fits with our results here. A number of theoretical studies point to satellite IAs being dominated by tidal torque induced radial alignments within their halos \citep{knebe08,faltenbacher07,pereira08}, which scale rapidly with separation, and tend to wash out on very large scales. There is also now evidence from various observations on both cluster and galaxy scales supporting the same picture \citealt{sifon15,singh15,huang18}). This, again, is consistent with \citet{johnston18} and \citet{fortuna20}, who suggest
satellite shapes are effectively random on sufficiently large scales. 
Centrals, on the other hand, tend to align with the host halo, and so trace the large scale correlations in the background large scale structure \citep{catelan01,kiessling15}. Although not shown in Figure \ref{fig:results:nla:cs_split}, it is also worth noting that the central galaxies show a clear monotonic increase in IA amplitude with redshift, a trend which is not replicated in satellite galaxies. 

\begin{figure}
\includegraphics[width=0.96\columnwidth]{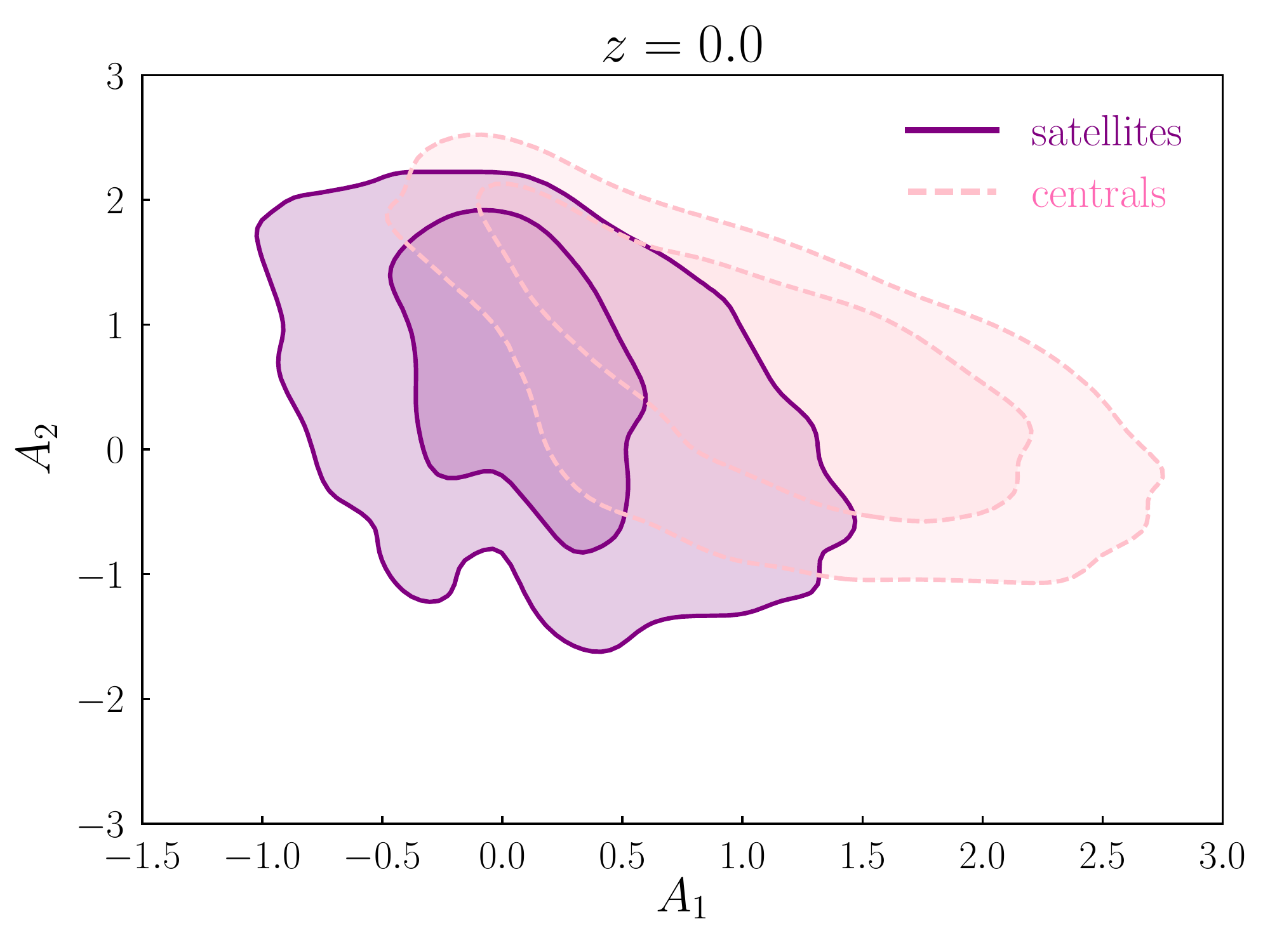}
\caption{TATT model constraints from \tng~at $z=0$,
after decomposing
the catalogue into central and satellite galaxies. The metric used to define
the two classes is outlined in Section \ref{sec:results:sc_split}. 
As in Figure \ref{fig:results:nla:cs_split}, the density sample used here
is the full, unsplit catalogue, and the satellite/central split is imposed
only on the shape sample. 
}\label{fig:results:tatt:cs_split}
\end{figure}

We next repeat our analysis of the 4 
split data vectors, but now using the three parameter TATT model instead of NLA. 
Figure~\ref{fig:results:tatt:cs_split} shows the marginalised parameter constraints at four redshifts. As in the simpler fits above, the central IA signal is stronger than that in satellite galaxies by a factor of a few, although
the constraints are degraded to the extent that it is difficult to draw meaningful predictions from this.
Again, there is no clear evidence of non-zero favoured values of either the tidal torquing amplitude $A_2$, or the density term $A_\delta = b_{\rm TA} A_1$, in either the satellite or the central population.

\begin{figure*}
\includegraphics[width=1.5\columnwidth]{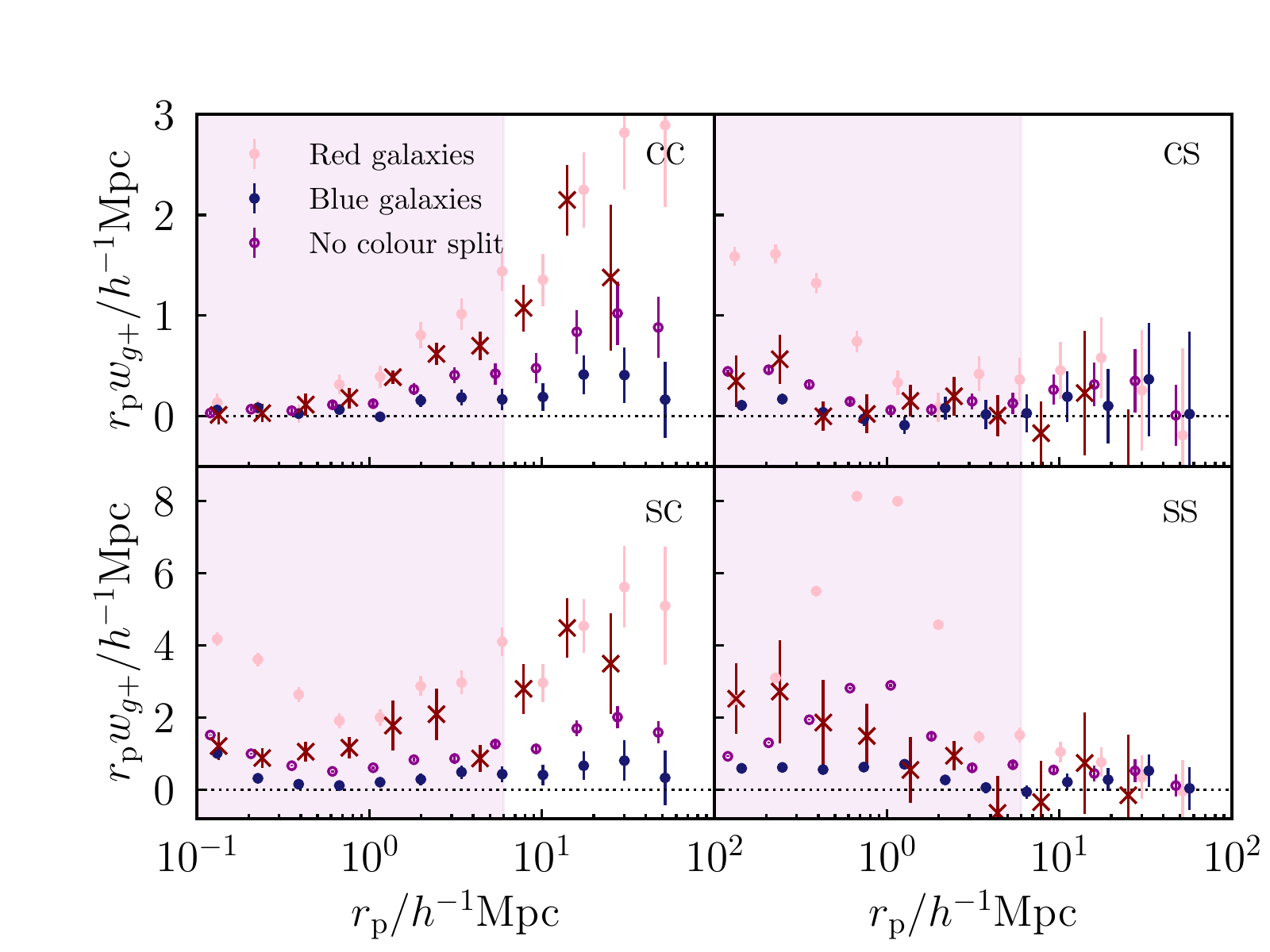}
\includegraphics[width=1.5\columnwidth]{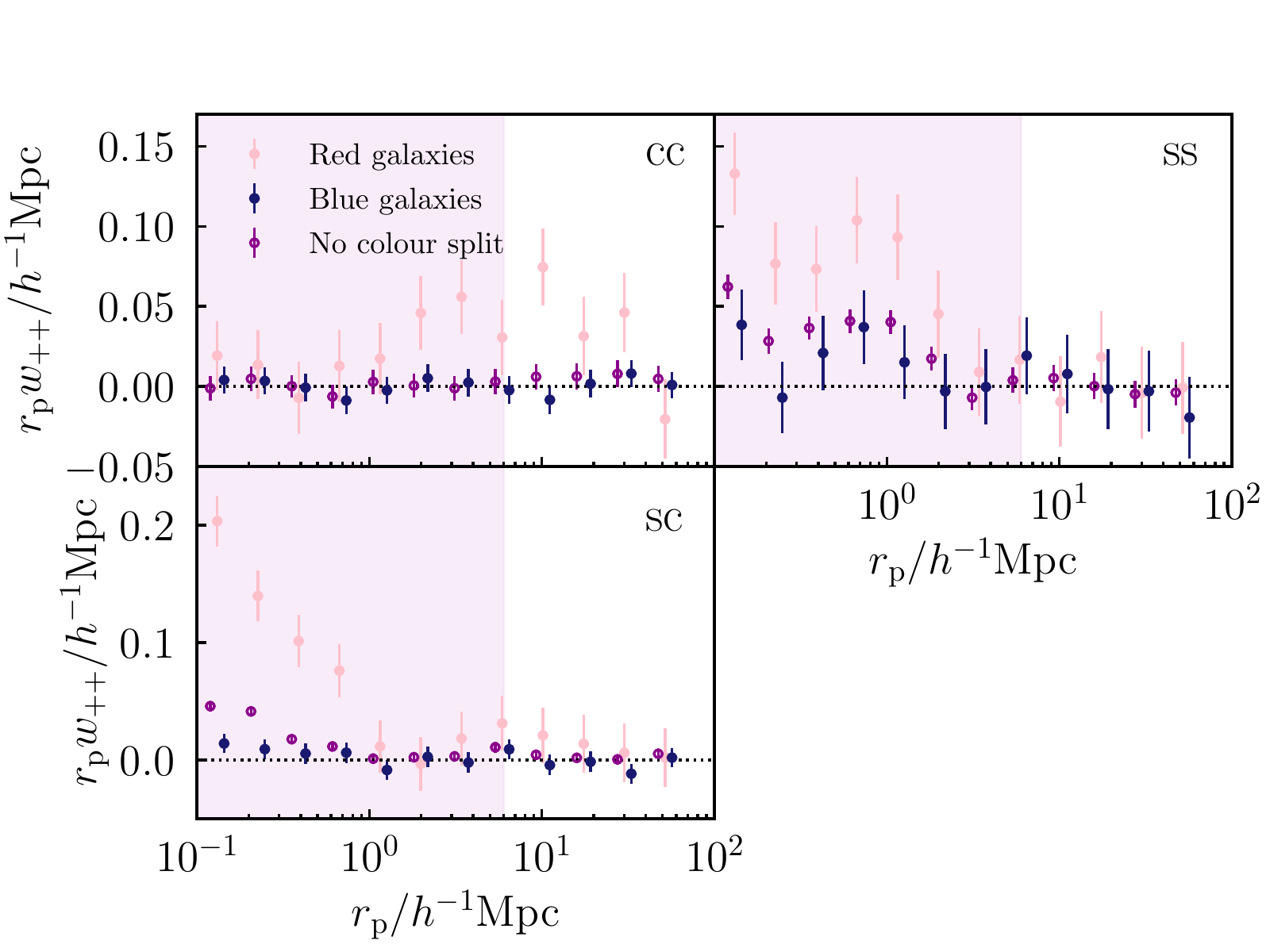}
\caption{\emph{Upper}: Projected galaxy-shape correlation functions $w_{g+}$, for various subsamples of the $z=0$ \tng~sample. Shown clockwise from the top right are the central autocorrelation; the satellite shape - central position correlation; the satellite autocorrelation; and the central shape - satellite position cross correlation. In addition to the satellite/central split, galaxies are split by colour (shown by the different colour points).
For reference, the shaded region shows scales excluded in the fiducial analysis.
The red crosses are analogous split measurements on real KiDS+GAMA data 
(\citealt{johnston18}; see their Fig. 7).
\emph{Lower}: The same, but for shape-shape correlations $w_{++}$.
}\label{fig:results:wgp_cs_split}
\end{figure*}

We also present the correlation functions of centrals and satellites in Figure \ref{fig:results:wgp_cs_split}. This is a worthwhile exercise for a variety of reasons, not least that there is information on the small scale IA signal missed in the large scale fits. While we cannot, at the present time, fit the IA signal on scales $<\sim 1\mpc$, the qualitative comparison can be instructive. Given that there is some evidence that they behave differently, we consider blue and red satellites/centrals separately here. We also drop the full sample density tracer, and instead use one of the four (red/blue, satellite/central) subsamples. The motivation here is that, while on large scales, the density tracer is effectively just that: a probe of the large scale matter distribution multiplied by a linear galaxy bias, on scales approaching the one halo regime this no longer holds. 
Figure \ref{fig:results:wgp_cs_split} shows these new data vectors. As shown we measure both $w_{g+}$ and $w_{++}$, and recompute the covariance matrices with the appropriate densities. For reference, the dark red crosses also show the equivalent satellite/central red galaxy $w_{g+}$ corrlations from KiDS$\times$GAMA here (c.f. \citealt{johnston18} Fig. 7, red points/band). 
On large scales at least, our \tng~red sample is consistent with their measurements. 
There are a few interesting features here to note, however. Firstly, we see a relatively strong red galaxy 1h contribution on scales $<1\mpc$. Although the general trends match the real data, with $ss$ and to a lesser extent $cs$ exhibiting strong scale dependent IAs in this regime, the magnitude is somewhat higher in our sample.  
This is particularly interesting, given that our sample characteristics are similar ($\langle L \rangle /L_0 = 0.91$ and $0.34$ for our red and blue samples respectively, compared with their $\sim0.99$ and $0.50$).
As discussed briefly in Section \ref{sec:results:colour_split}, we observe a persistent non-zero IA signal in blue galaxies on large scales; here we can see it is dominated by the $cc$ correlation, with a smaller contribution from $sc$. 
Also notable is that, contrary to what has often been assumed, the large scale satellite correlations do not appear to vanish on large scales. Focusing on the right hand panels, the purple and pink points are consistently positive and non zero. While small compared with the red central terms, and consistent with the dark red points from GAMA, there appears to be a detectable signal at the precision allowed by \tng. 

Now considering the lower panel, we see shape-shape correlations involving satellites do indeed appear to be zero on large scales, irrespective of colour. Indeed, the large scale $w_{++}$ is driven primarily by the $cc$ component, with all other subsets of the data apparently consistent with null signal at $>6\mpc$. As before, we see no significant 1h $cc$ term, down to $\sim 0.1 \mpc$ (a result which should be true by construction, since each halo contains only one central galaxy).

\section{Direct IA Constraints from 3D Fields}\label{sec:daff}
\label{sec:daff}

In earlier sections, we set out an analysis based on measuring and modelling the two-point functions of intrinsic 
galaxy shapes. This is the most common method for deriving information about intrinsic alignments from data, be it simulated or real (see e.g. \citealt{hirata07,singh15,chisari15}).
This section outlines an alternative approach, which exploits the fact that cosmological simulations allow direct access to the underlying matter field. The basic idea is that, with a suitable choice of smoothing scale, 
one can measure the components of Eq.~\eqref{eq:theory:expansion} directly and perform a linear fit to obtain constraints on the various amplitudes. Bypassing two-point correlations in this way has several advantages, not least that
it is potentially less susceptible to noise.

The method for obtaining the tidal tensor and the intrinsic shape field is described in Section~\ref{sec:measurements:tidal}. In brief, the process involves pixelising the simulation volume at given redshift, and so building smoothed 3D shape and density fields. We can then compute the tidal field by Fourier transforming the density field (see Eq. \eqref{eq:fourier_tidal}). One important thing to bear in mind is that we are free to choose the pixel scale, a choice which has some bearing on the physical interpretation of the result. 

With these ingredients in hand, we can proceed to fit for the amplitudes in Eq.~\eqref{eq:theory:expansion}.
By varying $C_1,C_2$ and $b_{\rm TA}$, we seek to minimise

\begin{multline}
\chi^2(\mathbf{p}_{\rm IA}) =  
\sum_{i,j,p} \\
\left [ 
\gamma^{I}_{ij,p} - 
\gamma^{I, model}_{ij,p}(\mathbf{p}_{\rm IA}) \right ] 
\mathbf{C}_{ij,p}^{-1} 
\left [
\gamma_{ij,p}^{I} - 
\gamma_{ij,p}^{I,model}(\mathbf{p}_{\rm IA}) \right ].
\end{multline}

\noindent
Here the IA model parameters are $\mathbf{p}_{\rm IA} = (C_1,C_2,b_{\rm TA})$. The indices $i,j$ indicate an element of the $3\times3$ shape tensor, and $p$ identifies a pixel, within which galaxy shapes are averaged. 
The theory prediction $\gamma^{I, model}_{ij,p}$ is obtained by evaluating Eq. \eqref{eq:theory:expansion}.
We will refer to this technique for constraining IA parameters, in contrast to the earlier two-point methodology, as Direct Alignment Field Fitting (DAFF).  
As in the two-point analysis, the likelihood sampling is performed using \blockfont{CosmoSIS}, using \blockfont{Multinest}; one can find the modules for this here: 
\url{https://github.com/ssamuroff/direct\_ia_theory/tree/master/likelihood/field\_fit}.

\begin{figure}
\includegraphics[width=\columnwidth]{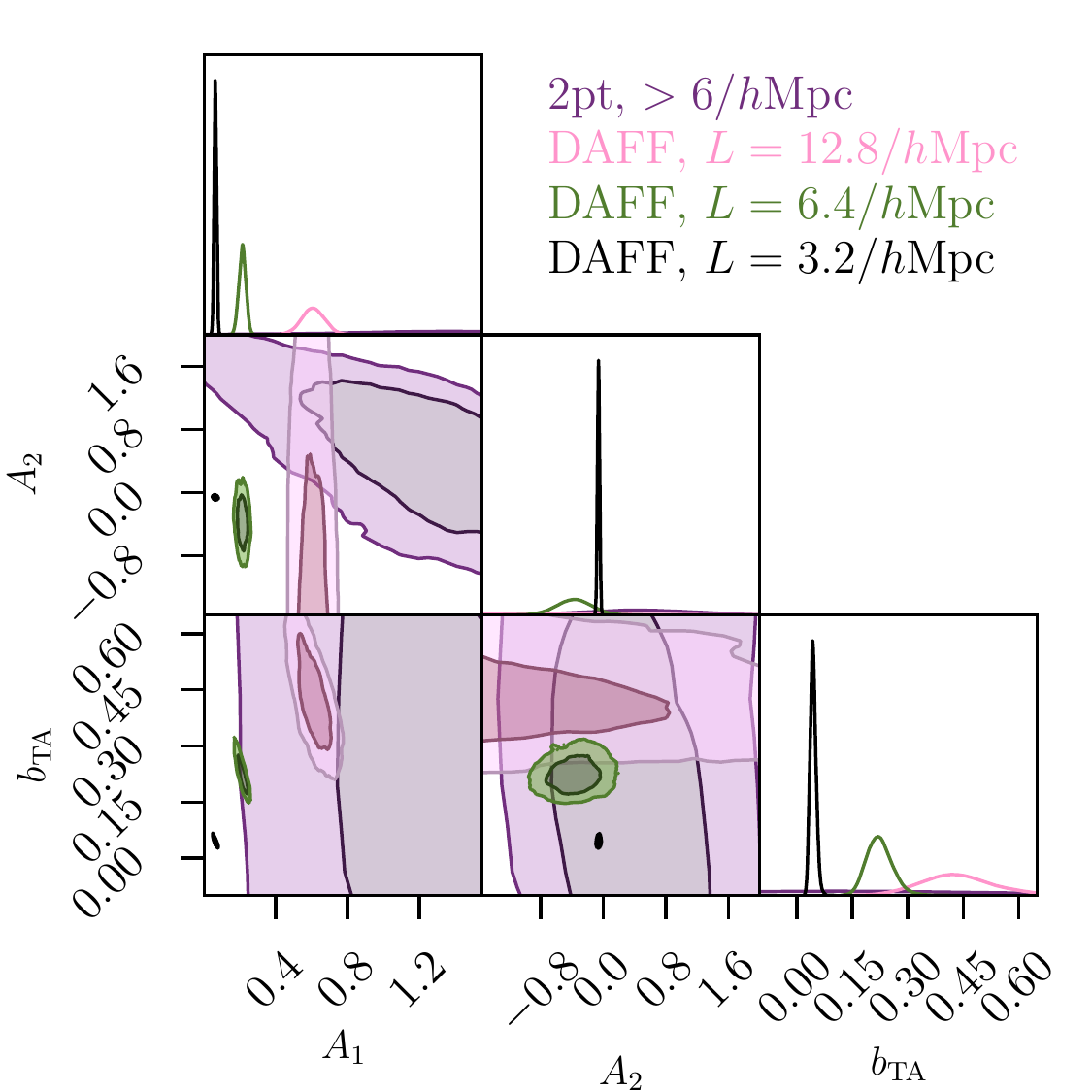}
\caption{Posterior IA parameter constraints obtained via the DAFF
fits described in Section \ref{sec:daff}.
Shown are results using three different pixel scales (indicated in the legend);
for reference, we also include the equivalent 1D constraints from
$w_{gg}+w_{g+}+w_{++}$, with a scale cut-off at $r_\mathrm{p}>6 h^{-1}$ Mpc.
}\label{fig:1pt:cornerplot}
\end{figure}

Our DAFF TATT constraints, using a range of pixel scales, are presented in Figure~\ref{fig:1pt:cornerplot}. For reference, the light purple contours also show the equivalent $z=0$ TATT model posterior from the \tng two-point analysis. 

Notably, on smaller smoothing scales particularly, there is a significant gain in signal-to-noise. Although, perhaps unsurprisingly, $L>12\mpc$ offers little information on the higher order IA contributions, with a suitable choice of scale, the TATT posterior volume is reduced quite considerably. Although this is highly promising in terms of the DAFF method's future utility, we should point out a few caveats in the comparison. 

It is perhaps worth remembering here that 
these are not independent datasets. The underlying galaxy field, 
and the shape noise are the same in each, albeit smoothed on different scales. 
This is also true of the matter tidal field. For such comparisons, it is difficult to gauge the significance of parameter shifts, given that the confidence contours do not account for these correlations.  

The second consideration, which
muddies the comparison, is that of the scales probed. The earlier analyses of
$w_{g+}$ and $w_{++}$ have an explicit window of sensitivity determined by our choice of
scale cuts, $6<\rp<68\mpc$. Within that window, however, all scales are fit simultaneously 
(albeit with unequal weight). The TATT implementation used in the two-point fits does not set an explicit smoothing scale, which some previous incarnations of NLA have, to suppress galaxy-scale fluctuations; rather the filter is included as an implicit element of the model. The various IA amplitudes are effectively renormalised to account for the impact of small scale processes on mid-to-large scale modes (see \citealt{blazek17}, Section~F for discussion).
The DAFF approach, in contrast, \emph{does} include a smoothing scale, in a way that is inherent and unavoidable. The various fields are explicitly pixelised, and averaged on a fixed scale with cubic 
pixels\footnote{The use of cubic pixels is an explicit modelling choice in our DAFF pipeline. 
One could conceivably apply e.g.\ Gaussian smoothing on top of the pixelisation.}.
While this allows some level of control over the physical scales probed, it makes direct comparison with two-point results difficult. 

One can, and people historically have, adopt a method similar to DAFF in the case of
galaxy bias, as discussed in some detail by \citet{desjacques18} (see Section~4.2, pages 85-93).
In order to correctly interpret the results there are 
corrections of the order of $\sigma^2_L$,
(i.e. the variance of linear density field on scale $L$),
which convert between  
an N-point bias, and that of the moments/scatter.
These corrections are complicated to compute, and are the focus of ongoing work. While it is necessary to have
a robust estimate of these terms in order to use the DAFF method to make constraints to a precision of better than a factor of a few, we set out here only to present a proof of concept, and so defer
calculation of these (order of unity on scales $\sim 6\mpc$) additional terms to a future work. 
That said, these corrections should alter both the centering of the IA posteriors and the width by roughly the same factor, such that the signal to noise is approximately conserved. Based on this reasoning, we expect an improvement in the signal-to-noise (posterior mean divided by the $1\sigma$ marginalised error) on $A_1$ of a factor of $\sim 4$ relative to the two-point constraints. 

With these caveats firmly in mind, it is apparent that when one goes to smaller smoothing scales, 
the favoured $A_2$ starts to deviate from zero. While the level of significance
is still only $\sim 2-3\sigma$ at most, this pattern makes sense.

\section{Discussion \& Conclusions}\label{sec:conclusion}
We have presented a detailed study of galaxy intrinsic alignments using two-point measurements from three of the most recent public hydrodynamic simulations. 
After halo mass reweighting of the samples to simplify the comparison, we find \tng~and \illustris~agree well within their respective uncertainties, although \mbii~favours a somewhat stronger IA signal.
Our key results are summarised below.
\begin{itemize}
  \item{We analysed each sample using the NLA model, which assumes linearity in the tidal field. All three of the simulations consider show strong evidence for a non-zero NLA amplitude. The results from the three are consistent within $\sim2 \sigma$, although \mbii~consistently displays a slightly stronger IA signal relative to the other two across all snapshots. }
  \item{For the first time, we fit the more complex TATT model to these various simulated data sets. 
  On scales $\rp>6\mpc$ we find no clear indication of $A_2, b_{\rm TA} \neq 0$ in \tng~($A_2=0.4\pm0.6$), and so no strong evidence for deviation from the NLA scenario, at least within the (relatively large) statistical uncertainty of the measurement.  
  \mbii, on the other hand, shows a mild preference for negative values ($A_2=-2.3\pm1.0$). 
  In all cases, the best fitting $A_1$ from the TATT fits is consistent with the amplitude from the NLA only fits on the same data, albeit with greater uncertainty. There is also some level of degeneracy between the two TATT amplitudes, such that combinations with non-zero $A_2$, combined with a slightly reduced but still non-zero $A_2$ are also allowed. } 
  \item{ We discussed a series of fits extending to smaller scales. We justified this by comparing galaxy-galaxy and matter-matter correlations, finding the linear bias assumption to hold in our case down to $\sim1\mpc$. Even with these relaxed cuts, we do not report a statistically significant detection of $A_2$ or $b_{\rm TA}$ in \tng.}
  \item{We presented a colour split IA analysis on \tng, an exercise enabled by the relatively realistic bimodal colour distribution it exhibits. As expected, the red sample displays a strong alignment signal across redshifts. Our results on blue galaxies are consistent with observations at low redshifts; at higher redshifts, $z>0.5$, where direct constraints on real data are lacking, we detect a non-zero IA signal $A_1\sim2$. We explored the origins of this blue IA feature, reporting that it persists even in relatively faint blue subsamples (down to $M_* < 1.9\times10^9 h^{-1}\mathrm{M}_\odot$), and is generated almost entirely by blue centrals. }
  \item{We examined the luminosity dependence in red galaxies, reporting results consistent with the those of \citet{johnston18}; the red \tng~sample favours a marginally shallower slope than \citet{joachimi11} and \citet{singh15}, whose fits are dominated by brighter galaxies. We have also reported the constraints on the luminosity dependence of the quadratic tidal torquing amplitude $A_2$. In the TATT fits, $A_1$ and $A_2$ exhibit consistent luminosity evolution, which is significant at the level of $1-2\sigma$.}
  \item{We presented a similar exercise for blue galaxies, which extend into the faint regime where there is a relative paucity of constraints from real data. Our fits prefer a power law index $\beta_1\sim0.24$, which is consistent with the equivalent fit to the red sample. In the TATT parameter space the constraining power is degraded, yielding $\beta_1$ and $\beta_2$ values consistent with no luminosity dependence.} 
  \item{We also fitted disaggregated central and satellite correlations. On large scales, our fits favour a weak but non-vanishing, satellite shape alignment signal. Centrals show a stronger signal, and a similar trend with redshift to the mixed sample. The TT and density weighting components of the TATT model do not differ systematically between satellites and centrals, with both $A_2$ and $b_{\rm TA}$ consistent with zero. At the correlation function level, we see the satellite alignment signal is dominated by the red $cs$ correlation; it persists in the mixed sample, on $\rp>6\mpc$, albeit subdominant by a factor of several to the red central terms. }
  \item{We have outlined a new method for recovering intrinsic alignment information from hydrodynamic simulations, which we refer to as DAFF. Although the results depend significantly on the choice of smoothing scale (in part because we have omitted small corrections, the computation of which is left for future work), the approach potentially offers a significant boost in constraining power relative to an equivalent two-point analysis, and greater control over the physical scales probed.}
\end{itemize}

\noindent
This work is one of a relatively small number that focus on deriving parameteric IA constraints from hydrodynamic simulations \citep{codis15a, chisari15, tenneti15, hilbert17, bhowmick19};
it is the first to attempt a comprehensive analysis of directly comparable samples from multiple simulations, including the current state of the art (\tng). Unlike most previous studies, we perform a simultaneous analysis, modelling a joint $w_{gg}+w_{g+}+w_{++}$ data vector. Our analysis also includes an analytic covariance matrix, which is both numerically stable and avoids the (potentially limiting) assumptions of internal estimators such as jackknife.

Although our findings are a building block in our understanding of the behaviour of large scale galaxy alignments, we urge caution in applying our findings directly to cosmological measurements.
That is, our samples are comparable with each other, but are not tailored to match the more complex selection redshift-dependent function of a typical lensing shape sample used in cosmic shear and galaxy-galaxy lensing measurements. Another useful exercise would be to use the observed trends with luminosity, colour and galaxy type to extrapolate out a mock IA signal, more representative of the contamination in real lensing data; in turn, this can be used to test our IA models in a cosmological context. This is a relatively straightforward extension of the results presented here, and is the focus of future work.  

The novel DAFF method, which does not involve measuring two-point correlations, is to our knowledge the first implementation in the literature. Although an analogous idea exists in the literature for galaxy bias (see \citealt{desjacques18}, Sec~4.2 and the references therein), it has never been discussed in the context of IAs before. The analysis on the $z=0$ \tng~snapshot should be seen as a proof-of-concept exercise; while promising, there are still gaps in our interpretation (see Section~\ref{sec:daff}), which are the subject of ongoing work, but beyond the scope of the current 
paper.

It is now well established that intrinsic alignments exist in the Universe, and must be accounted for at some level to avoid biasing cosmological analyses based on cosmic shear and galaxy-galaxy lensing. IAs have been included in cosmic shear analyses for as long as shear has been a competative cosmological probe \citep{heymans13,svcosmology,jee16,y1cosmicshear,hildebrandt18,hikage19,hamana19,chang19,asgari20}. Only recently, however, have the lensing data been of sufficient volume to potentially incur biases due to model insufficiency (see \citealt{krause15},
and the Stage IV forecasts of \citealt{fortuna20} and the upcoming tests in the context of DES Y3 \citealt{secco20}; see also \citealt{joachimi20} for an interesting counter discussion). Developing a fuller understanding of intrinsic alignments, then, will be crucial for, arguably, the current generation of cosmological surveys, and certainly the next. 
The current paper is one small step in this direction, providing the first detailed analysis at the level of model constraints on the best available cosmological hydrodynamic simulations. Our results, of course, come with a number of caveats. Most notably, our selection function is not intended to accurately match current or future lensing surveys. This is in part because recreating the complex redshift-dependent selection function in a real lensing sample, which would typically be based on a number of correlated observables, is a difficult task;
it is also, however, a function of our aim in this study. We wish to understand the behaviour of IAs at a physical level, in order to feed into understanding IAs and model building efforts, rather than make a detailed prediction or robustness test for a particular survey. The behaviour of intrinsic alignments on small physical scales is an important topic for future investigation, and one that could conceivably be addressed using hydrodynamic simulations; indeed, due to the larger number of measureable modes, the signal to noise on small scales is relatively high. 
The TATT approach allows some hope of pushing to smaller scales (though not into the regime of $<\sim 1 \mpc$, where one would need an explicit model for 1h alignment contributions). Unfortunately, a number of other poorly-understood effects enter on small scales, particularly nonlinear galaxy bias and baryonic physics. In order to pursue IA constraints on such scales, it is likely that one would need to consider both higher-order bias terms and the interplay with the higher-order IA terms.

\section{Acknowledgements}

We are grateful to Duncan Campbell, Maria Cristina Fortuna, 
Fran\c{c}ois Lanusse, Ami Choi, Scott Dodelson
and the DES WLWG 
for valuable discussions at various points in the writing of this paper.  
We would also like to thank Sukhdeep Singh for lending his
code to help validate our pipeline and Harry Johnston for 
kindly allowing us access to his KiDS+GAMA measurements. 
This work made use of the Coma cluster, which is hosted
by the McWilliams Center, at CMU.
We also used resources of the National Energy Research Scientific Computing Center (NERSC), a U.S. Department of Energy Office of Science User Facility operated under Contract No. DE-AC02-05CH11231.
SS and RM are supported by the US National Science Foundation (NSF) 
under Grant No.\ 1716131.

\bibliographystyle{mnras_2author}
\bibliography{refs}

\appendix

\section{Pipeline and Covariance Matrix Validation}\label{app:validation}

The likelihood pipeline used in this work is built from public code,
developed within the \blockfont{CosmoSIS} framework.
During the process of developing this code base, we implemented a series
of validation exercises, intended to ensure our results are both accurate 
and repeatable.

\begin{figure}
\includegraphics[width=1\columnwidth]{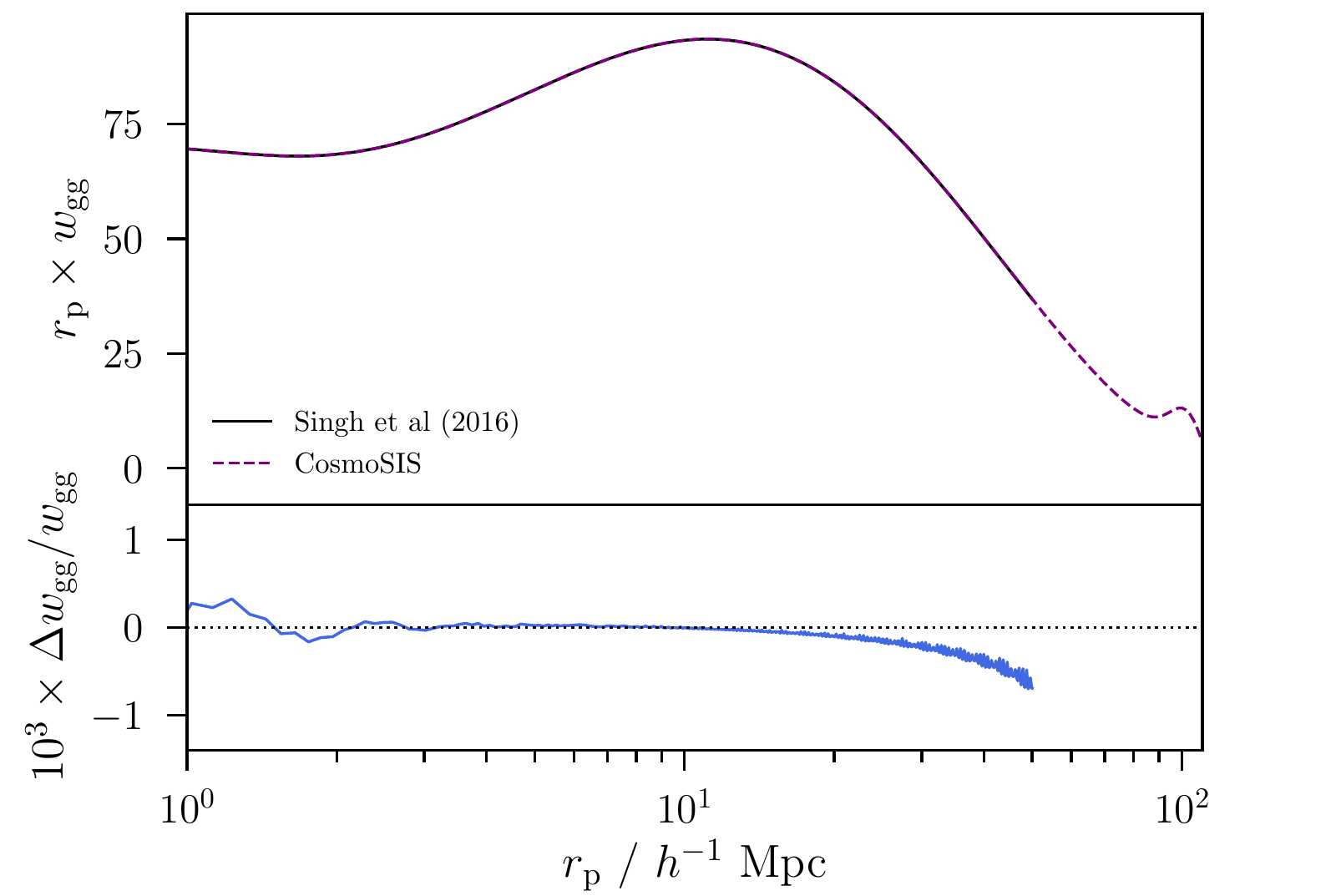}
\includegraphics[width=1\columnwidth]{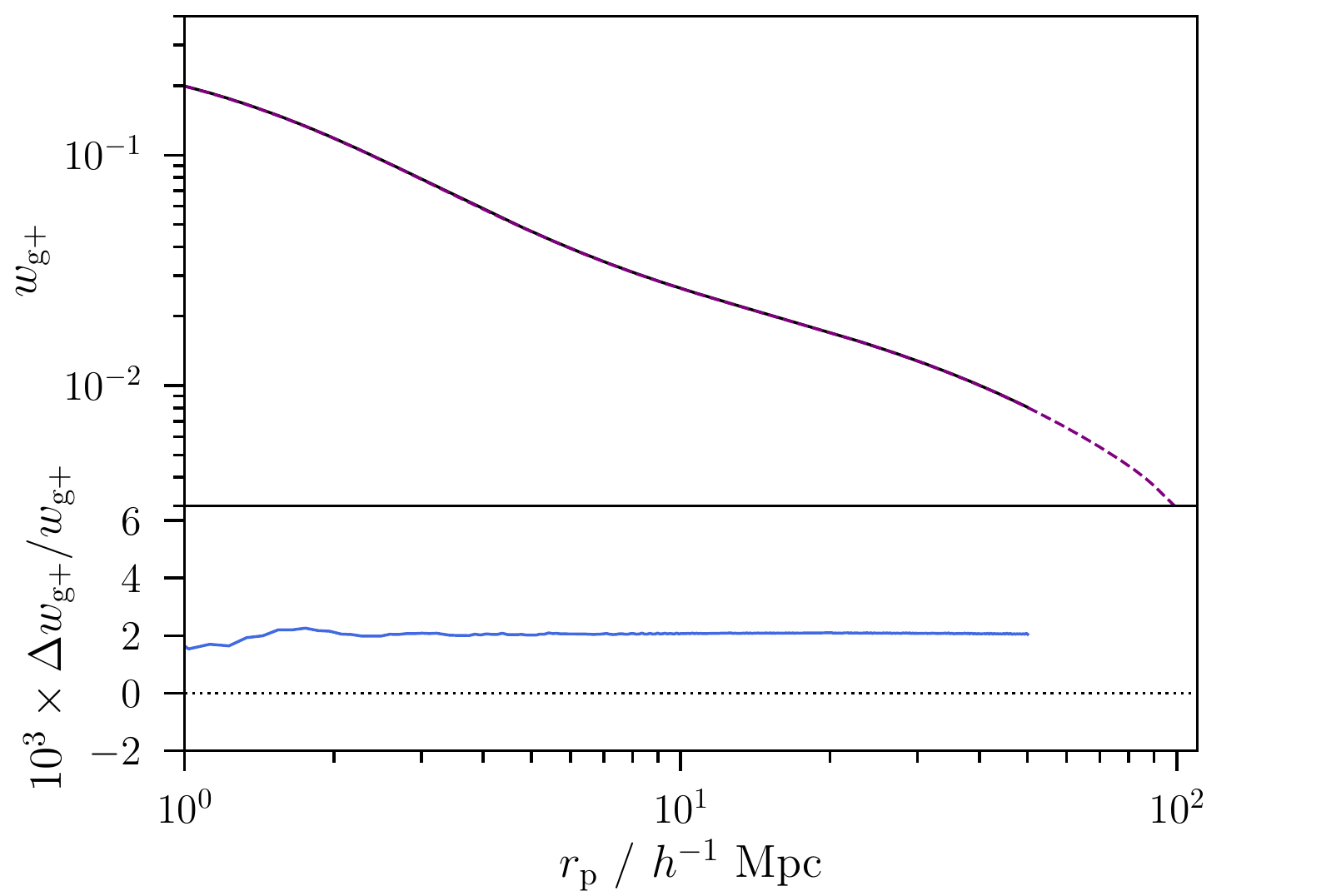}
\includegraphics[width=1\columnwidth]{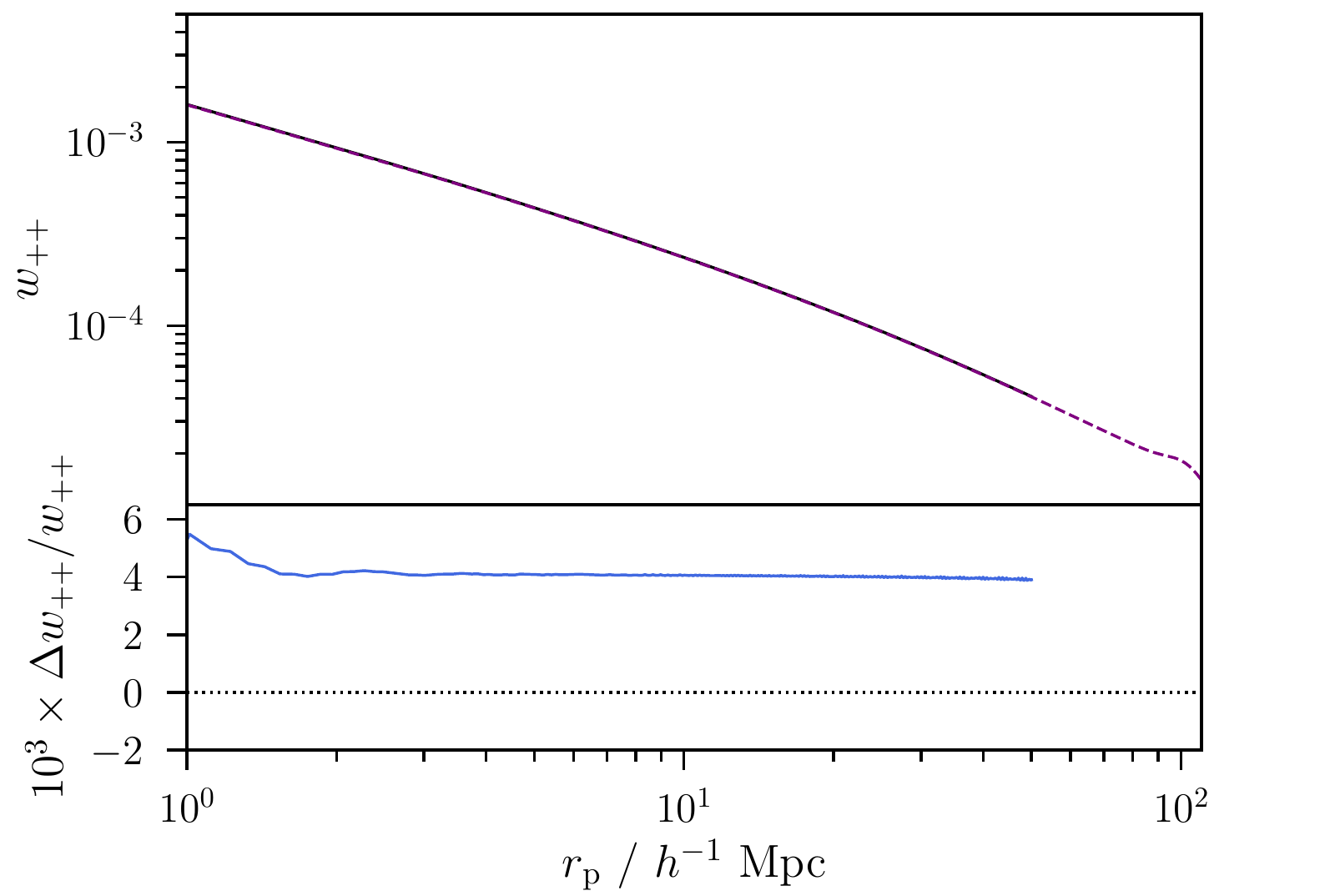}
\caption{A comparison of theory data vectors produced by two independent codes.
The dashed purple and black (solid) lines show the outputs 
of the \blockfont{CosmoSIS} module produced for this work, and an external
theory code used in \citet{singh15}.
In the lower panel we show the fractional residual between the two.  
}\label{fig:validation:datavecs}
\end{figure}

The first step in this process is a data vector-level comparison between different
theory codes.
We generate a nonlinear matter power spectrum using \blockfont{CosmoSIS}, 
which is then fed into (a) our theory pipeline, which is used
for inference in this work,
and (b) an external code developed by an independent group, and used in
\citet{singh15}. The two codes produce projected correlations
$w_{gg}(r_{\rm p})$, $w_{g+}(r_{\rm p})$ and $w_{++}(r_{\rm p})$.
The $r_{\rm p}$ sampling is slightly different, and so we interpolate to
a comparable set of values.
The result is shown in Figure \ref{fig:validation:datavecs};
we can see here that the two agree relatively well.
Though the residuals in the two IA correlations are non-zero
and roughly scale-independent, the difference is comfortably within $\sim0.5\%$. 

Though it is reassuring that the two codes are consistent with each other
at some (relatively sensible) set of input parameter values,
this is not in iteslf a rigorous demonstration that our pipeline is unbiased.
Using the \citet{singh15} code we then generate a fiducial data vector, $w_{gg}$, $w_{g+}$, $w_{++}$ at four redshifts
$z=(0.0,0.30,0.625,1.00)$. Using these mock data, we run our inference pipeline with the fiducial
(analytic) covariance matrices obtained through the process described in Section \ref{sec:measurements:cov}. We report that we can recover the input parameters to comfortably within $0.5\sigma$.

We also compare our analytic covariance matrix with an alternative, obtained by jackknife resampling. The jackknife covariance is generated by dividing the \tng~box into $4^3=64$ sub-volumes, and iteratively remeasuring our data vector. Although the comparison is useful as a cross test, it is worth bearing in mind that the jackknife estimator relies on a number of assumptions that do not strictly hold in our case \citep{hartlap07}. That is, although order-of-magnitude differences are not expected, we have first principles reasons to trust our fiducial covariance.   

\begin{figure*}
\includegraphics[width=1.5\columnwidth]{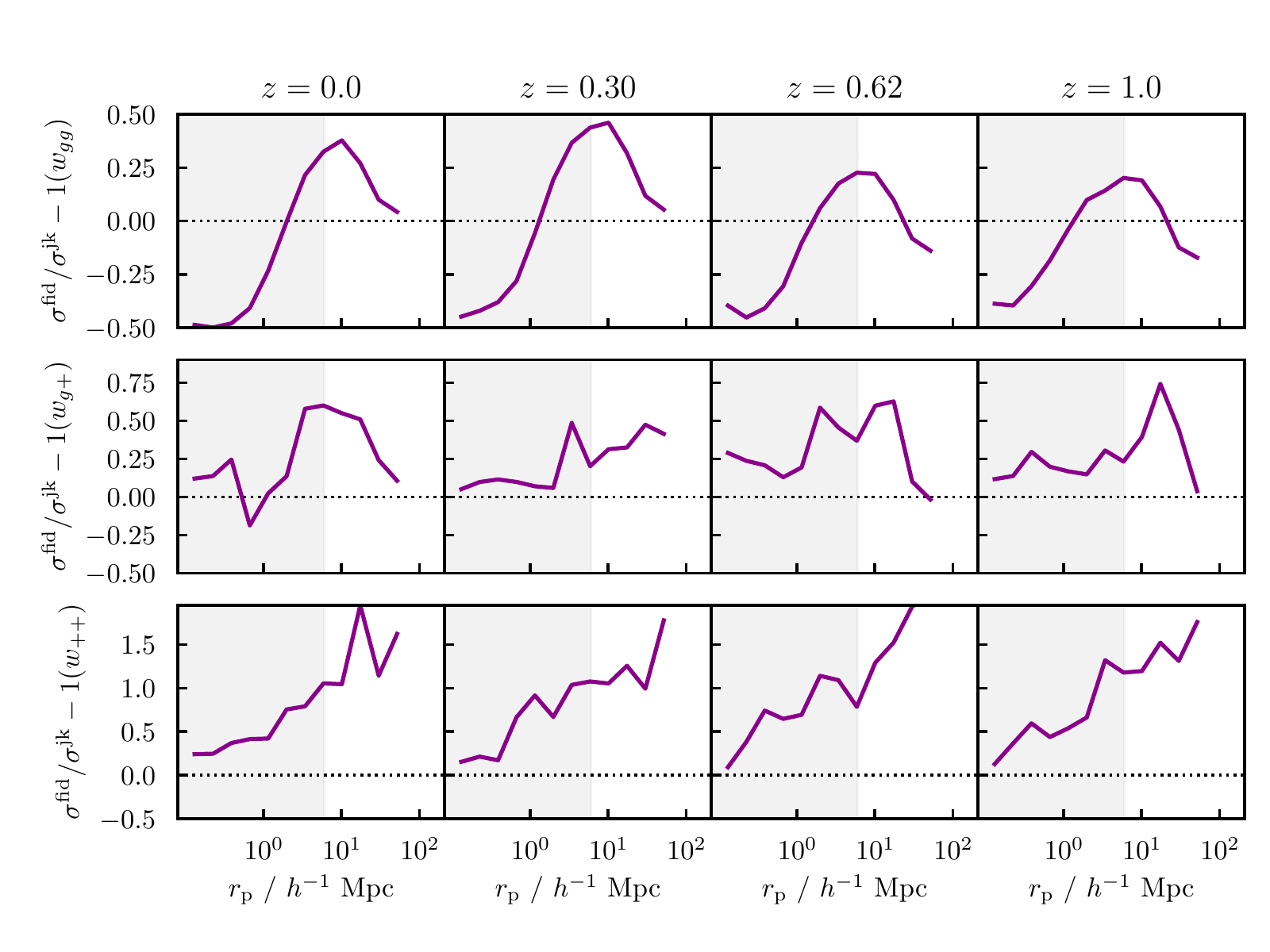}
\caption{A comparison of jackknife and analytic covariance matrices for \tng.
Here we show the square root of the diagonal of the two covariances, for
four snapshots and three two-point functions, as labelled. 
}\label{fig:app:covariance_comaprison}
\end{figure*}

The numerical comparison of the diagonals can be seen in Figure~\ref{fig:app:covariance_comaprison}. 
As can be seen, the differences are significant, on all scales considered. In most cases 
(all but $w_{gg}$ on small scales), jackknife tends to underestimate the uncertainties at the level
of $25\%$ or more.    

\section{Posterior Constraints from \mbii~\& \illustris}\label{app:posteriors}

\begin{figure}
\includegraphics[width=\columnwidth]{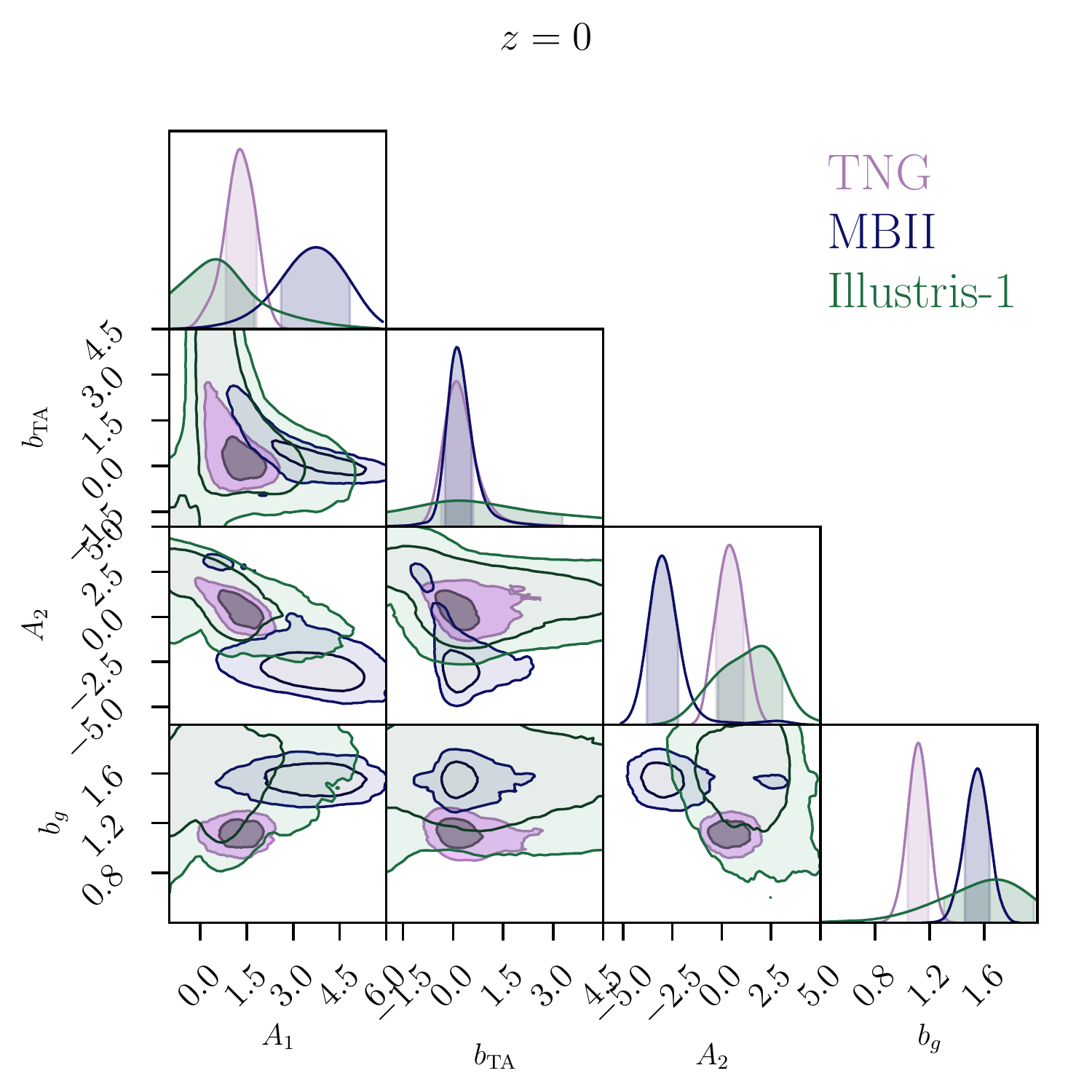}
\caption{Posterior TATT model constraints from the lowest redshift snapshot
of \tng, \mbii~and \illustris. 
}\label{fig:app:post1}
\end{figure}

\begin{figure}
\includegraphics[width=\columnwidth]{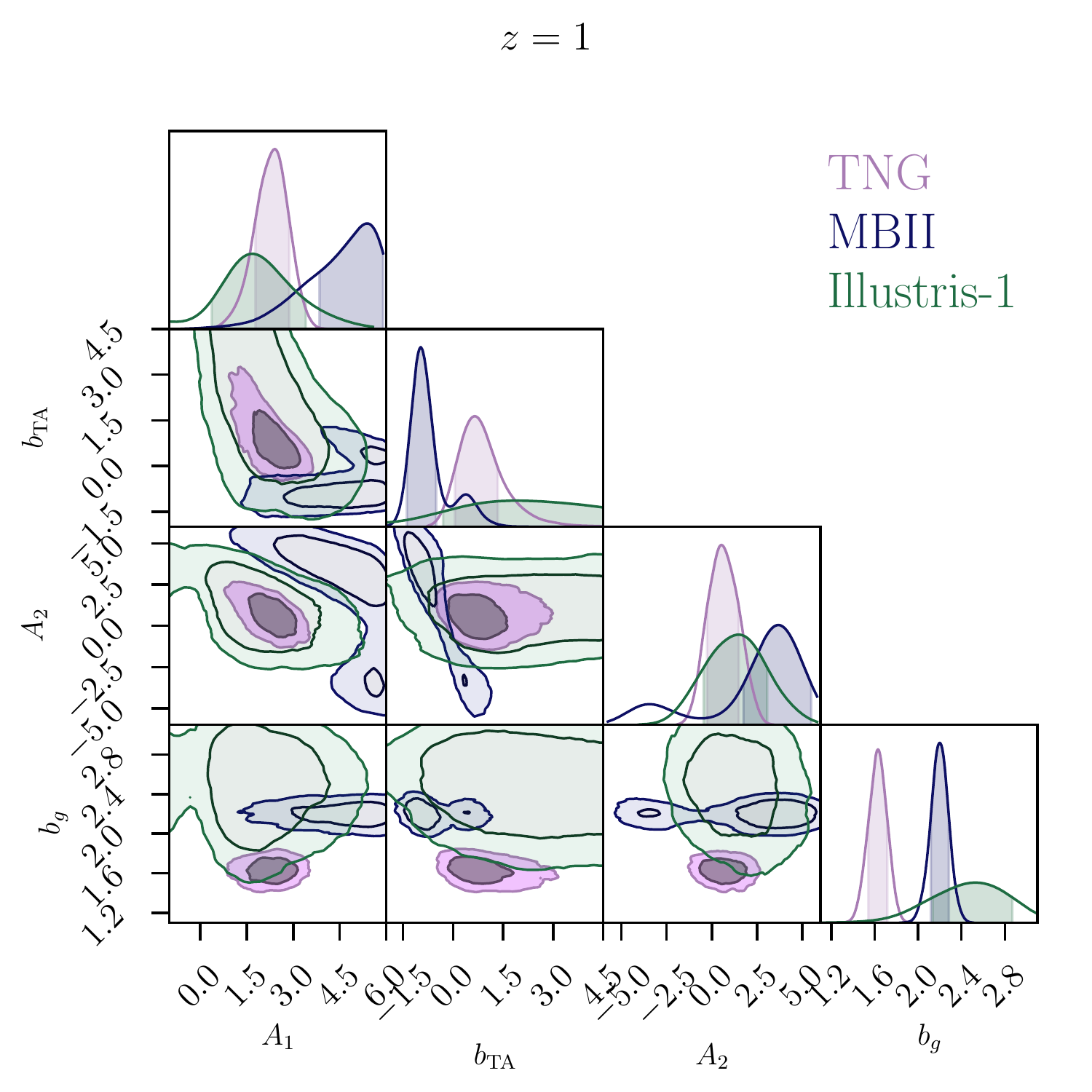}
\caption{The same as \ref{fig:app:post1}, but at $z=1$.
}\label{fig:app:post2}
\end{figure}

In this appendix we present the full posterior constraints on our three simulated samples.
In the main body of this work we presented only a selection of these to 
emphasise our most interesting findings. 
For completeness, they are shown in Figures~\ref{fig:app:post1} and \ref{fig:app:post2}.
This represents the baseline TATT analysis on our three simulations at $z=0$ and $z=1$. 
As described, the fiducial analysis has four free parameters ($A_1,A_2,b_{\rm TA},b_g$),
includes the joint data vector $w_{gg}+w_{g+}+w_{++}$, 
and scale cuts $\rp>6\mpc$ for all correlation functions.

\begin{table}
\begin{tabular}{ccccc}
\hline
Simulation        & Redshift      & $A_1$ & $A_2$ & $b_{\rm TA}$    \\
\hline
\hline
TNG               &  $0.0$        & $1.29\pm0.49$ & $0.32\pm0.65$ & $0.21\pm0.86$   \\
TNG               &  $0.3$        & $1.76\pm0.51$ & $0.65\pm0.76$   & $0.15\pm0.55$    \\
TNG               &  $0.62$       & $1.64\pm0.50$ & $0.65\pm0.79$   & $0.55\pm0.80$    \\
TNG               &  $1.0$        & $2.26\pm0.53$ & $0.62\pm0.80$   & $0.85\pm0.75$    \\
MBII              &  $0.0$        & $3.57\pm1.07$ & $-2.76\pm1.14$  & $0.26\pm0.63$    \\
MBII              &  $0.3$        & $3.24\pm1.27$ & $0.47\pm1.82$   & $-0.28\pm0.75$   \\
MBII              &  $0.62$       & $4.09\pm1.15$ & $1.85\pm2.26$   & $-0.94\pm0.58$   \\
MBII              &  $1.0$        & $4.53\pm1.15$ & $2.44\pm2.73$  & $-0.70\pm0.61$    \\
Illustris         &  $0.0$        & $0.48\pm1.47$ & $1.29\pm1.49$   & $-0.05\pm2.83$   \\
Illustris         &  $0.3$        & $3.07 \pm1.77$ & $-0.40\pm1.53$   & $-1.03\pm1.24$   \\
Illustris         &  $0.62$       & $1.03\pm2.11$ & $2.10\pm1.72$   & $-0.09\pm3.62$    \\
Illustris         &  $1.0$        & $1.32\pm1.90$ & $1.34\pm1.61$   & $1.34\pm3.14$    \\
\hline
\end{tabular}
\caption{The best-fitting TATT parameters and $1\sigma$ posterior uncertainties from all 
samples/redshifts considered in this work.
}\label{tab:results:fullpost}
\end{table}

\section{Impact of Galaxy Weights}\label{app:weights}
To allow a meaningful comparison of galaxy samples from the three simulations included in this paper,
we derive a set of galaxy weights for our \mbii~and \illustris~catalogues. The idea here is to 
weight the galaxy samples such that the distribution of host halo masses match
exactly. The weight assigned to galaxy $i$ from simulation $X$ is then: 

\begin{equation}
w^{X}_i = p^{\rm TNG}(M^j_{h}) / p^{X}(M^j_{h}),
\end{equation}

\noindent
where $p^{\rm X}(M_{h})$ is the normalised histogram of host halo masses in
simulation $X$, and $j$ is a mass bin to which galaxy $i$ belongs.

\begin{figure}
\includegraphics[width=\columnwidth]{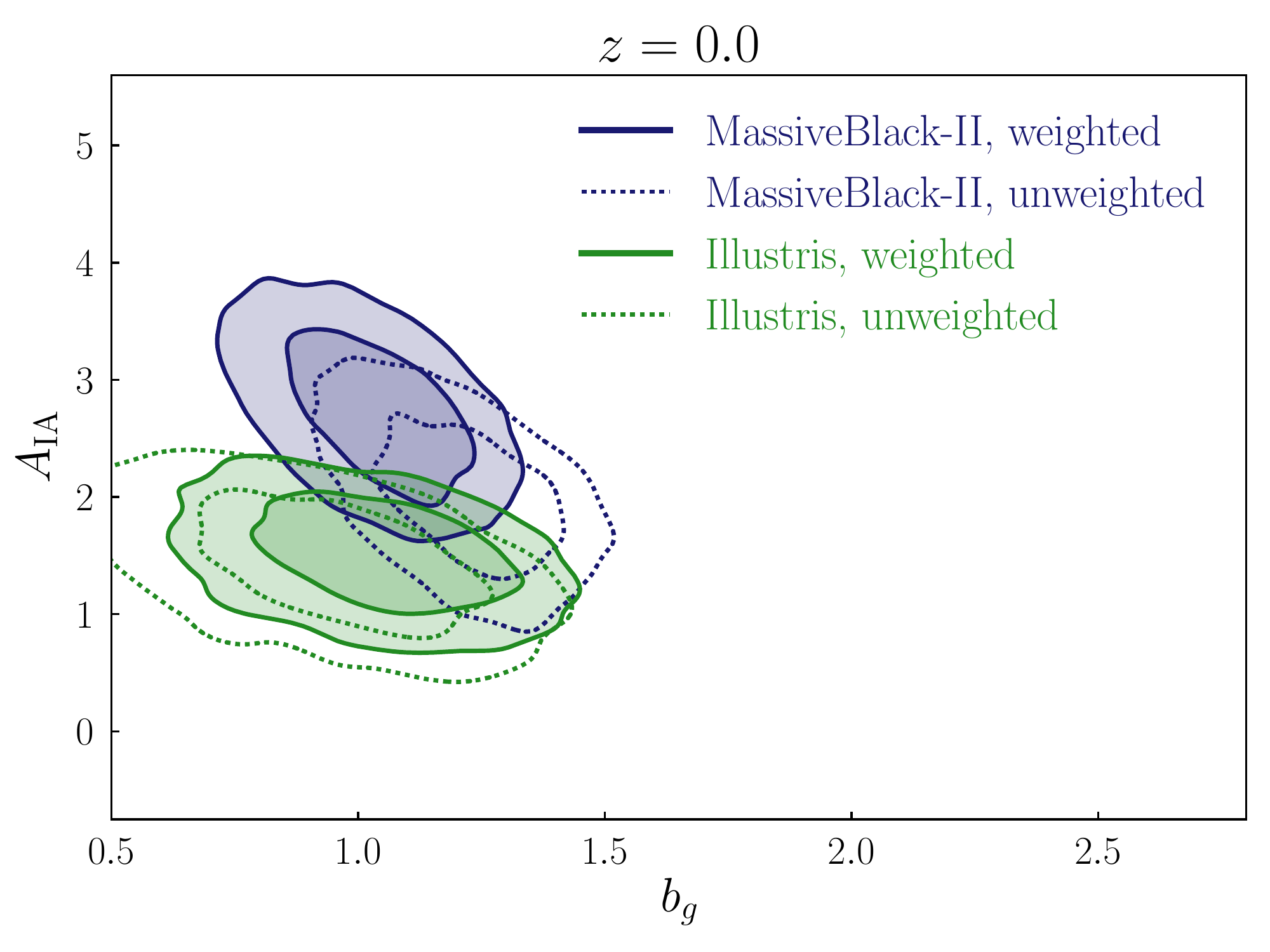}
\includegraphics[width=\columnwidth]{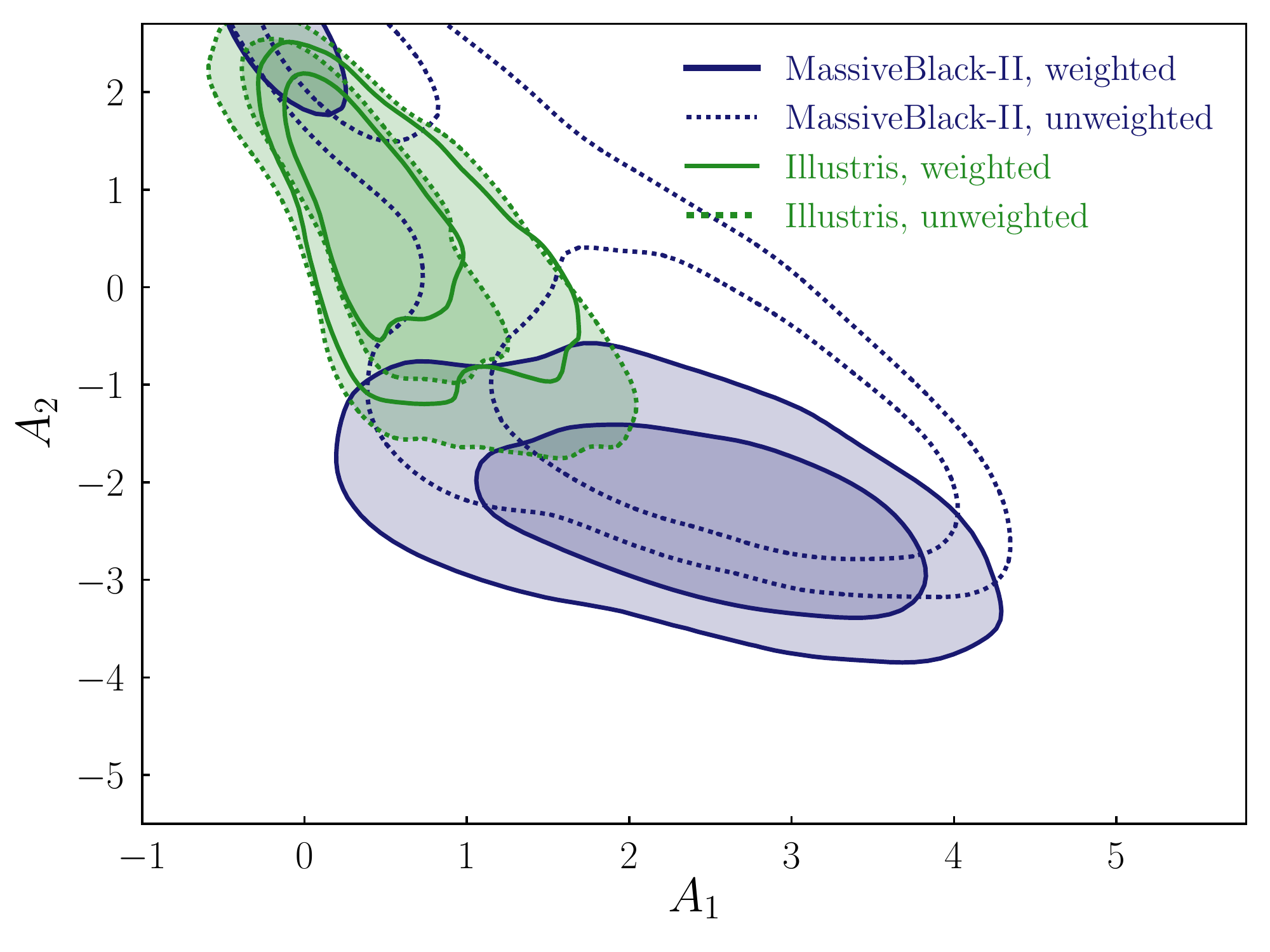}
\caption{Demonstration of the impact of galaxy reweighting on the 
posterior IA constraints.
\emph{Upper}: The $z=0$ constraints on the NLA model amplitude and linear
galaxy bias from \mbii~and \illustris, with and without weighting.
By construction \tng~is unaffected by weighting.
\emph{Lower}: The same, but using the TATT instead of NLA model.
}\label{fig:app:weights}
\end{figure}

The impact of this weighting on our IA constraints is shown in
Figure \ref{fig:app:weights}. 
Although for clarity we show only the contours from the lowest
redshift, we find very similar behaviour in the other three snapshots. 
It has been established elsewhere that there is a relatively tight relation 
between host halo mass and bias, at least on large physical scales;
it is, then, intuitively correct that the weighting should bring the 
galaxy bias constraints (upper panel) from the two simulations
into relatively close agreement.

\section{Validity of the Linear Galaxy Bias Approximation}\label{app:bias}

\begin{figure}
\includegraphics[width=\columnwidth]{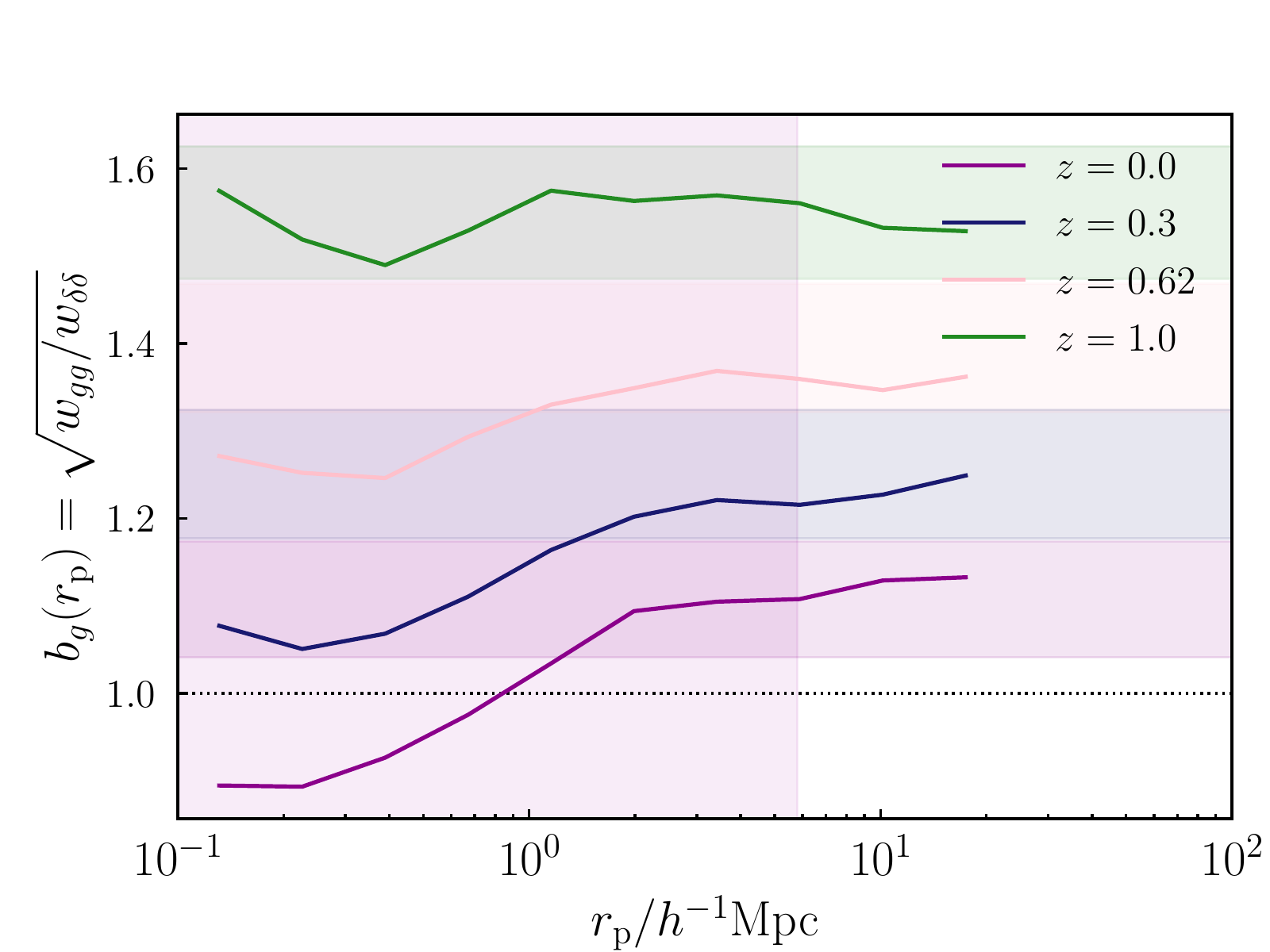}
\caption{Galaxy bias as a function of physical scale. 
The bias is estimated as the ratio of the matter-matter and galaxy-galaxy
projected correlations. The horizontal shaded bands show the best fitting
linear bias values and the $1\sigma$ uncertainties, as obtained from
fits to the large scale $w_{gg}$ correlations.
}\label{fig:app:bias}
\end{figure}

Since our perturbative TATT model includes higher-order terms, there is some value in seeking to push to slightly smaller scales. It is also true, however, that as one does so, one eventually enters the regime in which nonlinear galaxy bias also starts to become relevant. Such higher order bias contributions, and the cross-IA terms, are complex to model and not fully implemented in our analysis; if it exists, then, we would ideally like to identify a range of scales below our fiducal cut at $\rp=6\mpc$, on which the linear bias approximation is valid (or, at least, deviations from it are subdominant to other uncertainties).

We can obtain an estimate for the effective scale-dependent galaxy bias in \tng~as the ratio of the 
galaxy-galaxy projected correlation, and the matter-matter equivalent:
\begin{equation}
b'_{g}(\rp | z) = \sqrt{\frac{w_{gg}}{w_{\delta \delta}}}.
\end{equation} 
\noindent
We refer to this as an ``effective" bias because, strictly speaking, the galaxy bias is defined in terms of the 3D density field $b_g = \delta_g/\delta$ (or equivalently in terms of 3D power spectra). Converting from 3D power spectra to projected correlations $w_{gg}$ and $w_{g+}$ involves an integral over $k$ (e.g. Equations~\eqref{eq:wgp_basic}
and \eqref{eq:wgg_basic}), and if $b_g$ is scale dependent, it no longer separates cleanly from that integral. What we measure, then, is an effective bias $b'_g$, which is not quite the same as the true 3D galaxy bias $b_g(k)$.

The discrete snapshots in the simulation allow a relatively clean
measurement of $b'_g$ at a given redshift. 
For this exercise, we use the measured $w_{gg}$ correlation in a particular 
snapshot. Although, of course, this includes some level of statistical noise, 
the signal-to-noise is relatively high.
For the matter-matter part it is sufficient to use the theory prediction at
the input \tng~cosmology.
While \blockfont{halofit} is subject to its own uncertainties on small scales
($\sim5\%$ at $k<1\mpc$; \citealt{takahashi12}), we do not expect them to affect alter the conclusions
of our approximate calculations.

The resulting effective scale-dependent bias estimates are shown in Figure \ref{fig:app:bias}.
While the fiducial cut at 6\mpc~(the pink shaded region) does indeed effectively
exclude scales on which the bias cannot be captured by a single coefficient,
we can also see that it is relatively conservative. That is, there is a region from $\rp \sim 1\mpc$ upwards, in which (within uncertainties)
the bias is linear, but which are excluded by the fiducial cut. On the basis of these results, we carry out fits (see Section \ref{sec:results:scales}) with lower
cutoffs as low as $1\mpc$. 

At $z=0$, we see that the linear bias assumption begins to break down in our \tng~sample at $1 \mpc$. 
It is worth bearing in mind that in three dimensions, since one is measuring the actual galaxy bias, rather than $b'_g$, the approximation likely becomes invalid at some larger scale. By nature, the projected correlations at given \rp~mix contributions from larger 3D separations, which bring them closer to linear bias. The breaking scale seems to shift downwards at high $z$, which is perhaps as a result of the growth of structure (i.e. bias is closer to linear at high redshift at a given \rp).

\end{document}